\documentstyle[epsf,graphics]{mn}

\newcommand{\Msolar}{\mbox{\,$\rm M_{\odot}$}}        
\newcommand{\asec}{^{\prime\prime}}
\title[Quasars, their hosts and black holes]{Quasars, their host galaxies, and their
central black holes}
\author[J.S. Dunlop et al.]
{J.S. Dunlop$^1$\thanks{Email: jsd@roe.ac.uk}, R.J. McLure$^2$, M.J. Kukula$^1$, S.A. Baum$^3$, C.P. O'Dea$^3$ \&
\newauthor D.H. Hughes$^4$ \\
$^1$Institute for Astronomy, University of Edinburgh, Royal Observatory, 
Blackford Hill, Edinburgh, EH9~3HJ, UK \\
$^2$Department of Physics, University of Oxford, Nuclear \& Astrophysics
Laboratory, Keble Road, Oxford OX1 3RH, UK \\
$^3$Space Telescope Science Institute, 3700 San Martin Drive, Baltimore,
MD 21218, USA \\
$^4$Instituto Nacional de Astrof\'{i}sica, \'{O}ptical y Electr\'{o}nica (INAOE),
Apartado Postal 51 y216, 72000 Puebla, Pue., Mexico
}

\date{Accepted for publication in MNRAS}

\begin{document}

\maketitle

\begin{abstract}
We present the final results from our deep HST imaging study of the host 
galaxies of radio-quiet quasars (RQQs), radio-loud quasars (RLQs) and radio 
galaxies (RGs). We describe and analyze new WFPC2 $R$-band observations for 
14 objects which, when combined with the first tranche of HST imaging 
reported in McLure et al (1999), provide a complete and consistent set of 
deep, red, line-free images for statistically-matched samples of 13 RQQs, 10 
RLQs and 10 RGs in the redshift band $0.1<z<0.25$. We also report the results 
of new deep VLA imaging which has yielded a 5 GHz detection of all but one of 
the 33 AGN in our sample.

Careful modelling of our images, aided by a high dynamic-range point-spread 
function, has allowed us to determine accurately the morphology, luminosity,
scale-length and axial ratio of {\it every} host galaxy in our sample. Armed 
with this information we have undertaken a detailed comparison of the 
properties of the hosts of these 3 types of powerful AGN, both internally, 
and with the galaxy population in general. 

We find that spheroidal hosts become more prevalent with increasing nuclear 
luminosity such that, for nuclear luminosities $M_V < -23.5$, the hosts of 
both radio-loud {\it and} radio-quiet AGN are virtually all massive 
ellipticals. Moreover, we demonstrate that the basic properties of these hosts 
are indistinguishable from those of quiescent, evolved, low-redshift 
ellipticals of comparable mass. This result rules out the possibility that 
radio-loudness is determined by host-galaxy morphology, and also sets severe 
constraints on evolutionary schemes which attempt to link low-z ULIRGs with 
RQQs. 

Instead, we show that our results are as expected given the relationship 
between black-hole and spheroid mass established for nearby galaxies, 
and apply this relation to estimate the mass of the black hole in each 
object. The results
agree remarkably well with completely-independent estimates based on 
nuclear emission-line widths; all the quasars in our sample have $M_{bh} > 5 
\times 10^8~{\rm M_{\odot}}$ while the radio-loud objects are confined to 
$M_{bh}>10^9~{\rm M_{\odot}}$. This apparent mass-threshold difference,
which provides a natural explanation for why RQQs outnumber RLQs by a 
factor of 10, appears to reflect 
the existence of a minimum and maximum level of 
black-hole radio output which is a strong function of black-hole mass 
($\propto M_{bh}^{2-2.5}$). Finally, we use our results to estimate the 
fraction of massive spheroids/black-holes which produce quasar-level activity.
This fraction is $\simeq 0.1$\% at the present day, rising to $> 10$\% 
at $z \simeq 2-3$.

\end{abstract}

\begin{keywords}
galaxies: active -- galaxies: photometry -- infrared: galaxies --
quasars: general -- black hole physics
\end{keywords}

\section{Introduction}

Studies of the host galaxies of low-redshift quasars can enable us to 
define the subset of the present-day galaxy population which is capable 
of producing quasar-level nuclear activity. This is of obvious importance for 
constraining physical models of quasar evolution (Small \& Blandford 1992;
Haehnelt \& Rees 1993; Kauffman \& Haehnelt 2000), and for exploring the 
connection between black-hole and galaxy formation (Silk \& Rees 1998, 
Fabian 1999, Franceschini et al. 1999, 
Granato et al. 2001, Kormendy \&  Gebhardt 2001). Such 
observations are also of value for testing unified models of radio-loud AGN 
(e.g. Peacock 1987, Barthel 1989, Urry \& Padovani 1995), constraining 
possible evolutionary links between ULIRGs and quasars 
(Sanders \& Mirabel 1996), 
exploring the origin of radio-loudness (Blandford 2000), and as a means to 
estimate the masses of the central black holes which power the active nuclei 
(McLure et al. 1999).

Our view of low-redshift quasar hosts has been clarified enormously over the 
last five years, primarily due to the angular resolution and dynamic range 
offered by the Hubble Space Telescope (HST). After some initial confusion, 
recent HST-based studies have 
now reached agreement that the hosts of all luminous quasars ($M_V < -23.5$) 
are bright galaxies with $L > L^{\star}$ (Bahcall et al. 1997,
McLure et al. 1999, McLeod \& McLeod 2001). However, it can be argued, 
(with considerable justification) that this much had already been established 
from earlier ground-based studies (e.g. Smith et al. 1986, V\'{e}ron-Cetty \& 
Woltjer 1990, Taylor et al. 1996). 

In fact, as first convincingly demonstrated by Disney et al. (1995), the major 
advance offered by the HST for the study of quasar hosts is that it allows 
host galaxies to be mapped out over sufficient angular and dynamic range for 
a de Vaucouleurs $r^{1/4}$-law spheroidal component to be clearly 
distinguished from an exponential disc, at least for redshifts  $z < 0.5$. 
This is not to suggest that AGN host-galaxy morphological discrimination has 
proved impossible from the ground. Indeed for lower-luminosity AGN at 
$z < 0.1$, such as Seyfert galaxies, ground-based imaging has proved 
perfectly adequate for this task (e.g. Hunt et al. 1999) and 
in fact some early ground-based attempts to determine the morphology of 
low-redshift quasar hosts have also proved to be robust (e.g. Smith et 
al. 1986). However, to ensure an unbiassed comparison of RQQ and RLQ hosts it 
is necessary to study host galaxies at $z > 0.15$ and to be able to determine 
host-galaxy morphologies for quasars with luminosities up to $M_V < -26$. Even 
by moving to the infrared to minimize nuclear:host ratio, Taylor et al. (1996)
found that this could not be reliably achieved with typical ground-based 
seeing.

Nevertheless, great care needs to be taken to 
extract the full benefit of HST imaging of quasar hosts. In particular, 
deep observations are 
required to detect the extended 
low surface-brightness emission of even a massive host galaxy at $z \simeq 
0.2$ to a radius of several arcsec from the nucleus.
Unfortunately however, this inevitably leads to saturation 
of the nucleus, making accurate 
characterization of the luminosity of the central source impossible. This is 
crucial because, at the depths of interest for reliable host-galaxy 
characterization, scattered light in the WFPC2 PSF still makes a significant 
contribution to surface brightness out to an angular radius 
$\simeq 10$ arcsec (McLure, Dunlop \& Kukula 2000).
As demonstrated by McLeod \& Rieke (1995), these problems of surface 
brightness bias, saturation, and inadequate knowledge of the large-angle 
properties of the true WFPC2 PSF, can explain much of the confusion 
produced by the first studies of quasar hosts undertaken 
after the correction of the HST optics with COSTAR (e.g. Bahcall, Kirhakos 
\& Schneider 1994).

In this paper we present the final results from our 34-orbit Cycle-6
imaging study of quasar hosts, which was carefully  designed to avoid these
problems. Specifically, we acquired images of each quasar spanning a wide 
range of integration times (to allow an unsaturated, high dynamic-range image 
of each object to be constructed) and devoted an entire orbit to the 
construction of the necessary high dynamic-range PSF (via observations of a 
star of similar colour to the quasar nuclei, imaged at the same location on 
the same WF chip). Results from the first half of this
programme were reported in McLure et al. (1999), where 
images for 19 objects from our 33-source sample were presented, modelled and 
analyzed. Here we present and model the images for 
the 14 targets which were observed in the latter half of 1998 and in 1999, and 
then summarize and discuss the results derived from the analysis of the 
completed sample. The results presented in this paper thus complete, extend 
and in several cases supercede those presented in McLure et al. (1999) (e.g.
estimated black-hole masses for all objects are now calculated using 
more recent estimates of the black-hole:spheroid mass relation, yielding 
significantly lower values than were calculated by McLure et al. based on 
the relation presented by Magorrian et al. (1998)).

Several other substantial studies of low-redshift quasar hosts have now been 
undertaken with the HST (e.g. Bahcall, Kirkhados \& Schneider 1997; 
Hooper, Impey \& Foltz 1997; Boyce et al. 1998, McLeod \& McLeod 2001). 
However, one unique feature of the present study is the deliberate focus on 
a comparison of the hosts of the three main classes of powerful AGN, namely 
radio-quiet quasars (RQQs), radio-loud quasars (RLQs) and radio galaxies 
(RGs). Moreover, we have ensured that this comparison can be performed in an 
unbiassed manner by confining our sample to a narrow range in redshift 
($0.1 < z < 0.25$) and requiring that the individual sub-samples are matched 
in terms of their luminosity distributions (optical luminosity in the case 
of the RQQ and RLQ sub-samples, and radio luminosity in the case of the RLQ 
and RG sub-samples - see Dunlop et al. (1993), McLure et al. (1999) and 
Section 2 for further details). Another strength of this study is the wealth 
of pre-existing data at other wavelengths, as detailed in Section 2. This 
has allowed us to maximise the impact of the HST imaging (through, for 
example, the determination of $R-K$ colours for all the host galaxies). 
Finally it is worth emphasizing that in this study we have sought to extract 
the properties of the stellar population which dominates the mass of the host 
galaxy. Thus, while we do include a statistical analysis of the prevalence of 
peculiar features such as tidal tails, we have endevoured to minimize the 
distorting effect of the more transient activity by insisting on line-free 
imaging longward of the 4000\AA\ break, and masking out obvious assymmetries 
prior to modelling the host morphology. This approach, coupled with the 
careful design of our observations, is 
the most likely explanation for why the results presented in this paper are 
generally cleaner, and more homogenous than the results of many other recent 
studies.

The layout of this paper is as follows. In Section 2 we review the main 
properties of the matched RG, RLQ and RQQ samples, and summarize the 
wealth of supporting ground-based
multi-frequency data which now exists for these objects. In 
Section 3 we give details of the HST observations
and briefly review the process of data reduction and PSF determination 
(McLure et al. 1999; McLure, 
Dunlop \& Kukula 2000). Then, in Section 4 we give a brief description of the
new VLA observations of the RQQs in our sample which escaped previous
radio detection by Kukula et al. (1998), and present new 5~GHz flux 
densities and positions. In Section 5 we present the new HST data and 
briefly summarize the approach taken to modelling the WFPC2 images;
the images and 2-dimensional model fits for the 14 new objects are provided 
in Appendix A (along with brief notes on each source), with observed and 
fitted luminosity profiles presented in Appendix B. Next, in Section 6 we
analyze the results of modelling the HST images of the complete 
sample, and assess the implications of our results for unified 
models of AGN. In Section 7 we explore how quasar hosts relate to 
the general population of massive galaxies, and demonstrate via the Kormendy 
relation that the quasar hosts are drawn from the population of
massive `boxy' ellipticals which are found predominantly in cluster 
environments. We also include a comparison
of the environments and interaction statistics for quasar hosts and brightest
cluster galaxies. In Section 8 we exploit our 
results to investigate the properties of the central engines which power 
these luminous AGN, deriving estimates of black-hole mass and Eddington 
ratio, and exploring possible clues to the origin of radio-loudness. Finally, 
in Section 9 we briefly discuss the implications of our results for the 
cosmological evolution of quasars and massive galaxies. Our conclusions are 
summarized in Section 10.

For ease of comparison with previous work we adopt an Einstein-de Sitter 
universe with $H_0 = 50~{\rm km s^{-1} Mpc^{-1}}$ for the calculation of 
physical quantities throughout this paper.

\section{Sample and associated observations}

The HST imaging observations reported here complete the imaging of 
the full sample of 33 objects (10 RLQs, 13 RQQs and 10 RGs) 
defined for this study. This 
sample was selected from the slightly  larger (40-source) sample imaged in 
the infrared by Dunlop et al. (1993) and Taylor et al. (1996) through the  
imposition of the slightly more restrictive redshift limits $0.1 < z < 0.25$. 
As described in McLure et al. (1999), this restriction in redshift range 
ensures that our $R$-band imaging through the F675W filter is not 
contaminated by the presence of strong emission lines such as [O{\sc iii}], 
or H$\alpha$. 
The main effect of this additional redshift restriction is to exclude a small 
number of objects from the original sample of Dunlop et al. (1993) which have 
$0.25 < z < 0.35$. However, this has not significantly 
compromised the original statistical merits of this sample, namely that the 
RLQ and RQQ sub-samples are matched in terms of optical luminosity, and that 
the RLQ and RG samples are matched in terms of radio luminosity and radio 
spectral index (Dunlop et al. 1993).

Of the 33 sources in this sample, 19 were observed during the first
year of our Cycle 6 allocation. The observations and analysis of these objects
were presented by McLure et al. (1999). The remaining 14 objects for 
which the observations are presented and analyzed in this paper are listed 
in Table 1, along with the dates on which they were observed with the HST.

Also included in this paper are new, deep VLA observations of a subset of our 
RQQ sample. These 4.8~GHz observations, the results of which are presented in 
Section 4, go a factor of 3 deeper than the observations of Kukula et al. 
(1998) which were utilised by McLure et al. (1999). The important outcome of 
these observations is that we have now detected all but one of the RQQs 
in this sample at radio wavelengths. These new radio detections, coupled with 
completion of the HST observations, have allowed us to re-investigate and 
clarify a number of the relations between optical and radio properties which 
could only be tentatively explored by McLure et al. (1999).

\begin{table}
\centering
\caption{Observing dates for the objects presented in this paper. Note that, 
despite its archive designation as 3C59, 0204$+$292 is in fact now classified 
as an RQQ (see Taylor et al. 1996, and the radio luminosity quoted in 
Table 3).}
\begin{tabular}{llcl}
\hline \hline
Object  & $HST$ Archive & Type & Observing \\ 
        & designation   &      & date \\ \hline
0307$+$169  & 3C79        & RG  & Jul 10 1998 \\
0230$-$027  & PKS0230$-$027 & RG  & Sep 25 1998 \\
1342$-$016  & 1342$-$016    & RG  & Dec 10 1998 \\
1215$+$013  & 1215$+$013    & RG  & Jan 02 1999 \\
1215$-$033  & 1215$-$033    & RG  & Jan 06 1999 \\
1330$+$022  & 1330$+$022    & RG  & Apr 13 1999 \\
1217$+$023  & PKS1217$+$02  & RLQ & Jul 07 1998 \\
1020$-$103  & PKS1020$-$103 & RLQ & Jul 12 1998 \\
2135$-$147  & PKS2135$-$14  & RLQ & Oct 19 1998 \\
2355$-$082  & PKS2355$-$082 & RLQ & Oct 19 1998 \\
0204$+$292  & 3C59          & RQQ & Jul 09 1998 \\
1549$+$203  & 1E15498$+$203 & RQQ & Sep 03 1998 \\
2215$-$037  & EX2215$-$037  & RQQ & Sep 26 1998 \\
0052$+$251  & PG0052$+$251  & RQQ & Nov 06 1998 \\
\hline \hline
\end{tabular}

\end{table}

We have also obtained improved infrared (UKIRT $K$-band) images for a small 
number of the more luminous quasars in our sample since the publication of 
McLure et al. (1999). These have been published in McLure, Dunlop \& Kukula 
(2000), but the results of modelling these images are utilised in this paper
to assist in the improved analysis of galaxy colours which is presented in 
Section 6.6.

Finally we note that H$\beta$ emission-line spectroscopy of the RLQ and RQQ
samples discussed here has recently been completed by McLure \& Dunlop (2001).
The results of this spectroscopic study are referred to in the discussion of 
black-hole mass estimation presented in Section 8.

\section{HST Observations}

The observations were made with the Wide Field \& Planetary Camera 2 ({\it 
WFPC2}; Trauger et al. 1994) on the Hubble Space Telescope through the F675W 
filter. The filter spans a wavelength range of 877\AA\, from 6275.5 to 
7152.5\AA\, roughly equivalent to standard $R$ band. This filter was selected 
in preference to a wider filter because it allowed both [O{\sc iii}] and 
H$\alpha$ emission lines to be excluded from the band-pass for source 
redshifts in the range $0.1 < z < 0.25$, and thus ensured that a clean 
measure could be made of the level of continuum light emitted by the quasar 
host galaxy at wavelengths longwards of the 4000\AA\ break.
As in McLure et al. (1999) target sources were  
centred on the WF2 chip, which was chosen in preference to WF1 or WF3 
because of its marginally superior performance over the period immediatly 
prior to our observations.

Deep sensitive images of the host galaxies
are obviously desirable, but the necessary long, background-limited exposures 
inevitably mean that, in the case of the quasars, the central source becomes 
saturated. From such saturated images it is extremely difficult to reliably 
disentangle host-galaxy emission from the contribution of the PSF-convolved 
nuclear source.

Slightly different strategies were therefore adopted for the quasar and 
radio-galaxy samples. For the quasars, exposures of 5, 26 and 3$\times$600 
seconds were taken. The short exposures ensured that at least one unsaturated
image of each quasar would be obtained, thus enabling an accurate measurement 
of the central flux density. The three 600-second exposures each yielded a 
3$\sigma$ surface-brightness sensitivity of 
$\mu_R = 23.8$\,mag.\,arcsec$^{-2}$\,px${-1}$, and their comparison facilitated reliable cosmic ray removal 
using standard IRAF tasks. With azimuthal averaging, the combined 1800-second 
image of each quasar allows extended emission to be traced reliably down to a 
surface-brightness level of $\mu_R > 26$ mag. arcsec$^{-2}$.

For the radio galaxies there was little danger of saturation and so short
exposures were not required. Three 700-second exposures were therefore
obtained for each radio galaxy. Any remaining time in the orbit was filled 
with a shorter exposure of flexible length (usually 40 to 100 seconds).
Calibration was carried out using the standard pipeline.

As described in McLure et al. (1999) we devoted one orbit of our allotted 
HST time to constructing a deep, unsaturated stellar PSF using the F657W 
filter, with the star centred on exactly the same part of the WF2 chip as the 
target objects.
Full details (magnitude, colour etc) of the chosen star can be found in 
McLure et al. (1999), along with a detailed description of how we constructed 
a properly-sampled PSF of the required large dynamic range using dithered 
observations of a series of exposures ranging from 0.23 to 160 seconds.  

Here we simply re-emphasize that the resulting high dynamic range PSF, 
tailored as closely as possible to match our quasar observations, has proved 
to be absolutely crucial in allowing the reliable extraction of host-galaxy 
parameters from the WFPC2 images.

\section{New radio observations of RQQs}

Kukula et al. (1998) reported radio continuum observations of a sample
of 27 low-redshift radio-quiet quasars, including all of the RQQs in the 
current sample, made with the Very Large Array (VLA) at 1.4, 4.8 and 8.4~GHz. 
Although flux densities 
were obtained for the majority of the sample, six objects 
remained undetected at all three frequencies (the flux-density 
limit of the survey 
was $\simeq 0.2$~mJy at 4.8~GHz).

To remedy this situation additional high-sensitivity observations were made 
of these six RQQs in August 1999 (05/08/99) at 4.8~GHz, again using the VLA 
in its high-resolution `A' configuration. Data reduction was carried out 
using the standard procedure within {\sc AIPS}. The angular resolution of the 
maps is $\sim 0.4''$ (FWHM) at this frequency, corresponding to a physical 
size of $\sim 1.5$~kpc at the redshift of the quasars - thus neatly 
encompassing the nuclear region of the quasar host.

The results of these observations are listed in Table 2. Radio sources are 
detected within $\sim 1''$ of all bar one of the RQQs (positional accuracies 
are estimated to be within $\sim 100$~mas). Due to scheduling problems 
1549$+$203 could only be allocated half the time given to the other five RQQs 
and the map of this object suffers from a correspondingly higher noise level.

\begin{table*}
\centering 
 \caption{High-sensitivity VLA observations of the six 
 previously-undetected RQQs in our sample. 
Optical positions were measured from
 Digitised Sky Survey plates using the {\sc stsdas} package in {\sc
 iraf}. All radio observations were made in $C$-band (4.8~GHz) with
 the VLA in A-configuration (angular resolution $\simeq 0.4''$) on
 August 5 1999. The uncertainties in the measured flux densities are
 given as 3 times the rms noise in the image. Radio positions are
 accurate to within 100~mas.}
\begin{tabular}{@{}lcccccccc@{}}
\hline \hline
            &       &       &\multicolumn{2}{c}{Optical Position (J2000)}&\multicolumn{2}{c}{Radio Position (J2000)}& \multicolumn{1}{c}{4.8~GHz Flux} & \multicolumn{1}{c}{log$(P_{4.8GHz}$/} \\ 
Quasar      &  $z$  &  $M_{V}$ & RA ($h~m~s$)&Dec ($^{\circ}~'~''$)& 
RA ($h~m~s$)&Dec ($^{\circ}~'~''$)& \multicolumn{1}{c}{Density /mJy}& \multicolumn{1}{c}{/W~Hz$^{-1}$sr$^{-1}$)}\\ \hline
0244$+$194    & 0.176 & $-23.55$ & 02 47 40.85 & $+$19 40 57.8 & 02 47 40.84 & $+$19 40 57.8 &  $0.18\pm0.06$ & ~~21.3   \\
PG 0923$+$201 & 0.190 & $-24.56$ & 09 25 54.71 & $+$19 54 04.4 & 09 25 54.74 & $+$19 54 05.0 &  $0.14\pm0.06$ & ~~21.3   \\
PG 0953$+$414 & 0.239 & $-25.36$ & 09 56 52.35 & $+$41 15 22.5 & 09 56 52.39 & $+$41 15 22.2 &  $0.25\pm0.07$ & ~~21.7   \\
1549$+$203    & 0.250 & $-24.51$ & 15 52 02.36 & $+$20 14 00.5 &     --      &        --     &  $<0.12$       &$<$21.4   \\
2215$-$037    & 0.241 & $-23.73$ & 22 17 47.77 & $-$03 32 38.8 & 22 17 47.72 & $-$03 32 38.5 &  $0.13\pm0.08$ & ~~21.4   \\
2344$+$184    & 0.138 & $-23.76$ & 23 47 25.71 & $+$18 44 50.8 & 23 47 25.77 & $+$18 44 50.7 &  $0.19\pm0.08$ & ~~21.1   \\
\\ \hline \hline

\end{tabular}

\label{RQQtab}
\end{table*}

\section{New HST Results}

The WFPC2 F675W images, two-dimensional model fits, and the model-substracted
residual images of the 14 {\it new} objects are presented in Appendix A
in Figs A1--A14, along with brief notes on each individual source.
Comparable images and notes for the other 19 objects in the
sample can be found in McLure et al. (1999).
The observed luminosity profiles for the 14 {\it new} 
objects are presented in Appendix B in Figs B1--B14, along with 
the best-fitting model profiles extracted from the two-dimensional model
fits.

Full details of the two-dimensional model procedure which
we have used to determine the properties of the host galaxies can be found
in McLure, Dunlop \& Kukula (2000), along with the results of
extensive tests of its ability to reclaim host-galaxy parameters
from simulated data based on a wide range of host-galaxy:nucleus 
combinations at different redshifts. As emphasized in McLure, Dunlop \& Kukula
(2000), the success of this modelling depends on an accurate high 
dynamic
range PSF, and the construction of an accurate error frame for
each quasar image.

In brief, the modelling of the HST images was carried out in
three separate stages. The first stage involves assessing how well the data 
can be reproduced {\it assuming} that the host galaxy is {\it either} an
elliptical galaxy (with a surface-brightness distribution described by a de 
Vaucouleurs $r^{1/4}$-law) or a pure exponential disc. The remaining five 
parameters (host-galaxy position angle, host-galaxy axial ratio, host-galaxy 
scale-length, host-galaxy luminosity, and nuclear luminosity) are then 
adjusted until, when convolved with the PSF, the model best
fits the data as determined by $\chi$-squared minimization (note that it is 
not assumed {\it a priori} that the radio galaxies have a negligible nuclear
component). Then, if one assumed galaxy morphology yields a significantly
better fit than the other, we can say that the galaxy is {\it better}
described by a de Vaucouleurs law or by an exponential disc. 
As with all the modelling performed on the HST sample, once the minimum 
$\chi^{2}$ solution had been found, the modelling code was repeatedly
re-started from close to the minimum $\chi^{2}$ solution, in order
 to ensure that the solution was stable. 

The results of applying this procedure to the new HST images are given
in Table 3, alongside the results already determined by McLure et al. (1999).
The striking feature of these results, now confirmed with the complete
sample, is that with the exception of 3 RQQs (0052$+$251 and the two 
lowest-luminosity RQQs 0257$+$024 and 2344$+$184) every single host galaxy is
better described by a de Vaucouleurs law.

In our second approach we have removed the requirement of assuming that
the host galaxy can be described as either a pure $r^{1/4}$-law or exponential
disc, and allow a sixth parameter $\beta$ (where the luminosity profile
of the galaxy is given by $I(r) \propto exp(-r^{\beta})$) to vary continuously.
Thus $\beta = 1$ should result if the galaxy is best described by 
a pure exponential disc, and $\beta = 0.25$ should result if the galaxy really does follow a pure de Vaucouleurs law, but {\it all} values of $\beta$
are available to the program if this results in an improved quality of
fit. The results of applying this procedure
to the HST images are given in Table 4 (again for the complete sample) and 
illustrated in Fig 1. These results are discussed in more detail in Section
6.1.

Finally, an examination of Table 4 and Fig 1 reveals that
whereas a very clean preference for $\beta = 0.25$ is displayed by the RGs
and RLQ hosts, a few of the RQQ hosts (in particular, as mentioned above, 
0052+251 and the two lowest-luminosity RQQs 0257+024 and 2344+184)
have best-fit $\beta$ values which are intermediate between the values of 
0.25 or 1.0 expected for pure elliptical or disc hosts. 
For this reason it was decided that the RQQ hosts should all be re-modelled 
with a 9 free-parameter fit, which allowed for
the combination of both disc and bulge contributions to the host's
surface-brightness distribution. For nine out of 
the 13 RQQs this extra freedom still resulted in no significant 
disc component. However, for four objects
(the above-mentioned three RQQs plus 0157$+$001) this procedure 
produced a significantly-improved model fit, and it is the $L_{host}$ and 
$L_{nuc}$ values from these combined fits which are used in all of the 
subsequent analyses. These four bulge-disc combinations are also noted in 
Table 3, and in that table (and in subsequent analysis) it is 
the scale-length, axial ratio and position angle of the dominant 
component which is adopted for these combined-fit objects.
 
The luminosity profiles, extracted from the
two-dimensional model fits are presented for the 14 new objects 
in Appendix B. The profiles are followed out to a radius of 10$\asec$,
which is representative of the typical outer radii used in the
modelling ($\langle r \rangle =11\asec$).

\begin{table*}
\centering
\caption{\small The outcome of attempting to model the AGN host
galaxies as either an exponential disc or a de Vaucouleurs spheroid.
Source name and redshift are given in the first two columns, with the 
logarithm of radio luminosity given in column 3.
The preferred host-galaxy morphology is given in column 4, with the
$\Delta\chi^{2}$ between the chosen and alternative model listed in
column 5.  In column 6\, $r_{1/2}$ is given irrespective of the chosen
host morphology.  Column 7 lists $\mu_{1/2}$ in units of $R$ mag
arcsec$^{-2}$.  Columns 8 and 9 list the integrated apparent
magnitudes of the host galaxy and fitted nuclear component converted
from F675W to Cousins $R$-band, while column 10 gives the ratio of
integrated galaxy and nuclear luminosities.  Columns 11 and 12 give the
axial ratio and position angle (east of north)  
of the best-fiting host-galaxy model.}
\begin{tabular}{ccclrcccclcr}
\hline
\hline
Source & $z$ &log$(P_{4.8GHz}/$&Host& $\Delta\chi^{2}$ & $r_{1/2}$/kpc  &$\mu_{1/2}$&$R_{host}$
& $R_{nuc}$  & $L_{nuc}/L_{host}$ & $b/a$ & PA/$^{\circ}$ \\
& & ${\rm W Hz^{-1} sr^{-1}})$&Morphology&&&&&&&&\\
\hline
{\bf RG}&&&&&&&&&&&\\
0230$-$027&0.239&24.8&Elliptical&5900    &\phantom{1}7.7&21.8&17.5&&&0.95&113\\

0307$+$169&0.256&25.5&Elliptical&3500    &\phantom{1}9.4&21.4&17.2&20.9&\phantom{1}0.03&1.00&13\\

0345$+$337&0.244&25.5&Elliptical&2400&13.1&23.3&18.0&21.1&\phantom{1}0.06&0.70&99\\

0917$+$459&0.174&25.7&Elliptical&33000&21.9&23.0&16.1&19.4&\phantom{1}0.05&0.76&36\\

0958$+$291&0.185&25.3&Elliptical&7800    &\phantom{1}8.5&22.0&17.1&18.5&\phantom{1}0.27&0.95&45\\

1215$-$033&0.184&24.1&Elliptical&9300    &\phantom{1}8.5&22.0&17.1&22.3&\phantom{1}0.008&0.87&60\\

1215$+$013&0.118&24.0&Elliptical&14000    &\phantom{1}4.7&21.0&16.5&19.9&\phantom{1}0.05&0.94&142\\

1330$+$022&0.215&25.4&Elliptical& 7400   &15.7&22.9&17.1&19.5&\phantom{1}0.11&0.79&79\\

1342$-$016&0.167&24.4&Elliptical& 29000   &23.3&22.9&15.6&21.8&\phantom{1}0.003&0.93&96\\

2141$+$279&0.215&25.2&Elliptical&8500&24.8&23.5&16.7&25.6&\phantom{1}0.0003&0.74&148\\

\hline
{\bf RLQ}&&&&&&&&&&&\\
0137$+$012&0.258&25.2&Elliptical&5100&14.2&22.6&17.2&17.3&\phantom{1}0.8&0.85&35\\

0736$+$017&0.191&25.4&Elliptical&8900&13.3&22.9&16.9&16.2&\phantom{1}1.9&0.97&13\\

1004$+$130&0.240&24.9&Elliptical&500&\phantom{1}8.2&21.5&16.9&15.0&\phantom{1}5.8&0.94&29\\

1020$-$103&0.197&24.7&Elliptical&4200   &\phantom{1}7.1&20.8&17.2&16.8&\phantom{1}1.4&0.73&46\\

1217$+$023&0.240&25.9&Elliptical&2400   &11.1&21.7&17.3&16.3&\phantom{1}2.5&0.8&16\\

2135$-$147&0.200&25.3&Elliptical&2700   &11.6&22.7&17.2&16.2&\phantom{1}2.5&0.95&72\\

2141$+$175&0.213&24.8&Elliptical&570&\phantom{1}8.2&21.2&17.3&15.9&\phantom{1}3.7&0.47&118\\

2247$+$140&0.237&25.3&Elliptical&8100&13.5&22.4&17.2&16.9&\phantom{1}1.3&0.63&118\\

2349$-$014&0.173&24.9&Elliptical&13000&19.2&22.7&15.9&16.0&\phantom{1}0.9&0.89&45\\

2355$-$082&0.210&24.5&Elliptical&3000     &10.4&22.0&17.1&17.4&\phantom{1}0.77&0.73&177\\
\hline

{\bf RQQ}&&&&&&&&&&&\\
0052$+$251&0.154&21.6&Bulge/Disc&2500    &3.2&20.8&16.7&15.4&\phantom{1}3.2&0.61&117\\

0054$+$144&0.171&21.9&Elliptical&6100&10.4&21.7&16.6&15.5&\phantom{1}2.7&0.61&108\\

0157$+$001&0.164&22.8&Bulge/Disc&3900    &15.5&22.0&15.6&16.2&\phantom{1}0.57&0.88&116\\

0204$+$292&0.109&22.7&Elliptical&33000    &\phantom{1}8.8&20.9&15.9&16.0&\phantom{1}0.89&0.73&71\\

0244$+$194&0.176&21.3&Elliptical&2700&\phantom{1}9.3&22.7&17.5&16.8&\phantom{1}1.9&0.92&77\\

0257$+$024&0.115&22.2&Disc/Bulge&10000    &11.7&21.7&15.9&21.0&\phantom{1}0.009&0.88&134\\

0923$+$201&0.190&21.3&Elliptical&1700&\phantom{1}8.2&22.1&17.2&15.7&\phantom{1}4.2&0.98&141\\

0953$+$415&0.239&21.7&Elliptical&90&\phantom{1}7.6&22.4&18.2&15.2&15.4&0.86&115\\

1012$+$008&0.185&22.0&Elliptical&1100&28.7&23.8&16.6&16.2&\phantom{1}1.5&0.64&109\\

1549$+$203&0.250&$<$21.4\phantom{$<$}&Elliptical&510    &\phantom{1}5.0&22.2&18.9&16.8&\phantom{1}6.5&0.88&34\\

1635$+$119&0.146&23.0&Elliptical&35000&\phantom{1}7.6&21.6&16.8&18.1&\phantom{1}0.3&0.69&179\\

2215$-$037&0.241&21.4&Elliptical& 1500    &\phantom{1}6.7&21.4&17.5&18.2&\phantom{1}0.56&0.84&88\\

2344$+$184&0.138&21.1&Disc/Bulge& 8900   &17.5&23.8&16.8&20.3&\phantom{1}0.04&0.67&103\\
\hline \hline
\end{tabular}

\end{table*}

\section{Analysis of the full  sample}

\subsection{Host-Galaxy Morphologies}

Examination of the results presented in Tables 3 and 4 confirms that the 
somewhat complex observing strategy outlined in Section 3 
has successfully allowed the determination of 
host-galaxy morphology for all 33 objects in the sample. Also
immediately apparent from Table 3 is that
the huge amount of information available to the modelling code 
has not only allowed a
clear morphological preference to be made, but can formally exclude the
 alternative host in all cases ($\Delta \chi^{2}=25.7$ corresponds to
a 99.99\% confidence level for a 5 parameter fit).

The results from the modelling of the RG and RLQ sub-samples are in
good agreement with orientation-based unification, with all 20 objects
found to have elliptical host galaxies. A perhaps more striking
feature of these results is the extent to which the classic $r^{1/4}$
de Vaucouleurs law provides a near-perfect description of the host
galaxies of the radio-loud objects (see Fig 1).
 As can be seen from Table 4 the best-fitting $\beta$ values 
for the combined  RG and RLQ sub-samples all lie in the narrow range
$0.19<\beta<0.26$, with the RG sample 
alone displaying an even narrower spread, 
$0.21<\beta<0.25$. 
This conclusion is further strengthened by a 
comparison of the beta histogram for the three sub-samples with that 
from the $\beta$-modelling 
tests performed by McLure, Dunlop \& Kukula (2000). The application of the
Kolmogorov-Smirnov test to the $\beta$ distribution of the 29 objects
which were found to have single-component elliptical host galaxies,
and those resulting from the $\beta$-model testing, returns a probability
of $p=0.23$. This confirms that, as far as the modelling code is
concerned, any differences between these hosts and pure elliptical
galaxies are not statistically significant. The best match between the 
test results and the actual data is for the RG sub-sample ($p=0.39$),
perhaps as would be expected considering the lack of a dominant
point-source contribution.

\begin{table}
\centering
\caption{\small The outcome of the variable-$\beta$ modelling.  Column
2 lists the host morphology of the best fitting `fixed $\beta$' model
(results of which are given in Table 3).  The best-fitting values for
the $\beta$ profile parameter are given in column 3. Column 4 gives 
$\Delta\chi^{2}$, which  quantifies the improvement in fit 
offered by 
the variable-$\beta$ model over that already achieved 
with the best-fitting pure disc or elliptical
model. Simulations indicate that $\beta$ can be reclaimed to within a 
typical uncertainty of $0.01 - 0.02$.}
\begin{tabular}{clcr}
\hline
\hline
Source & Host& $\beta$&$\Delta\chi^{2}$ \\

\hline
{\bf RG}&&&\\

0230$-$027&Elliptical&0.25&0\\

0307$+$169&Elliptical&0.21&52\\

0345$+$337&Elliptical&0.25&2\\

0917$+$459&Elliptical&0.23&264\\

0958$+$291&Elliptical&0.25&16\\

1215$-$033&Elliptical&0.24&0\\

1215$+$013&Elliptical&0.25&3\\

1330$+$022&Elliptical&0.24&16\\

1342$-$016&Elliptical&0.23&90\\

2141$+$279&Elliptical&0.25&2.2\\
\hline
{\bf RLQ}&&&\\
0137$+$012&Elliptical& 0.19&126\\

0736$+$017&Elliptical& 0.19&239\\

1004$+$130&Elliptical& 0.25&5\\

1020$-$103&Elliptical&0.19&132\\

1217$+$023&Elliptical&0.26&2\\

2135$-$147&Elliptical&0.25&0\\  

2141$+$175&Elliptical& 0.28&23\\

2247$+$140&Elliptical& 0.25&17\\

2349$-$014&Elliptical&0.26&10\\

2355$-$082&Elliptical&0.26&383\\
\hline
{\bf RQQ}&&&\\
0052$+$251&Bulge/Disc&1.09&267\\

0054+144&Elliptical&0.25&3\\

0157+001&Bulge/Disc&0.24&133\\

0204$+$292&Elliptical&0.24&216\\

0244+194&Elliptical&0.22&47\\

0257+024&Disc/Bulge&0.75&2792\\

0923+201&Elliptical&0.30&44\\

0953+415&Elliptical&0.27&9\\

1012+008&Elliptical&0.38&102\\

1549$+$203&Elliptical&0.25&0\\

1635+119&Elliptical&0.18&550\\

2215$-$037&Elliptical&0.25&0\\

2344+184&Disc/Bulge&0.43&1044\\
\hline \hline
\end{tabular}

\end{table}

\label{morphs}
\begin{figure}
\vspace{14cm}
\centering
\setlength{\unitlength}{1mm}
\includegraphics{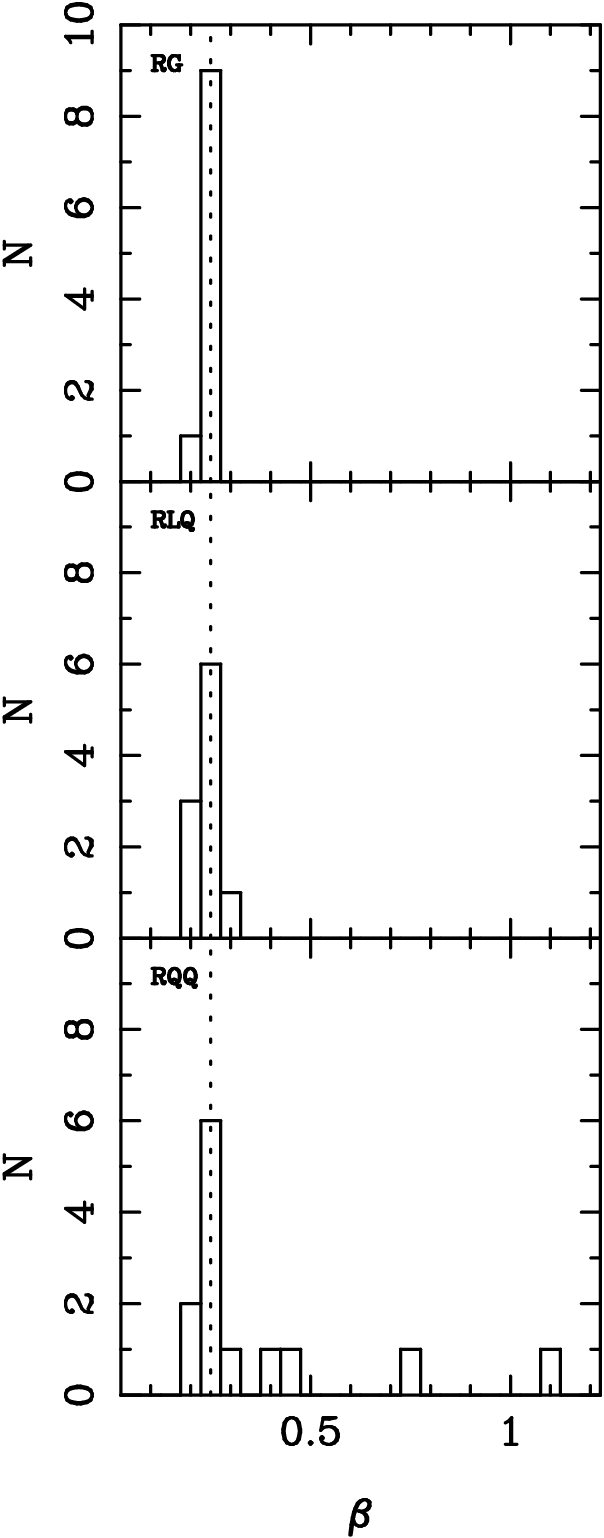}
\caption{Sub-sample histograms of the best-fit $\beta$ values from the
 variable-$\beta$ modelling. The dotted line lies at $\beta=0.25$,
corresponding to a perfect de Vaucouleurs law.}
\label{betahist}
\end{figure}

Considering the long-standing belief that RQQs are often located
in disc galaxies, the results from modelling of the RQQ sub-sample 
are quite unambiguous, with 9 of the 13 host galaxies showing no evidence 
for any disc component. Of the four objects which are best matched by
a combined disc/bulge model, the luminosities of 0157+001 and 0052+251
are dominated by their bulge components, which respectively account
for 83\%\ and 71\%\ of the total host luminosity. This means that the number of
bulge-dominated RQQ hosts is 11 out of the 13 objects.

It is interesting to note that the two most disc-dominated
galaxies, 0257+024 and 
2344+184, are the hosts of 
by far the lowest luminosity AGN in the entire 33-object
sample. In fact, converting their total luminosity (host+nucleus)
from the model fits for these two objects 
to the equivalent absolute $V$-magnitudes gives $M_{V}=-22.6$ and
$M_{V}=-21.9$ respectively (assuming $V-R=0.8$ at $z=0$, Fukugita at al.
1995). Given that for this study the adopted
quasar/Seyfert borderline is $M_{V}=-23.0$, it is clear that these two
objects are not actually {\it bona fide} RQQs. The clear
implication from this result is that all {\it true} quasars, with 
$M_{V}<-23.0$, reside in luminous bulge-dominated hosts, irrespective of 
their radio power. 

The morphological determinations for the RQQ host galaxies have
placed on a firm footing the suggestion made previously by Taylor at al.
(1996) and McLeod \& Rieke (1995) that the probability of a RQQ
having an early-type host was an increasing function of the quasar
luminosity. Unlike the previous ground-based studies, the high
resolution and temporally stable PSF offered by HST has permitted the
confirmation of what were hitherto necessarily tentative conclusions, due to
the uncertainties introduced by ground-based seeing conditions. Therefore, a
strong conclusion from the morphological analysis of from the new HST
images is that the radio luminosity of an AGN is not directly
related to host-galaxy morphology.
The relationship between host morphology and AGN luminosity is explored further
in Section 7.

\subsection{Host-Galaxy and AGN Luminosities}

The host and nuclear
luminosities are presented in the form of integrated absolute Cousins
$R$-band magnitudes in Table 5, and in Figs 2 and 3. These have been 
derived from the apparent 
magnitudes listed in Table 3 which have been calculated by integrating 
the best-fit model components to infinite radius, and adopting
the F675W flight-system zero-point given by Holtzman et al. (1995). The 
similarity between the F675W filter and the standard Cousins $R$-band
is such that, with the use of this zero-point, the difference is
 of the order $\pm0.05$ magnitudes. The conversion from apparent to
absolute magnitudes has been performed using 
$k$-corrections assuming spectral indices of $\alpha=1.5$ and
$\alpha=0.2$ for the host and nucleus respectively ($f_{\nu} \propto
\nu^{-\alpha}$). 

For all but 4 of the sources, the values of $M_R$(host) listed in Table 5 
can be regarded as equivalent to the value of $M_R$(bulge). For the 
4 objects which benefitted from a combined disc+bulge model the 
best-fitting values of $M_R$(bulge) are as follows:
0052$+$251: $M_R$(bulge) $= -22.95$, 
0157$+$001: $M_R$(bulge) $= -24.32$, 
0257$+$024: $M_R$(bulge) $= -21.48$, 
2344$+$184: $M_R$(bulge) $= -22.44$.
 
\begin{figure}
\vspace{14cm}
\centering
\setlength{\unitlength}{1mm}
\includegraphics{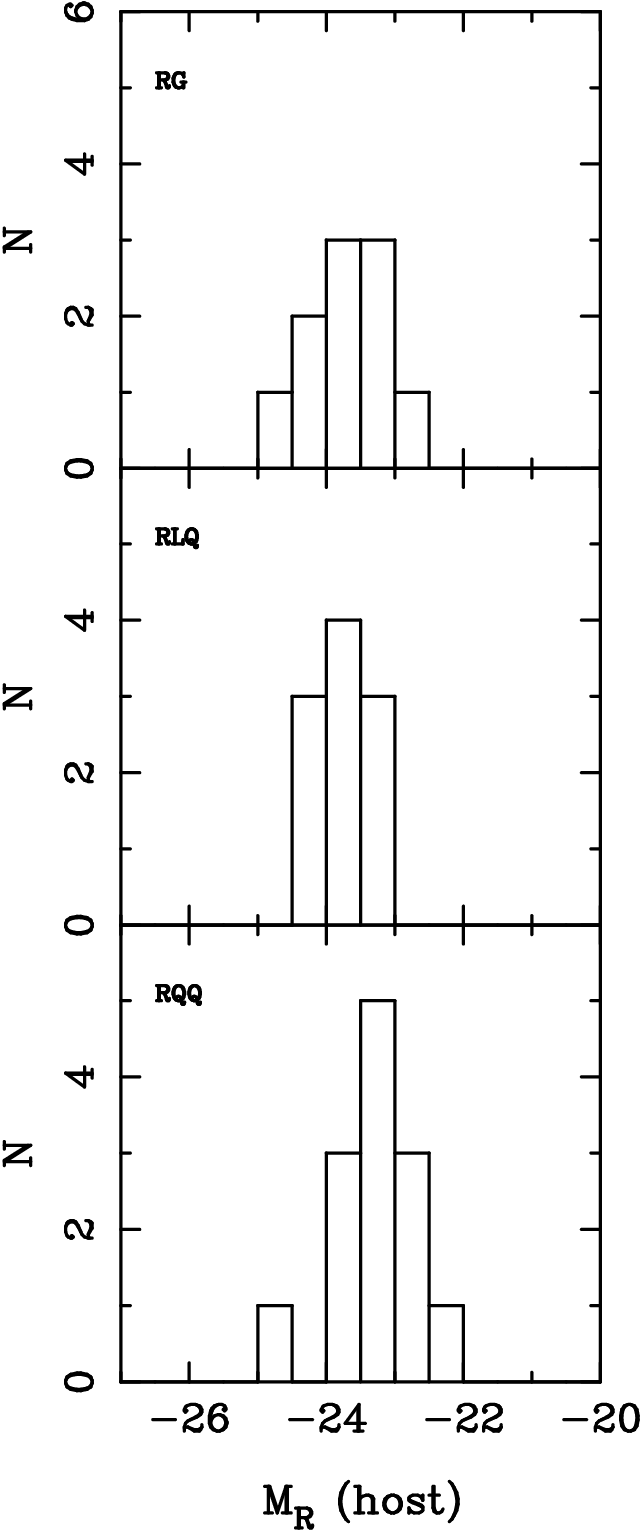}
\caption{Sub-sample histograms of the host-galaxy, 
absolute, integrated $R$-Cousins magnitudes.}
\label{hostmags}
\end{figure}
\begin{figure}
\vspace{14cm}
\centering
\setlength{\unitlength}{1mm}
\includegraphics{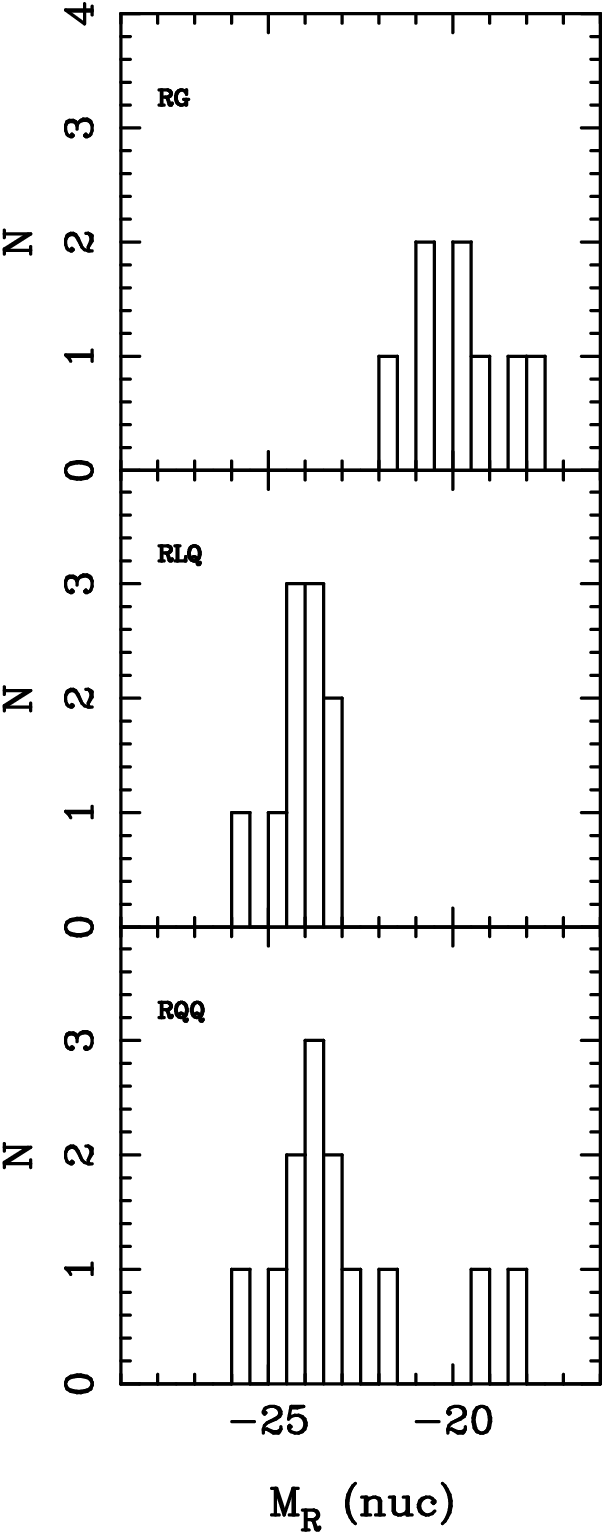}
\caption{Sub-sample histograms of the nuclear absolute
integrated $R$-Cousins magnitudes.}
\label{nucmags}
\end{figure}

\begin{table*}
\caption{\small Absolute magnitudes ($M_R$), and optical-infrared ($R-K$) colours of the best-fitting 
host galaxy and nuclear component for each AGN.
Column 2 gives the $R$-band absolute magnitudes ($M_R$) of the
total host galaxy derived from the current modelling of the HST data, 
assuming a spectral index of 
$\alpha=1.5$ (where $f_{\nu}\,\propto\,\nu^{-\alpha}$).
Column 3 gives the $R$-band absolute magnitudes ($M_R$) of the nuclear
component as derived from the current modelling of the HST data assuming
a spectral index of $\alpha=0.2$.
Columns 4 and 5 list the $R-K$ colours of the host galaxy and nuclear
component respectively. These colours were derived by combining 
12-arcsec aperture $R$-band photometry from our HST-based models 
with the 12-arcsec 
aperture $K$-band photometry derived by Taylor {\it et al.} (1996) and McLure, Dunlop \& Kukula (2000), to
minimize the uncertainty introduced by errors in constraining the galaxy
scale-lengths at $K$ (see text for further details).}
\begin{tabular}{ccccc}
\hline
\hline
Source & $M_{R}(host)$ &$M_{R}(nuc)$ & $(R-K)_{Host}$ 
&$(R-K)_{Nuc}$\\
\hline
{\bf RG}&&&&\\
0230$-$027&$-$23.55&-&2.1&\phantom{1}-\\

0307$+$169&$-$23.96&$-$19.94&2.1&\phantom{1}6.7 \\

0345$+$337&$-$23.05&$-$19.63&3.6&\phantom{1}5.8\\

0917$+$459&$-$24.20&$-$20.65&3.4&\phantom{1}5.0\\

0958$+$291&$-$23.36&$-$21.70&1.9&\phantom{1}4.5\\

1215$+$013&$-$22.85&$-$19.35&2.5&\phantom{1}6.9\\

1215$-$033&$-$23.29&$-$17.87&2.6&\phantom{1}5.1\\

1330$+$022&$-$23.70&$-$21.00&2.8&\phantom{1}5.0\\

1342$-$016&$-$24.58&$-$18.35&2.6&\phantom{1}6.0\\

2141$+$279&$-$24.09&$-$14.94&3.3&\phantom{1}10.4\\

\hline
{\bf RLQ}&&&&\\
0137$+$012&$-$24.04&$-$23.53&2.8&\phantom{1}3.1\\

0736$+$017&$-$23.58&$-$24.01&3.2&\phantom{1}3.2\\

1004$+$130&$-$24.10&$-$25.70&3.0&\phantom{1}2.0\\

1020$-$103&$-$23.36&$-$23.47&2.3&\phantom{1}3.6\\

1217$+$023&$-$23.71&$-$24.40&3.2&\phantom{1}3.2\\

2135$-$147&$-$23.37&$-$24.09&2.5&\phantom{1}4.0\\

2141$+$175&$-$23.43&$-$24.52&2.7&\phantom{1}1.9\\

2247$+$140&$-$23.80&$-$23.80&2.8&\phantom{1}2.6\\

2349$-$014&$-$24.32&$-$24.00&2.9&\phantom{1}3.6\\

2355$-$082&$-$23.62&$-$23.06&2.7&\phantom{1}3.0\\
\hline
{\bf RQQ}&&&&\\
0052$+$251&$-$23.33&$-$24.39&2.5&\phantom{1}2.0\\

0054$+$144&$-$23.61&$-$24.49&3.1&\phantom{1}1.8\\

0157$+$001&$-$24.52&$-$23.70&2.8&\phantom{1}2.8\\

0204$+$292&$-$23.30&$-$23.03&2.5&\phantom{1}3.2\\

0244$+$194&$-$22.77&$-$23.24&2.3&\phantom{1}3.2\\

0257$+$024&$-$23.46&$-$19.69&2.6&\phantom{1}6.7\\

0923$+$201&$-$23.25&$-$24.57&3.2&\phantom{1}3.2\\

0953$+$415&$-$22.86&$-$25.52&3.0&\phantom{1}2.5\\

1012$+$008&$-$23.78&$-$23.95&3.2&\phantom{1}2.2\\

1549$+$203&$-$22.26&$-$23.98&3.4&\phantom{1}3.1\\

1635$+$119&$-$23.05&$-$21.54&3.3&\phantom{1}3.7\\

2215$-$037&$-$23.52&$-$22.58&2.7&\phantom{1}4.4\\

2344$+$184&$-$22.97&$-$20.30&2.6&\phantom{1}5.7\\

\hline \hline

\end{tabular}

\end{table*}

\subsubsection{Host-Galaxy Luminosities}

The mean and median integrated absolute magnitudes
of the best-fit host galaxies in each sub-sample are:

\begin{tabbing}
$\langle M_{R} \rangle =-23.53\pm0.09$ \hspace{0.5cm} \= median=$-$23.52 \hspace{0.5cm} \= (ALL)\\
$\langle M_{R} \rangle =-23.66\pm0.16$ \> median=$-$23.63 \>(RG)\\
$\langle M_{R} \rangle =-23.73\pm0.10$ \> median=$-$23.67 \>(RLQ)\\
$\langle M_{R} \rangle =-23.28\pm0.15$ \> median=$-$23.30 \>(RQQ)
\end{tabbing}

\noindent
Two features of these results merit comment. First,
the agreement between the absolute magnitudes of the host galaxies of
the RG and RLQ sub-samples can be seen to be extremely good, with the
median figures differing by only 0.04 magnitudes. This can be
interpreted as strong evidence in favour of orientation-based
radio-loud unification (see also Section 6.7). The second obvious
feature of these results is that these new HST images appear
to confirm the traditional finding that the hosts of RQQs are less
luminous than those of RLQs, although the median difference of 0.37
magnitudes is a factor of two smaller than the difference typically 
claimed (e.g. Kirhakos et al. 1999).  This does not
seem to be an artifact of the inclusion
of RQQ hosts with a substantial disc component, since if attention
is confined to the 9 RQQs with solid elliptical model fits the values
remain virtually unchanged, i.e.:

\begin{tabbing}
$\langle M_{R} \rangle =-23.30\pm0.17$ \hspace{0.5cm} \= median=-23.30
\hspace{0.5cm} \= (RQQ) 
\end{tabbing}

\noindent
A large variation in the luminosity
difference between RQQ and RLQ host galaxies has been reported in the
literature. Differences have ranged from RQQ hosts being fainter
than their RLQ counterparts by $0.7\rightarrow1.0$ magnitudes in optical
studies (eg. Smith et al. (1986), V\'{e}ron-Cetty \&\ Woltjer (1990),
Bahcall et al. (1997)), to no formal luminosity difference being detected
from the near-infrared imaging of this HST sample (Taylor et al. 1996). It is
therefore interesting to compare the difference in luminosity detected
here with the results of the other two recent HST $R$-band imaging
studies of Hooper et al. (1997) and Boyce et al. (1998), both of which also
used two-dimensional modelling to analyse the host galaxies. Below
we list the mean differences detected in the three
studies ($\langle M_{RQQ} \rangle - \langle M_{RLQ} \rangle $) with the 
associated standard error. In
calculating the figures for the Hooper et al. programme, where no attempt
was made to distinguish the host galaxy morphologies, the luminosities
of the best-fit $r^{1/4}$ model have been used.\\

\begin{tabbing}
$\Delta M_R = 0.43\pm0.20$ \hspace{1cm} \= This Work\\
$\Delta  M_R = 0.53\pm0.23$ \> Hooper et al. \\
$\Delta  M_R = 0.67\pm0.29$ \> Boyce et al. \\
\end{tabbing}

\noindent
The clear implication from these results is that the host galaxies of
RQQs are consistently fainter than those of RLQs by $\sim0.5$ magnitudes
in the $R$-band. Unlike in earlier studies, this difference can no
longer be attributed to the model fitting of RQQs producing disc fits,
since only 2 of the 26 RQQ host magnitudes included in the above
figures are derived from an exponential host model. The common bias of the RLQ
redshifts being consistently higher than those of the RQQs can also be
firmly rejected as a possible cause of the difference, with the two
quasar types having well-matched redshift distributions in all three
studies. At least for the results presented here, and those of Hooper
et al., the selection of the RLQ and RQQ quasar samples to have matched
optical magnitudes, also excludes the possibility that the host
magnitude difference is as a result of the RLQs being intrinsically
more luminous. However, it is possible that the host-galaxy results of
Boyce et al. may have been influenced by this effect, considering that
the RLQs in that study are intrinsically $0.6\pm 0.8$ magnitudes
brighter than the RQQs, although, as can be seen from the large error
associated with this difference, the overlap of the total quasar
luminosities is significantly larger than that of the hosts. It is worth noting
at this point that although the results of Taylor et al. certainly do
not formally support a difference in host magnitudes (with $\Delta M_K
= 0.4\pm0.3$), they are in fact perfectly consistent with the three sets
of $R$--band HST results. The question of whether there is any detectable 
correlation between the host-galaxy and nuclear luminosities is explored 
below in Section 6.2.3.

A general result that can be taken from the luminosities of the
best-fitting host galaxies is that, with the exception of the two
Seyfert objects, all of them lie at the extreme end of the elliptical galaxy
luminosity function. Adopting $M_{R}^{\star}=-22.2$ (Lin et al. 1996),
after having converted the published value of $M_{R}^{\star}=-21.8$ to
an integrated magnitude, it can be seen that {\it all} of the  hosts have 
luminosities of $L\ge L^{\star}$, with 25 of the 33 having luminosities
$L\ge 2L^{\star}$. These results are in good agreement with those of
the previous $K$-band imaging study which also found all the hosts to be
brighter than $L_{K}^{\star}$. Independent support for this result
comes from the findings of Hooper et al. (1997) and Boyce et al. (1998), both 
of which used the F702W (wide R) filter on WFPC2. Using the Lin et al. value for
$M_{R}^{\star}$, Hooper et al. found 15 of their 16 ($z \simeq 0.4$) quasars 
to have $L \ge L^{\star}$, while Boyce et al. found all 11 of their 14, 
objects for which a host model was fitted, to have $L\ge L^{\star}$ 
(both sets of results having been converted to
our adopted cosmology).

\subsubsection{Nuclear Luminosities}

The luminosities of the best-fitting nuclear components are listed in Table 5
and illustrated in Fig 3, with sub-sample averages summarized below.
The radio galaxy 0230$-$027 has been excluded from these summary
statistics as the modelling code detected no unresolved component in 
this object.

\begin{tabbing}
$\langle M_{R} \rangle =-22.27\pm0.45$ \hspace{0.5cm} \= median=$-$23.36 \hspace{0.5cm} \= (ALL)\\
$\langle M_{R} \rangle =-19.27\pm0.64$ \> median=$-$19.64 \> (RG)\\
$\langle M_{R} \rangle =-24.07\pm0.22$ \> median=$-$24.01  \> (RLQ)\\
$\langle M_{R} \rangle =-22.95\pm0.57$ \> median=$-$23.70 \> (RQQ)
\end{tabbing}

\noindent
As expected, the unresolved nuclear components displayed by the RGs are
nearly two orders of magnitude weaker than those of the RLQs, in good
agreement with the orientation-based radio-loud unification scheme
first described by Peacock (1987) and Barthel (1989) (see Section 6.7). 
It would appear 
from the mean figures given 
above that there is a substantial difference in the 
nuclear components of the RQQ and RLQ quasar samples. However, the noticeable 
offset in the difference between the mean and median figures, combined 
with examination of
histogram shown in Fig 3, reveals that the
RQQ mean is being biased by the three objects which have 
total luminosities fainter than
M$_{V}=-23.0$. If these three objects are excluded from the RQQ
sample, the mean and median values of nuclear luminosity become: 

\begin{tabbing}
$\langle M_{R} \rangle =-23.95\pm0.26$ \hspace{0.5cm} \= median=$-$23.97
\hspace{0.5cm} \= (RQQ)
\end{tabbing}

\noindent
completely consistent with the equivalent RLQ figures. The
similarity between the nuclear components of the RQQ and RLQ
sub-samples is reassuring considering that the sample-selection process
was indeed originally designed to produce two quasar samples
well-matched in luminosity. This result also re-affirms that the measured
difference in the respective host magnitudes of the RQQs and RLQs {\it cannot}
be attributed to some bias arising from the RLQs in our sample
having, on average, more powerful nuclei.

\subsubsection{Quasar Host-Nuclear Luminosity Correlation}
\label{hostcor}

Because the measured difference between the RQQ and RLQ host-galaxy
magnitudes is substantially greater than the average difference between  their
respective nuclear components (excluding the three sub-luminous
objects) it is not expected that there should be a strong host-nuclear
luminosity correlation. This is indeed the case, and 
application of the Spearman Rank
correlation test confirms that there is no evidence for any
correlation, returning a probability $p=0.95$ that the null
hypothesis of no correlation is acceptable. 

The null result found here contrasts with the positive correlations
found previously by many authors (eg. Smith et al. (1986), Bahcall et al.
(1997), Hooper et al. (1997)). The suspicion often cast upon the reality
of a positive host-galaxy:nuclear correlation is that there
are two obvious selection effects at work. The first of these is that
it is obviously much more difficult to detect faint galaxies
which host brighter quasars. It seems clear that given the
difficulties associated with ground-based seeing, and the perils of
PSF-subtraction on HST data (eg. Bahcall et al. 1997), that this could
well be a contributing factor. In the present study there are 
several RQQs with bright
nuclear components, and relatively faint host galaxies (eg. 0953+415), 
providing confidence that the techniques employed here have been succesful in
overcoming this possible source of a false correlation. The other
selection effect which could contribute to a false positive result is that weak
AGN in bright host galaxies will not be classified as quasars. 

Although the results presented here for powerful low-$z$ quasars show
no evidence for a correlation, it does appear to exist for 
lower-luminosity Seyfert galaxies. 
In the case of Seyferts, the reduction in
both redshift and contrast between the host and nuclear components means
that it can be confidently assumed that all of the host galaxies are
being detected, removing one of the sources of bias mentioned
above. Indeed, the work of McLeod \&\ Rieke (1994) 
on two samples of low- and
high-power AGN at low-$z$ has led them to suggest that the relation
between host and nuclear luminosity is of the form that there is a
minimum host luminosity required to produce a particular quasar
luminosity. This would be consistent with both the positive correlation
found at lower AGN power, and the flat relation found here for higher
powered quasars. The host galaxies for the quasars studied here have
already been shown to be among the brightest known, and therefore
demand that the galaxy-quasar relation tails-off at high quasar
luminosity. The question of how
the nuclear and host-galaxy luminosities relate to central black-hole
mass and quasar fuelling rate is pursued in Section 8.

\subsection{Scale-lengths}

\begin{figure}
\vspace{14cm}
\centering
\setlength{\unitlength}{1mm}
\includegraphics{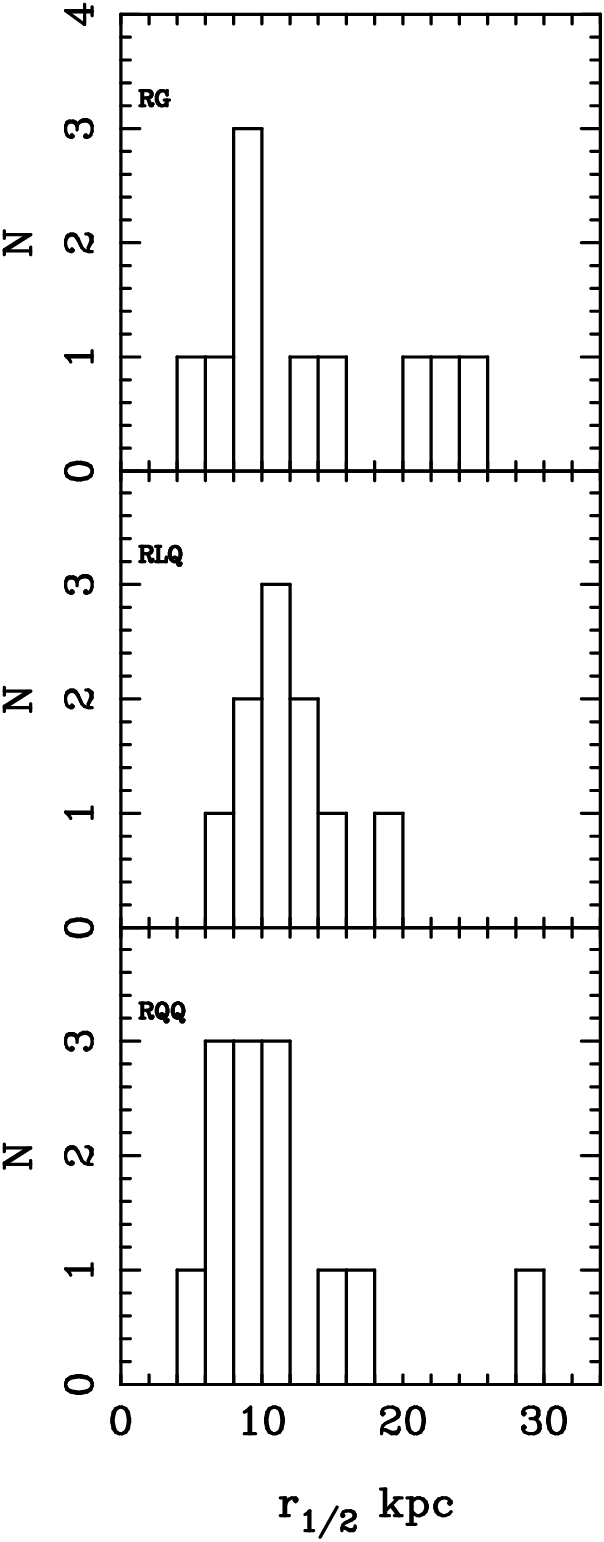}
\caption{Histograms of the best-fitting host-galaxy scale-lengths for the
three AGN sub-samples.}
\label{scalehist}
\end{figure}

The best-fitting values of the host-galaxy scale-lengths listed in 
Table 3 are displayed as histograms for the
separate sub-samples in Fig 4. The average figures for
the three sub-samples are :

\begin{tabbing}
$\langle r_{1/2} \rangle =12.23\pm1.00$ \hspace{0.5cm} \= median=10.45 \hspace{0.5cm} \= (ALL)\\
$\langle r_{1/2} \rangle =13.76\pm2.18$ \> median=11.27 \> (RG)\\
$\langle r_{1/2} \rangle =11.73\pm1.07$ \> median=11.28 \> (RLQ)\\
$\langle r_{1/2} \rangle =11.45\pm1.66$ \> median=\phantom{0}9.30  \> (RQQ)
\end{tabbing}

\noindent
Taking the median
figures, the RG and RLQ sub-samples can again be seen to be in
remarkable agreement, while the RQQ hosts are  $\sim20\%$ smaller.

\begin{figure}
\vspace{9cm}
\centering
\setlength{\unitlength}{1mm}
\includegraphics{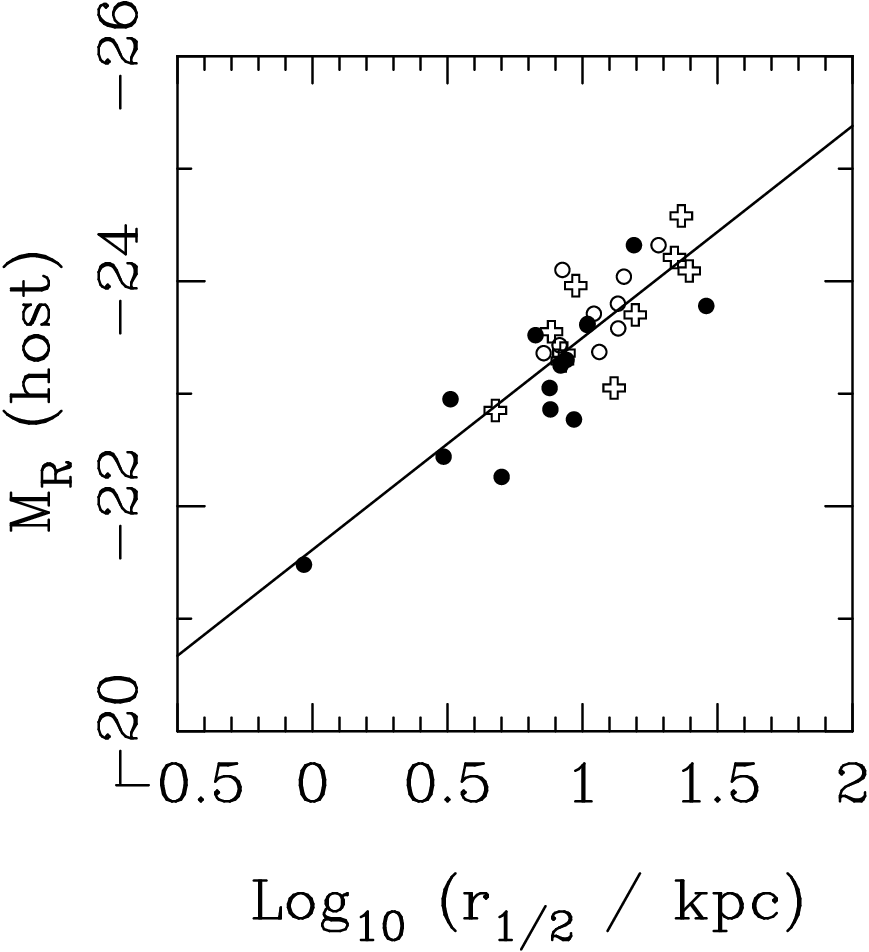}
\caption{Plot of absolute host magnitude against best-fitting host scale-length
for the RGs (crosses), RLQs (open circles) and RQQs (filled circles) in
the HST sample. The solid line is the
least-squares fit to the data, which has the form $L \propto
r^{0.75}$. For the four RQQs which required a 
combined disc/bulge fit, the best-fitting bulge parameters 
have been plotted.}
\label{l_r}
\end{figure}

Galaxy half-light radii of $\simeq10$ kpc mark out these radio galaxies
and quasar hosts as being giant elliptical galaxies.
In a study of the galaxies in the Virgo cluster,
Capaccioli et al. (1992) found that beyond a scale-length of $\sim 3$ kpc the
only galaxies to be found were extremely luminous ellipticals, or cD
type brightest cluster galaxies (BCG). The question of how the
properties of the host galaxies studied here relate to those of BCGs
is explored further in Section 7. In further support of the
radio-loud unification scheme the scale-length figures obtained here
for the RG and RLQ sub-samples can be compared with the $B$- and
$V$-band study of Smith \&\ Heckman (1990), who determined the
scale-lengths of 41 powerful radio galaxies in the redshift range
$0<z<0.26$. The median scale-length figure determined is equivalent to
$\sim17$ kpc in the cosmology adopted here, somewhat larger than the
results presented above. However, if the comparison is restricted to
the 22 objects studied by Smith \& Heckman which have 
strong optical emission lines, more likely to have the same FR
II morphology as the majority of radio-loud objects studied here, 
then the median
scale-length falls to 13 kpc, in excellent agreement with the figures
presented above. 

If the host scale-lengths are plotted against the integrated model
luminosities, as in Fig 5, a tight correlation is found, and 
a least-squares fit to these data
produces a relation of the form $L\propto r^{0.75}$. This is in impressively
close agreement with the corresponding result $L\propto r^{0.70}$ found 
for low-$z$ inactive ellipticals by Kormendy (1977). Given this
result, it is obviously of interest to ask whether the best-fit model
parameters for $\mu_{1/2}$ and $r_{1/2}$ given in Table 3 
do in fact yield a Kormendy relation of slope $\sim3$.

\subsection{The Kormendy Relation}
\label{korsec}
\begin{figure}
\vspace{8.7cm}
\centering
\setlength{\unitlength}{1mm}
\includegraphics{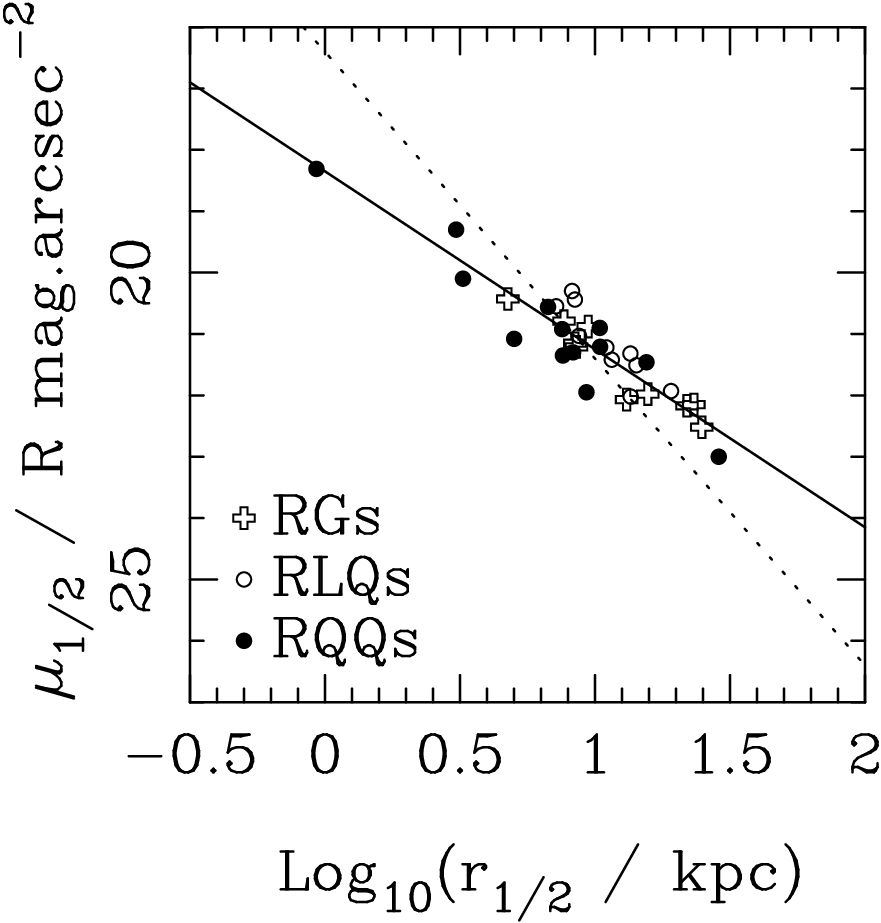}
\caption{The Kormendy relation followed by the hosts of all 33-objects
in the HST sample.  For the four RQQs which have a significant disc
component the best-fitting bulge component has been plotted.
The solid line is the least-squares fit to the data
which has a slope of 2.90, in excellent agreement with the slope of
2.95 found by Kormendy (1977) for inactive ellipticals in the
$B$-band. The dotted line has a slope of 5, indicative of 
what would be expected if 
the scale-lengths of the host galaxies had not been properly constrained
(see text)}
\label{kormendy}
\end{figure}

The $\mu_{1/2}-r_{1/2}$ relation for the host galaxies of all
33 objects is shown in Fig 6. For the four RQQ sources which have
combined disc/bulge fits it is the parameters for the bulge component
that have been plotted. The least-squares fit to the data (solid line)
has the form:  

\begin{tabbing}
$\mu_{1/2} = 2.90_{\pm0.22}\log_{10} r_{1/2}  + 18.35_{\pm 0.22}$ 
\hspace{1cm} \= (ALL)
\end{tabbing} 

\noindent
(errors are $\pm 1\sigma$) showing for the first
time from optical imaging, that the host galaxies
of quasars lie on this photometric projection of the fundamental plane. 
The individual fits to the $\mu_{1/2}-r_{1/2}$ relations for the 
individual sub-samples are:

\begin{tabbing}
$\mu_{1/2}=2.86_{\pm0.32}\log r_{1/2}+18.44_{\pm0.36}$ \hspace{1cm} \= (RG)\\
$\mu_{1/2}=3.98_{\pm0.71}\log r_{1/2}+17.02_{\pm0.75}$ \> (RLQ)\\
$\mu_{1/2}=2.99_{\pm0.34}\log r_{1/2}+18.39_{\pm0.30}$ \> (RQQ)
\end{tabbing}

It is readily apparent from the individual Kormendy relations that the
RG and RQQ sub-samples follow very similar relations, consistent with
each other in terms of slope and normalization. One encouraging 
aspect of this is that the best-fitting bulge components
for the RQQ objects which have combined disc/bulge model fits 
lie naturally on the
Kormendy relation defined by the other nine, single component, RQQ
hosts. With no restrictions placed on the range of parameter values
available to the modelling code, the fact that the fitted bulge components
have physically sensible values gives further confidence that the
introduction of the combined disc/bulge fits was justified, and does
not involve over-fitting of the data. 

It would
appear from the individual fits given above that the RLQ sub-sample 
follows a steeper slope, compared with the best-fit relation to the other
two sub-samples. It is also noticeable that the 1 $\sigma$ error
returned by the least-squares fitting procedure is more than twice
that returned from the fitting to the RG sub-sample (which also has
10 objects), suggesting that there may well be outliers within the RLQ
sub-sample which are biasing the fit.

This is indeed the case, with two outlying RLQ hosts 
(1004+130, 2141+175), one of which (1004+130) is the most nuclear-dominated 
RLQ in the sample, apparently distorting the fit due to the 
probable under-estimation of their host-galaxy scale-lengths.
If these two objects are removed the best-fitting relation for the 
RLQ hosts becomes:
\begin{equation}
\mu_{1/2} = 3.19_{\pm0.67}\log_{10} r_{1/2}  
+ 17.95_{\pm 0.72}
\end{equation}
\noindent

Thus, with the exclusion of the two outlying RLQ hosts, the three
Kormendy relations followed by the sub-samples are internally consistent 
with each
other and, perhaps more importantly, formally inconsistent with a
slope of 5 which, as
highlighted by Abraham et al. (1992), would be a clear indication that 
the data and/or the modelling
procedure employed to analyse the images is unable to break the
$\mu_{1/2}-r_{1/2}$ degeneracy.

We stress this point here because some authors 
have claimed to have demonstrated that AGN hosts follow a Kormendy 
relation, when in fact the derived relation simply reflects a failure
to constrain galaxy scale-lengths. Indeed some authors
(e.g. de Vries et al. 2000) have even mistakenly interpreted
a slope of 5 as evidence that AGN hosts do not follow the same Kormendy
relation as normal giant ellipticals. 

The reason that a `pseudo Kormendy relation' with a slope of 5 can be 
produced with inadequate data and/or modelling follows 
simply from the fact that the integrated luminosity of a
galaxy model follows the relation:

\begin{equation}
L_{int} \propto I_{1/2} r_{1/2}^2
\end{equation}

\noindent
where $I_{1/2}$ is the surface-brightness at $r_{1/2}$. Given that $\mu_{1/2}
\propto -2.5\log{I_{1/2}}$, it follows that if the modelling
procedure can successfully constrain the host luminosity, but not the
scale-length, then the apparent best-fitting values 
of $\mu_{1/2}$ and $r_{1/2}$ will be
randomly distributed along a relation obeying:

\begin{equation}
\mu_{1/2} \propto 5\log{r_{1/2}}
\end{equation}

\noindent
with appropriate normalization to fit the integrated luminosity. Due
to the fact that the host galaxies 
display a relatively small spread in luminosity
(mean $M_R = -$23.44, $\sigma=0.64$) it might be expected that, {\it if} our
analysis was unable to accurately determine the individual
host scale-lengths, the resulting $\mu_{1/2}-r_{1/2}$ relation would be well
fitted with a slope of 5 (as found by de Vries et al. 2000) 
and a normalization to match
M$_{R}=-23.44$. Such a relation has the form:

\begin{equation}
\mu_{1/2} = 5.0\log_{10} r_{1/2}  
+ 16.40
\end{equation}

\noindent
and is plotted as the dotted line in Fig 6. It is obvious
from Fig 6 that this relation is not consistent with the
data, giving confidence that the methods of
analysis employed here have allowed the accurate determination of the
host-galaxy scale-lengths.

The diagnostic power of Fig 6 is discussed in the context of recent 
studies of the fundamental plane in Section 7.2.

\subsection{Axial Ratios}   
\begin{figure}
\vspace{14cm}
\centering
\setlength{\unitlength}{1mm}
\includegraphics{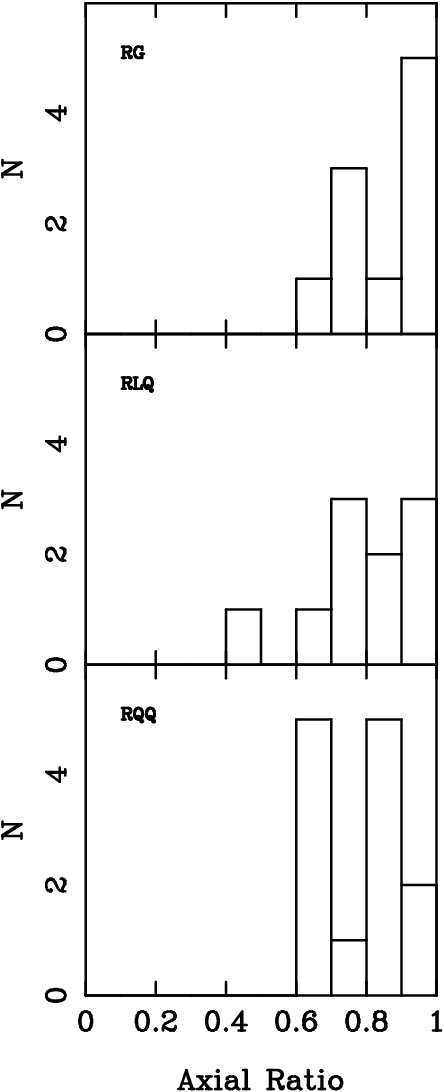}
\caption{The axial ratio distributions for the three host-galaxy
sub-samples as determined by the two-dimensional modelling.}
\label{axialhist}
\end{figure}
If the AGN host galaxies are indeed indistinguishable from massive
inactive ellipticals then they should display an axial ratio
distribution which is also identical. The axial ratio distribution of
normal ellipticals is well studied and known to peak at values of
 $b/a\ge0.8$ (Sandage, Freeman \& Stokes 1970; Ryden 1992). The
best-fitting axial ratios from the two-dimensional modelling are
listed in Table 3  and displayed in the form
of separate sub-sample histograms in Fig 7. The
corresponding mean and median values are: 

\begin{tabbing}
$\langle b/a \rangle =0.81\pm0.02$ \hspace{0.5cm} \= median=0.85 \hspace{0.5cm} \= (ALL)\\
$\langle b/a \rangle =0.86\pm0.03$ \> median=0.90 \> (RG)\\
$\langle b/a \rangle =0.80\pm0.05$ \> median=0.82 \> (RLQ)\\
$\langle b/a \rangle =0.78\pm0.03$ \> median=0.84 \> (RQQ)\\
\end{tabbing}

\noindent
It can be seen from these figures that the axial ratio distributions
of all three sub-samples are perfectly consistent with that expected
from a sample of elliptical galaxies. There is a slight suggestion
that the RGs have higher axial ratios than
average, but this is
not formally significant.

The axial ratio results from the two-dimensional modelling agree well
with the recently-published findings of Boyce et al. (1998) ,who found that
all 11 of the quasars from their 14-object ($z\simeq 0.3$)
 sample for which a model
fit was possible, displayed axial ratios with $b/a>0.65$. In contrast,
Hooper et al. (1997) found only 2 of their 16 ($z\sim0.45$) quasars
to have axial ratios with $b/a >0.6$. The reasons for this apparently
contradictory result probably lie in a combination of the higher
redshifts of the Hooper et al. objects, and their use of the PC instead
of the WF detectors. As has been explained, these two factors will
undoubtedly result in a reduced sensitivity to low surface-brightness
features, given that the exposure times used were the same as adopted
in this study. In fact, Hooper et al. noted their 
concern that their modelling technique of two-dimensional
cross-correlation could have been influenced by high surface-brightness 
features such as bars or tidal tails. Given
the clear result presented here, and the support of the Boyce et al.
results at similar redshifts, this seems the most likely 
explanation of the Hooper et al. result.

\subsection{Host-Galaxy Colours}

\label{colours}
\begin{figure}
\vspace{14cm}
\centering
\setlength{\unitlength}{1mm}
\includegraphics{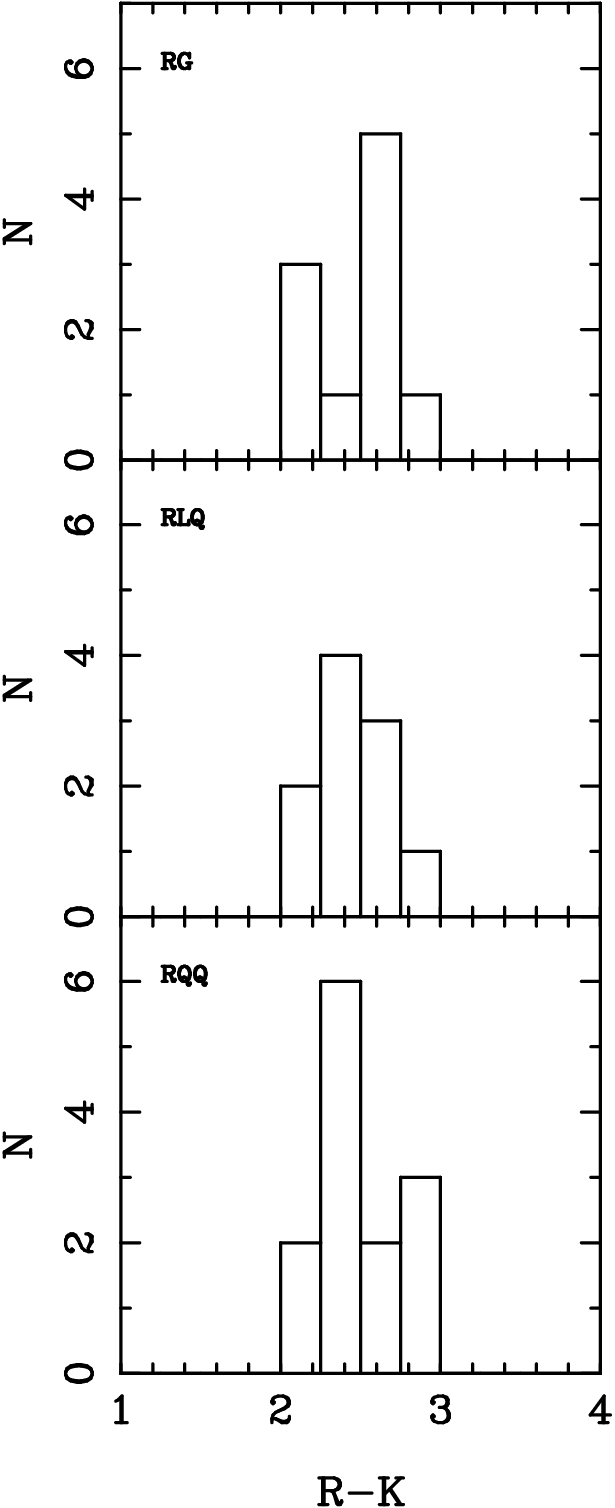}
\caption{The rest-frame $R-K$ colours for the three host-galaxy
sub-samples. The three sub-samples can be seen to be consistent with each
other, tightly distributed around a value of $R-K \simeq 2.5$.}
\label{colourhist}
\end{figure}

The results presented in the last five sub-sections strongly suggest that
in terms of morphology, luminosity, scale-length, Kormendy relation, 
and axial-ratio distribution the 
hosts of powerful AGN are identical to normal, massive, inactive
ellipticals. The one final parameter
which can be readily recovered from the modelling is host-galaxy
colour. The desire to obtain reliable optical-infrared colours for the
host galaxies was one of the original motivations for this HST imaging
study. Given that the results presented thus far suggest the host
galaxies are otherwise normal massive ellipticals, it is now even more
interesting to investigate whether the hosts also display the red
colours associated with old stellar populations, or whether they have
significantly bluer colours, indicative of either a generally young
stellar population, or of substantial secondary star formation
induced by interactions, the central AGN, or both. 

The simplest and most obvious way to calculate 
$R-K$ colours for the host galaxies is to combine the integrated optical
magnitudes presented in Table 5 with the integrated absolute $K$-band 
magnitudes derived for the sample by Taylor et al. (1996).
However, there are two potentially serious problems 
with this strategy. First, Taylor et al. did not always decide on 
the same host morphology as has now 
been revealed by the HST imaging, and so in some cases their 
quoted absolute $K$ magnitude is derived from, for example, 
a disc fit which originally 
appeared marginally preferable to a de Vaucouleurs law.
This obviously matters because the disc fits to the $K$-band data were
$0.5\rightarrow1.0$ magnitudes fainter than the equivalent elliptical
fits. Second, in the light of the new, better constrained models derived 
from the HST data, and the results of more recent tip-tilt infrared imaging 
reported by McLure, Dunlop \& Kukula (2000), it is now clear
that the half-light radii derived by Taylor et al. were systematically 
over-estimated (most likely due to the $\sim 1\asec$ seeing
and coarse spatial resolution ($0.62\asec$/pix) of the original K--band
observations - see also Simpson 
et al. 2000). 
Given that the
integrated luminosity of the standard de Vaucouleurs $r^{1/4}$ law is
proportional to $I_{1/2} r_{e}^2$, it is clear that scale-length errors of the
order of $1.5-2$ could seriously bias the integrated
$K$-band luminosity. 

Therefore, to obtain a robust value for the $R-K$ colour of each host from 
the existing data we proceeded as follows. First, the 
$K$-band luminosities were based on the best-fitting model
with the same morphology as the equivalent HST result. Second, the $R-K$
colours were based on $12\asec$-diameter 
aperture photometry performed on both the HST and $K$-band
model fits, to minimize the impact of scale-length errors on 
the derived colour, while at the same time including the vast majority 
($>75$\%) of host-galaxy light.

The results of these host-galaxy $R-K$ calculations are listed in Table 5,
with absolute $R-K$ histograms for the separate
sub-samples shown in Fig 8. The absolute colours have been
calculated assuming 
spectral indices of $\alpha=1.5 \ \& \ \alpha=0.0$ for the $R$- and
$K$-bands respectively. 

It is readily apparent from Fig 8 that the colour
distributions of the three sub-samples are consistent with each
other, with mean and median absolute colours:\\

\begin{tabbing}
$\langle R-K \rangle =2.48\pm 0.05$ \hspace{0.5cm} \= median=2.44 \hspace{0.5cm} \= (All)\\
$\langle R-K \rangle =2.47\pm 0.09$ \> median=2.56 \>(RG)\\
$\langle R-K \rangle =2.48\pm 0.07$ \> median=2.46 \>(RLQ)\\
$\langle R-K \rangle =2.48\pm 0.08$ \> median=2.41 \>(RQQ)\\
\end{tabbing}

\noindent
As demonstrated by Nolan et al. (2001) such colours are consistent 
with those expected of an evolved  stellar population of age $10-13$ Gyr.
The homogeneity of these colours provides a further strong indication that
the hosts of all three classes of AGN are derived from the same parent 
population of massive, well-evolved elliptical galaxies.
The inevitable
conclusion from this is that these galaxies, or at least the
bulk of their stellar
populations, must have formed at high redshift, a theme
which is revisited in Section 9. The fact that the
$R-K$ colours are so similar to passive ellipticals forces the further
conclusion that any star-formation associated with AGN activity must either
be dust enshrouded, confined to tidal features masked from the
modelling, or possibly restricted to the central $\sim 1$~kpc
of the host (in which case the additional emission would
almost certainly be attributed to the unresolved nuclear 
component). Large-scale
star-formation involving a substantial fraction of the mass of the 
host galaxy, is effectively ruled out by these results.
  
\subsection{Nuclear Colours: Implications for Unified Models of Radio-Loud AGN}

\begin{figure}
\vspace{8.7cm}
\centering
\setlength{\unitlength}{1mm}
\includegraphics{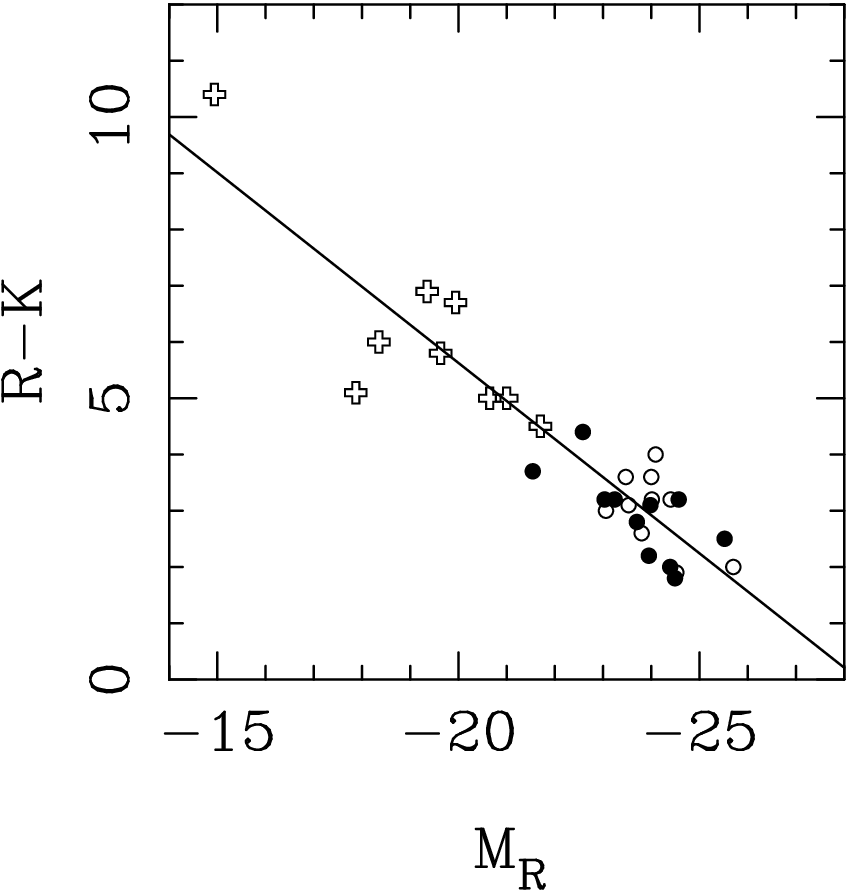}
\caption{Plot of apparent $R-K$ colour versus absolute $R$-band
luminosity of the fitted nuclear components of the objects in the HST
sample. Shown in the plot are the RG nuclei (crosses), RLQ nuclei
(open circles) and
RQQ nuclei (filled circles). Also shown is the least-squares fit to the data
which has a slope of $0.68\pm0.05$.}
\end{figure}

The $R-K$ colours of the fitted nuclei for all 33 objects (except the RG
0230$-$027 - see above) are plotted against absolute nuclear $R$ magnitude 
in Fig 9. It is clear from this plot that there is no significant 
difference between the colours of the nuclei of the RQQs or RLQs. More 
importantly, however, this plot allows us to explore the relationship between
the relatively weak nuclear components found in the RGs, and the (naturally)
brighter components found in the RLQs. 

This is of value because it allows us to perform another test of the 
proposed unification of RLQs and RGs by orientation 
(Peacock 1987, Barthel 1989). As discussed above, our host-galaxy results 
can already be viewed as providing strong support for such a picture, due to 
the fact that the basic properties (i.e. luminosity, size,
axial ratio, age etc) of the RGs and RLQ hosts in our matched samples are
indistinguishable. However, it must also be realised that identical 
RG and RLQ host galaxy properties would also be expected 
in a scenario where the two classes of radio-loud AGN are in fact 
linked by evolution in time, rather than by orientation.

Fig 9 offers the possibility of differentiating between
these two alternative explanations, by allowing us to test 
whether the $R-K$ colours of the RG nuclear
components are consistent with 
dust-obscured quasar nuclei. If it is assumed that the
reddening of the RG nuclei obeys a $\kappa_{\lambda} \propto \lambda^{-1}$
law, then the reddening vector which should connect the RGs and RLQs in Fig 
9 would be expected to be given by $E(R-K)=0.70A_{R}$.

In fact the slope of the least-squares fit in Fig 9 (solid line) is
$0.68\pm 0.05$, which corresponds to a dust reddening law 
$\kappa_{\lambda} \lambda^{-0.95\pm 0.15}$. 
Furthermore, if the fitted $R$--band
nuclear components of the RGs are de-reddened using a slope of $0.68$,
until each object has the mean colour of the RLQ nuclei ($R-K=3.0$), 
then the mean luminosity of the de-reddened RG nuclei is
$M_{R}=-23.90\pm 0.46$. It can be seen that this is perfectly consistent
with the previously determined mean nuclear luminosity of the RLQ sample,
$M_{R}=-23.73\pm0.10$. Given the clean nature of this result, there
seems little room for argument that the RG and RLQ sub-samples are
most consistent with being drawn from the same population of objects, 
viewed from different orientations, with every RG displaying 
evidence of a buried quasar nucleus.

\section{The relation of quasar hosts to `normal' galaxies}

\subsection{Dependence of Host Morphology on Nuclear Luminosity}

In the preceding section we presented a series of results which together 
lead to the inescapable conclusion that the hosts of virtually 
all powerful AGN are essentially normal, massive
ellipticals. This result is particularly clean for the radio-loud
objects in our sample. However, exceptions to any rule are often of great 
significance, and one should not ignore the fact that we have found two 
disc-dominated hosts in our RQQ sample, along with two 
bulge-dominated RQQ hosts which do, however, possess significant 
disc components. 

What might such exceptions to the general `perfect elliptical' rule be 
telling us? An important clue comes from the fact that, as discussed above,
the two disc-dominated hosts in our sample transpire to contain the two 
least-luminous 
nuclei. Indeed, in much of the preceding analysis, these two objects 
have had to be deliberately excluded from sample 
comparisons due to the fact that 
they are not actually luminous enough to be classed as quasars. However, 
the fact that these two radio-quiet low-luminosity interlopers are the only
RQQs in our sample with disc-dominated hosts strongly suggests that host 
morphology is a function of nuclear {\it optical} luminosity.

Our own samples do not span a sufficient range in optical luminosity to 
allow a statistical test of this hypothesis. However, we can explore this 
possibility, and place our results in a wider context, by combining them with 
those of Schade et al. (2000) who have studied the morphologies of the host 
galaxies of a large sample of X-ray selected AGN spanning a wider but lower 
range of optical luminosities. This we have done in Fig 10, which shows the 
ratio of bulge-to-total host luminosity plotted as a function of nuclear
optical power. 

This instructive diagram illustrates a number of important points. First, it 
shows that while pure spheroidal galaxies can host AGN with optical 
luminosities ranging over 
two orders of magnitude, very disc-dominated hosts start to die out for
$M_V (nuc) < -21$ and no disc-dominated galaxies appear capable 
of hosting nuclear emission more luminous than $M_V \simeq -23$ (close to
the traditional, albeit rather ad hoc, quasar:seyfert classification boundary).
Second, this plot confirms that the two most `discy' hosts uncovered in the 
present study do indeed lie in a region of the parameter space where 
disc-dominated host galaxies 
are quite common. Third, in the context of this diagram,
the persistence of significant discs around two of the more luminous RQQs in
our sample (0052+251 and 0157+001) seems perfectly natural, and certainly not
inconsistent with the clear trend towards more spheroid-dominated hosts 
with increasing nuclear output. 

Fig 10 clarifies the origin of much of the confusion which has surrounded 
the nature of RQQ hosts. In particular it is clear that studies of RQQ samples 
which extend to include significant numbers of objects with nuclei less 
luminous than $M_V \simeq -23$ are likely to uncover significant numbers of 
hosts with substantial discs, in contrast to the results of the present study.
This does indeed appear to be the case. For example, Bahcall et al. (1997) 
and Hamilton et al. (2002) have reported that approximately one third to one 
half of radio-quiet quasars lie in disc-dominated hosts. However, if one 
restricts their results to quasars with nuclear magnitudes $M_V < -23.5$
(for which we find 10 out of 11 RQQs lie in ellipticals), it
transpires that 6 out of the 7 remaining objects in the Bahcall et al. study
lie in ellipticals, while 17 out of the remaining 20 objects in the archival 
study of Hamilton et al. also have elliptical host galaxies.

\begin{figure}
\vspace{9cm}
\centering
\setlength{\unitlength}{1mm}
\includegraphics{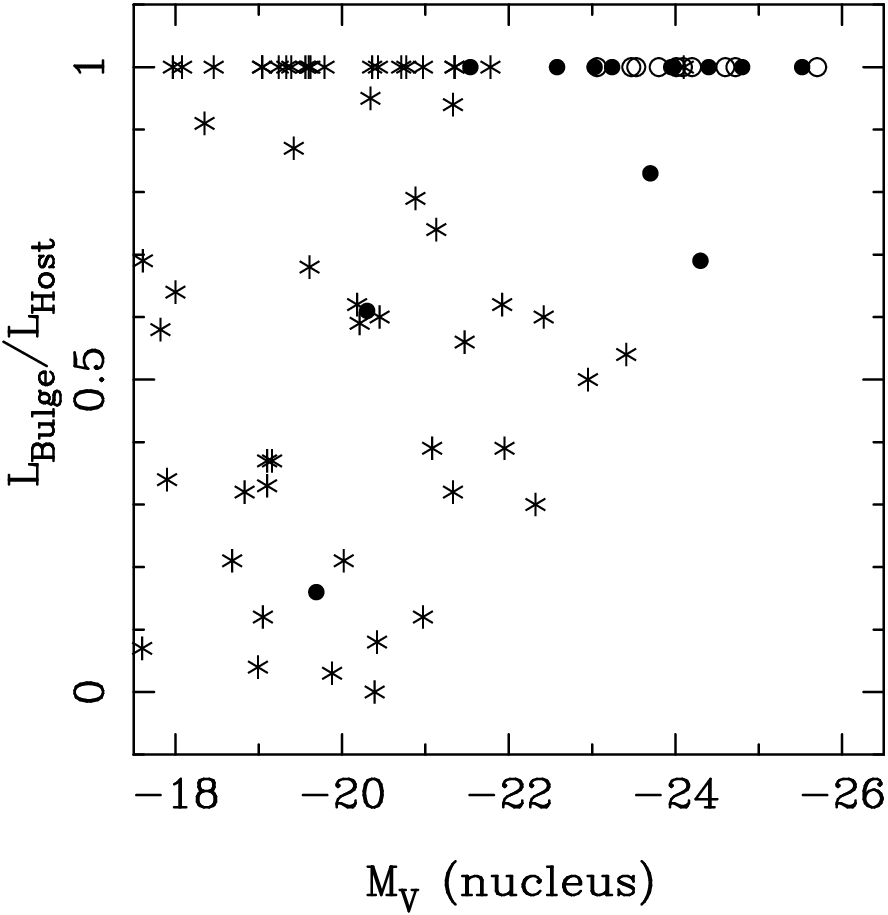}
\caption{Plot of relative contribution of the bulge/spheroidal 
component to the total luminosity of the host galaxy versus absolute $V$-band
luminosity of the fitted nuclear components. The plot shows the results
for our own HST sample (RLQs as open circles, RQQs as filled circles)
along with the results from Schade et al. (2000) for a larger sample of
X-ray selected AGN (stars). 
This plot illustrates very clearly how disc-dominated
host galaxies become increasingly rare with increasing nuclear power, as
is expected if more luminous AGN are powered by more massive black holes
which in turn are housed in more massive spheroids.}
\end{figure}

Fig 10 is also at least qualitatively as expected if AGN are in fact
drawn at random from the general galaxy population, subject to the constraint
that the host galaxy contains a black hole of sufficient mass to produce the
observed nuclear output (see also Wisotzki, Kuhlbrodt \& Jahnke 2001). As discussed further
in Section 8, a black-hole mass $m_{bh} > 5 \times 10^8~{\rm M_{\odot}}$
appears to be required to produce a quasar with $M_V < -23.5$. When this 
constraint is 
combined with the now apparently inescapable result that black-hole mass is 
proportional to spheroid mass $M_{bh} \simeq 0.001 - 0.003 M_{sph}$, 
then Fig 10 can be viewed as a simple 
manifestation of the fact that spheroids in the present-day universe with 
baryonic masses $> 3 \times 10^{11} {\rm M_{\odot}}$ are 
rarely accompanied by significant discs (and are thus classed as massive 
ellipticals).

This conclusion leads naturally to the question of why most massive black holes
in the present-day universe should be inactive, while a subset are receiving 
sufficient fuel to shine as quasars. Below we present the results of a series 
of statistical comparisons designed to decide whether there are in fact 
{\it any} significant observable differences between the hosts (and host 
environments) of quasars and comparably massive, but inactive, elliptical 
galaxies.

\subsection{Comparison with ULIRG merger remnants}

Before proceeding to compare quasar hosts with old passively-evolving massive
ellipticals, it is worth pausing to consider whether host galaxies with the
properties we observe could plausibly be produced as the outcome of a 
(relatively recent) violent merger of two disc galaxies. This might already 
seem fairly implausible given the series of results presented in Section 6,
in particular the red host-galaxy colours consistent with old evolved 
populations. Nevertheless, it is well known from both simulations 
(Barnes \& Hernquist 1992)
and observations that the red$\rightarrow$infrared light from 
ULIRGs often follows an 
$r^{1/4}$ law (Wright et al. 1990, Scoville et al. 2000) and it is possible 
that dust extinction in such objects could also produce red colours
(albeit such colours would be unlikely to mimic so well the 
properties of an evolved population, as discussed by Nolan et al. 2000). 
Moreover,
several authors have proposed an evolutionary scheme in which ULIRGs might 
evolve into radio-quiet quasars on a time-scale $< 1$~Gyr (Sanders \& 
Mirabel 1996).

Fortunately, the data on quasar hosts presented here are of sufficiently 
high quality to completely exclude the possibility that they are the recent 
outcome of the violent disc-disc mergers which appear to be the origin of most
ULIRGs in the low-redshift universe. While it is true that ULIRGs such as 
Arp220 have surface brightness profiles well-described by an $r^{1/4}$-law, 
recent observations (Genzel et al. 2001) have shown that such remnants lie in
a completely different region of the fundamental plane than that which we have shown is occupied by quasar hosts. Specifically, the effective radii of 
ULIRGs is typically 
$\simeq 1$~kpc, an order of magnitude smaller than the typical effective radii
of the quasar hosts as illustrated in Figs 4, 5 and 6. 

\begin{figure}
\vspace{9cm}
\centering
\setlength{\unitlength}{1mm}
\includegraphics{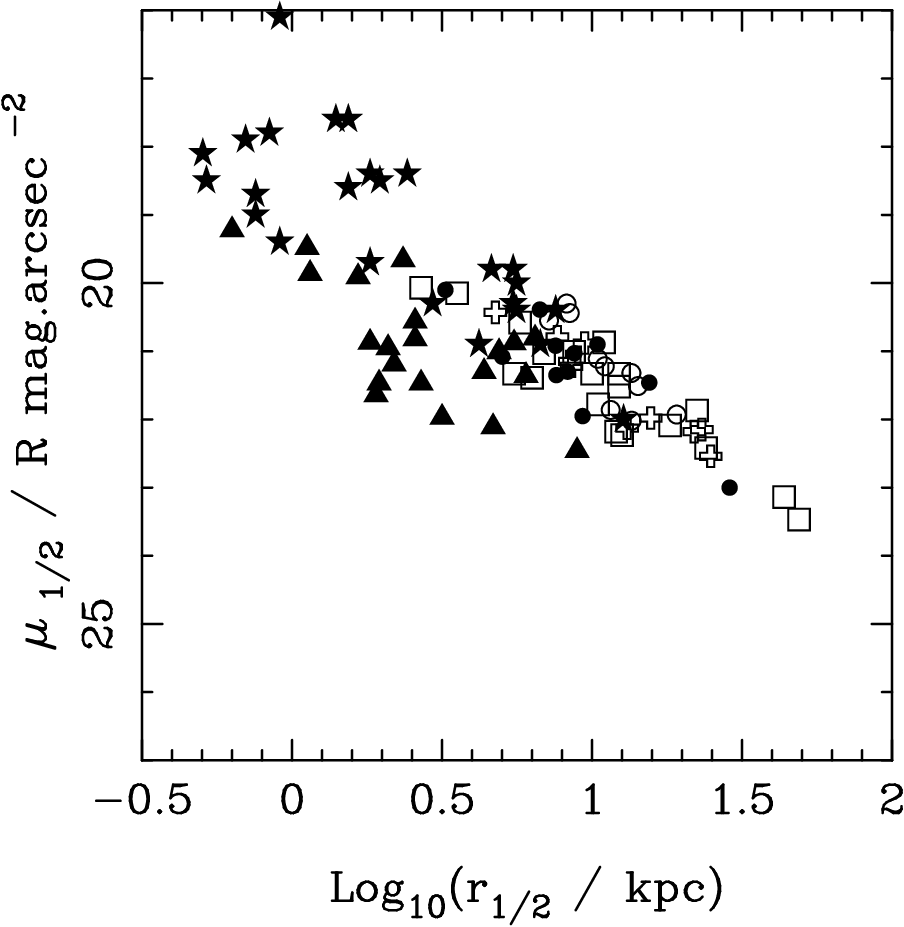}
\caption{A comparison of the properties of the AGN hosts with those 
displayed by various other types of spheroid on the photometric 
projection of the fundamental plane. Symbols for the quasar hosts 
and radio galaxies are as in Figs 5 and 6. The stars are the data 
for ULIRGs and LIRGs from Genzel et al. (2001) transformed from the infrared
to the $R$-band assuming $R-K$=2.5. 
Triangles and squares indicate the positions
of `discy' and `boxy' ellipticals from Faber et al. (1997) after conversion 
to $H_0 = 50~{\rm km s^{-1} Mpc^{-1}}$.}
\end{figure}

Indeed, one can go further and conclude that whereas ULIRGs can be 
the progenitors of the 
population of intermediate-mass ellipticals which display compact cores and
cusps (Faber et al. 1997), the quasar hosts lie in a region of the 
$\mu_e - r_e$ plane which is occupied by boxy, giant ellipticals with large 
cores, a large 
fraction of which lie in the centres of clusters (Faber et al.
1997). This is demonstrated
in Fig 11 where we have constructed a composite $R$-band Kormendy diagram,
combining our own results on the 31 bulge-dominated AGN hosts
with the data from Genzel et al. (2001)
on LIRGs and ULIRGs, and the data from 
Faber et al. (1997) on `discy' and `boxy' ellipticals.

This plot provides strong evidence that 
the quasar hosts belong to the class of large `boxy' ellipticals,
which various authors have suggested form at earlier times than their
lower-mass `discy' counterparts (Kormendy \& Bender 1996, 
Faber et al. 1997).

In summary, all the available evidence indicates that luminous low-redshift 
quasars are the result of the re-triggering of a massive black hole at the 
centre of an old evolved elliptical, and do not generally 
occur in the remnants 
of the recent disc-disc mergers which power ULIRGs. 
While there is one RQQ 
in our sample which is also a ULIRG (0157+001), 
in this case it is clear that at least
one of the merging galaxies already has a massive spheroid capable of 
containing a pre-existing massive black hole. Thus, while there is clearly
some overlap between the ULIRG and quasar phenomenon, present evidence 
suggests that, at least at low redshift, the ULIRG $\rightarrow$ 
quasar evolutionary
route can only apply to a fairly small subset of objects.

\subsection{Comparison with Brightest Cluster Galaxies}
\label{bcgcomp}

Having established that the hosts of quasars are massive ellipticals, 
it is then of interest to explore how their properties compare to those
of inactive comparably-massive ellipticals at similar redshift. In this 
sub-section we compare the properties of the quasar hosts, and in particular 
their scale-lengths, with those of bright galaxies found in rich clusters. This 
comparison is motivated, in part, by the location of the host galaxies 
on the $\mu_e - r_e$ projection of the fundamental plane as discussed
in the previous subsection. However, this motivation is further 
strengthened by the results of a study of the environments of the quasars and 
radio galaxies undertaken by McLure \& Dunlop (2000). 
McLure \& Dunlop (2000) used 
the full WFPC2 images of our AGN sample to measure the spatial clustering 
amplitude $(B_{gq})$. This quantity is inevitably rather
poorly constrained for each individual object, but, on average, they found that
all three classes of AGN typically inhabit environments as rich as Abell 
Class $\simeq 0$. Moreover, several objects appeared to lie in noticeably 
richer environments (Abell Class 1).

Given these clustering results, and the fact that the quasar hosts are clearly
comparable to many BCGs in terms of optical luminosity, it is interesting to
explore whether the sizes of the quasar hosts are as expected in the light 
of their apparent cluster environment. Strong circumstantial evidence that 
the quasar host galaxies are directly comparable to first-ranked cluster 
galaxies comes from a comparison of the host-galaxy Kormendy relation
with that found for cD galaxies by
Hamabe \& Kormendy (1987) in the $B$-band:
\begin{equation}
\mu_{1/2} = 2.94\log_{10} r_{1/2}  
+ 20.75
\end{equation}
\noindent
If a typical elliptical galaxy colour at $z=0.2$ of $B-R=2.4$ is
assumed (Fukugita et al. 1995), then the inferred $B$-band 
Kormendy relation for all 33
objects in the HST sample presented in Section 6.4 becomes:
\begin{equation}
\mu_{1/2} = 2.90\log_{10} r_{1/2}  
+ 20.75
\end{equation}
\noindent
which can be seen to be in excellent agreement with the 
Hamabe \&\ Kormendy result. 

\begin{table*}
\begin{center}
\caption{The model results for the four brightest galaxies in each 
of five clusters obtained from the HST archive, spanning a range in richness
from Abell Class 0 to Abell Class 4. The redshifts of the clusters
are 0.11, 0.19, 0.23, 0.175 \&\ 0.171 respectively. 
The images of A0018, A0103 and A2390 were taken through
the F702W (wide $R$) filter.
The images of A2390 and A1689
were taken through the F814 ($I$-band) filter.
The conversion between
$I$- and $R$-magnitudes has been performed assuming $R-I=0.8$
(Fukugita {\it et al.}). The final reduced images
had signal-to-noise comparable to the HST host-galaxy data.}
\begin{tabular}{ccccccccccc}
\hline
\hline
&Class&0&Class&0  &Class&1 &Class&4 &Class&4\\
&Abell&0018&Abell&0103 &Abell   &2390 &Abell &1689 &Abell&2218\\  
\hline
Galaxy &$r_{e}$/kpc&M$_{R}$&$r_{e}$/kpc&M$_{R}$&$r_{e}$/kpc&M$_{R}$&$r_{e}$/kpc&M$_{R}$&$r_{e}$/kpc&M$_{R}$\\
\hline
1&30.5&$-$22.65&24.9&$-$23.62&27.8&$-$24.52 &75.8&$-$25.33 &100.5&$-$25.51   \\
2&25.4&$-$22.50& \phantom{2}3.4&$-$22.74&21.2&$-$23.14 &59.0&$-$24.70 &\phantom{1}51.9&$-$24.54  \\
3&16.6&$-$21.47& \phantom{2}7.6&$-$22.69&12.1&$-$22.99 &18.7&$-$23.62 &\phantom{1}35.7&$-$23.68   \\
4&\phantom{1}5.2&$-$21.14 & \phantom{2}7.6&$-$22.69&\phantom{1}4.1&$-$22.21  &\phantom{1}7.6&$-$23.50  &\phantom{1}28.8&$-$23.29  \\
\hline
\hline
\end{tabular}
\end{center}

\label{abellres1}
\end{table*}

To explore further the connection between host cluster environment
and galaxy size, a sample of 51 first-ranked cluster galaxies 
(confined to the same redshift range as the HST sample 
but spanning a range of Abell classes) was constructed from the
study of Hoessel \&\ Schneider (1985).
Application of the KS
test to the scale-length distributions of the AGN host galaxies, and the
14 first-ranked cluster galaxies in this sample 
from Abell clusters of class 0 \&\ 1,
shows that the two are not significantly different ($p=0.22$). However,
the extension of the comparison sample to include the first-ranked galaxies
of Abell class 2 clusters (15 objects), shows these two distributions
to be different at the $2\sigma$ level ($p=0.011$). 

This is an very interesting result because it provides independent 
support for the results of the cluster analysis of McLure \& Dunlop (2000)
which also favour an environment more consistent with Abell class 0, or at 
most (in only a few cases) Abell class 1.
 
Finally, to check that the ground-based results discussed above 
can be confirmed by modelling of more recent HST data, and to minimize possible
systematic errors, we decided to model a series of HST images of 
bright cluster galaxies (at comparable redshift to our quasars) 
using the same 2-D modelling code used to model the AGN hosts.
HST WFPC2 images of five
Abell clusters with similar redshifts to the objects in the HST sample
were therefore retrieved from the HST archive 
facility\footnote{http://archive.stsci.edu}. Following the production of
reduced images of comparable
signal-to-noise to the host galaxy images, the four brightest galaxies
in each cluster were identified and modelled in identical fashion to
the AGN hosts. The results of the elliptical model fits to
these 20 galaxies are presented in Table 6. 

The scale-lengths and absolute magnitudes listed for the BCGs in 
Table 6 can be instructively compared with the corresponding 
quantities for the AGN hosts listed in Tables 3 and 5. 
First, it is clear 
that none of our AGN hosts is comparable in size to either the first or second ranked cluster galaxies in very rich, Abell Class 4 clusters.
However, the {\it largest} of the AGN hosts are comparable in size to
the 1st or 2nd ranked galaxies in the Abell Class 0 and Class 1 clusters. 
This more detailed galaxy-by-galaxy comparison thus provides further
support for the basic conclusions of McLure \& Dunlop (2000) already mentioned above. Indeed, at least within the radio-loud subsamples, there is a 
suggestion of a rather clean link between the biggest galaxies and the 
richest cluster environments. Specifically, the two radio galaxies found by 
McLure \& Dunlop (2000) to have the richest 
environments (1342$-$016 (see Fig 12) \& 
0917+459) transpire (see Table 3) to be two out of the three RGs with $r_e > 
20$~kpc. Moreover, both the values of $B_{gq}$ for these objects
($\simeq$1500 and $\simeq$1000) and their half-light radii both 
point towards the same conclusion (given Table 6), namely that these most
massive hosts are the first- or second- ranked BCGs in clusters of richness
Abell Class 0$\rightarrow$1. A similar remarkably clean correspondence 
can be found within the RLQ sample. The two RLQs found by 
McLure \& Dunlop (2000) to have the richest environments (2349$-$014 and 
0137$+$012) transpire (see Table 3) to be the only two RLQs with hosts
for which $r_e > 14$~kpc.
However, no such clean correspondence 
is evident within the RQQ sample. This may mean that 
RQQs, even when found in cluster environments, are less likely to be central cluster galaxies than their radio-loud counterparts. However, given the large 
errors on $B_{gq}$ such a conclusion may perhaps be premature.

In summary, this comparison of the properties of the hosts of powerful AGN
and bright cluster galaxies supports the basic conclusion reached by McLure \& Dunlop (2000) via the completely independent measure of clustering amplitude.
The larger AGN hosts have properties comparable to first or second ranked BCGs 
in Abell class 0$\rightarrow$1 clusters. Both radio-loud and radio-quiet 
quasars thus appear to occur in a range of
environments  which is certainly consistent with the range of environments
occupied by inactive ellipticals of comparable size and luminosity. While 
none of our host galaxies are comparable to the largest known BCGs, we must caution that this does not mean AGN avoid such environments
(as has sometimes been inferred for FRII radio sources
from the results of, for example, Prestage \& Peacock 1988).
In fact, because Abell Class 3 and 4 clusters are very rare, with a 
number density less than $< 1.6 \times 10^{-6}~{\rm h^{3} Mpc^{-3}}$ 
(Croft et al. 1997), even an all-sky survey covering 
the redshift band 
of this study ($0.1 < z < 0.25$) would be expected to contain $< 1500$
such rich clusters. Consequently, given that only 1 in $\simeq 1000 - 10000$ 
massive 
black holes appears to be active in the present-day universe
(see Section 9 and Wisotzki, Kuhlbrodt \& Jahnke 2001), we should not be surprised by the fact that no low-redshift quasar has yet been found to lie
within such a rich environment or hosted by an ultra-massive BCG.

\begin{figure}
\vspace{9cm}
\centering
\setlength{\unitlength}{1mm}
\includegraphics{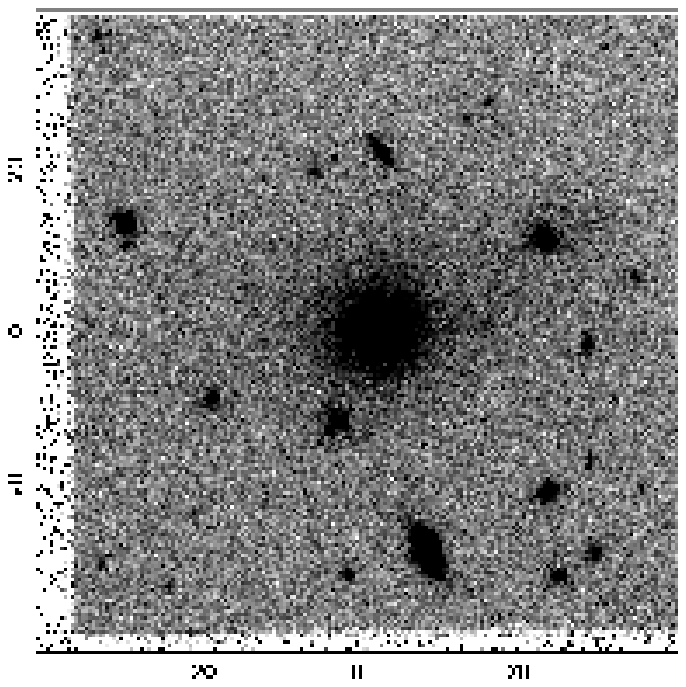}
\caption{The full WF2 image of the radio galaxy 
1342$-$016 covering an area of
$80\asec\times80\asec$. A large number of companion objects can be
seen which, if at the redshift of the quasar, would appear consistent
with a moderately rich cluster environment.}
\label{1342full}
\end{figure}

\subsection{Interactions} 
\label{interaction}

It is often stated in the literature (eg. Smith et al. 1986, Hutchings
\&\ Neff 1992, Bahcall et al. 1997) that morphological disturbance is a
common feature of the host galaxies of powerful AGN. In fact, 
at first sight,
the $R$-band images presented in Appendix A and in McLure et al.
(1999) show a
relatively low occurrence of morphological disturbance, with only three
objects (1012+008, 2349$-$014 \&\ 0157+001) undergoing obvious,
large-scale, interactions. However, as can be seen from the
model-subtracted images, the removal of the best-fitting axisymmetric
model for each host galaxy does reveal the presence of peculiarities 
such as excess flux, tidal tails and close companions, at
lower surface-brightness levels. It is tempting to conclude from this
that many of these AGN may have been triggered into action by recent 
interaction with companion objects. However, the true significance of
the levels of interaction seen in the AGN host galaxies can only be
judged by comparison with an equally-detailed investigation of the
morphologies of a control sample of comparable, inactive galaxies. Here 
we attempt to quantify the significance of such peculiarities by constructing
a control sample based on the twenty bright cluster galaxies taken
from the HST archive discussed in the previous subsection.

With the range of redshifts covered by the five Abell clusters it was
possible to construct 19 object pairs, each consisting of one of our AGN
sample paired with a suitable control, drawn from the 20 cluster
galaxies. Model-subtracted images of each of the galaxy pairs were
then assessed in a blind test for the occurrence of four possible
indicators of disturbance or interaction, i.e:
\begin{itemize}
\item{Residual tidal/spiral arm features,}
\item{Residual asymmetric flux,}
\item{Residual symmetric flux,}
\item{Companion object inside $10\asec$ radius,}
\end{itemize}

\noindent
where neither the identity of the AGN, nor which 
object  in each pair was active, was known to the decision
maker. The results of
this process are shown in Table 7.

The one clear result of
this test is that, taken as a group, the active galaxies 
display a greater occurrence of tidal/spiral arm features in
their model-subtracted images than the control sample. There is no
significant difference in the other three
indicators. It is noteworthy however, that of the six
occurrences of tidal/spiral arms, only one of these (2349$-$014) is a
radio-loud object. To test whether or not this indicates an inherent
difference between the radio-loud and radio-quiet AGN, the blind test
was repeated
for all 33 HST objects, without control galaxies, but still keeping
the identity of each AGN secret, in order to prevent any subconscious
bias. The results of this test are presented in Table 8.

\begin{table*}
\caption{The results of the blind test to determine any difference in
the prevalence of
features indicative of interaction or disturbance in the AGN sample
(active) compared to inactive galaxies, using 20 BCG as an
inactive  control sample (see text for details).}
\begin{tabular}{lcccc}
\hline
\hline
Object Type &Tidal/Spiral&Asymmetric Flux&Symmetric Flux&Companion\\
\hline
Active       &5&8&5&7 \\
\hline
Inactive     &0&10&7&9\\
\hline
\hline
\end{tabular}

\label{blind1}
\end{table*}

\begin{table*}
\caption{The results of the blind test to determine if any
differences exist in the occurrence of features indicative of
interaction or disturbance between the AGN sub-samples (see text 
for details).}
\begin{tabular}{lcccc}
\hline
\hline
Object Type &Tidal/Spiral&Asymmetric Flux&Symmetric Flux&Companion\\
\hline
RQQ &6&5&1&2\\
\hline
RLQ &1&6&2&4\\
\hline
RG &0&4&2&3\\
\hline
\hline
\end{tabular}

\label{blind2}
\end{table*}

Several features of the results presented in Table
8 merit comment. First, it can be seen that
there is no significant difference between the RG and RLQ
sub-samples, consistent with all the other results presented in this paper.
Second, it would initially appear that there is
a significant difference between the radio-loud objects and the RQQ
sub-sample in terms of the occurrence of tidal/spiral arm features.
 However, an examination of which of the RQQs are contributing to this
result shows that four of them are the objects which have combined
disc/bulge model fits. Therefore, if the comparison is restricted to include
only those RQQs with single-component elliptical host galaxies, this
apparent difference disappears.  

The results of these
blind comparison tests thus support two separate
conclusions. First, the finding that there are no detectable
differences between the elliptical AGN host galaxies and the BCG
control sample provides further evidence that these two classes of
object are directly comparable. Second, the apparently more frequent
occurrence of tidal/spiral arm features in the RQQ sub-sample is
only a manifestation of the previously-determined fact that a minority of
RQQ host galaxies, two of which are not actually true quasars, have
substantial disc components, and thus does not provide evidence 
for an inherent
difference in interaction rates between radio-quiet and radio-loud
objects.

Perhaps one should not be surprised by the lack of evidence 
for spectacular large-scale ULIRG-like interactions in our quasar 
sample, given that as described below,
the activation of a quasar requires the delivery of only 
$\simeq 1~{\rm M_{\odot}
yr^{-1}}$ to the central black hole.

\section{The AGN-host connection}

As discussed by McLure et al. (1999), our finding that the hosts of quasars 
are almost all spheroidal galaxies allows exploitation of the  
now well-established correlation between black-hole mass and spheroid mass
to estimate the masses of the central black holes in these objects, completely
independent of any of their observed {\it nuclear} properties. We can then 
use this information to explore which, if any, of the observed properties
of the AGN are linked to black-hole mass.

In this section we revisit this issue for a number of reasons. First, 
of course, our matched sub-samples 
are now complete. Second, in the intervening two years the application of 
3-integral models has resulted in a revising down of the constant of 
proportionality between $M_{bh}$ and $M_{sph}$ from 0.006 (Magorrian et al. 
1998) to $\simeq 0.003 - 0.001$ (Kormendy \& Gebhardt 2001). 
Third, independent estimates of the 
black-hole mass in a substantial number of our AGN have now 
been made on the basis of nuclear $H_{\beta}$ line-width by McLure \& Dunlop
(2001). Fourth, the new VLA data presented in Section 4 mean that we can now 
explore the relationship between black-hole mass and radio luminosity 
without being hampered by a large number of upper limits within the 
`radio-quiet' subsample. 

\subsection{The Black-Hole Spheroid Connection}

As summarized by Kormendy \& Gebhardt (2001), `reliable' black-hole mass 
estimates are now available for at least 37 
nearby galaxies. As a result of this now substantial sample, 
clear correlations
have been uncovered between black-hole mass and host-spheroid luminosity, and 
between black-hole mass and host stellar velocity dispersion. It has 
been claimed that the latter is 
the tighter correlation, indicating that the basic physical link is 
with spheroid mass. Since in the present study we do not possess 
measurements of host stellar velocity dispersions, we obviously 
have to estimate 
host spheroid mass on the basis of the spheroid luminosity returned by 
our 2-D modelling. However, a recent re-analysis of the bulge luminosities
of low-redshift galaxies indicates that the relationship between bulge 
luminosity and black-hole mass is in fact just as tight as that between 
black-hole mass and central galaxy 
velocity dispersion (McLure \& Dunlop 2002).

To estimate black-hole masses on the basis of host spheroid luminosities
we have therefore adopted the following relations, while acknowledging 
the significant scatter present in both.

\begin{equation}
M_{sph} = 0.00123 L^{1.31}
\end{equation}
\begin{equation}
M_{bh}=0.0025M_{sph}
\end{equation}

\noindent
consistent with the mass-to-light ratio relation for ellipticals
determined by Jorgensen, Franx \& Kj\ae rgaard (1996) and the 
results of Gebhardt et al. (2000). We note that while these conversion
factors are certainly also consistent with those quoted in the recent
review by Kormendy \& Gebhardt (2001), there is now a growing 
body of evidence that the true conversion factor between $M_{sph}$
and $M_{bh}$ may be a factor $\simeq 2$ lower than that adopted here
(Merritt \& Ferrarese 2001, McLure \& Dunlop 2002). If one wishes
to adopt this lower factor (i.e. $M_{bh}=0.0013M_{sph}$) then the 
estimates of black hole mass utilised in the analyses that 
follow in the remainder of this paper
can simply be divided by a factor of 2. Obviously none of the arguments
concerning relative masses are affected by this uncertainty, and indeed
none of the physical arguments are significantly
altered by such a further modest
reduction in the black-hole mass estimates.   

The results of applying these relations to calculate inferred bulge and 
black-hole masses are presented in Table 9. Also given in Table 9 
(column 4) are black-hole masses derived  for the majority of the quasars
in these samples by McLure \& Dunlop (2001) on the basis of assuming
that the $H_{\beta}$-producing broad-line clouds are in Keplerian 
orbits.

The new spheroid-based black-hole estimates are compared with the 
$H_{\beta}$-derived values in Fig 13. This diagram shows that, with few
exceptions, there is very good agreement ($\simeq 0.2$~dex) between the
two completely-independent estimates of black-hole mass. Indeed, given
the well-known uncertainties in both approaches, the level of agreement 
demonstrated by Fig 13 can be viewed as adding considerable credence
to the estimated values of black-hole mass, and indeed also provide 
support for the basic premise of a gravitationally-bound broad-line region
as assumed by McLure \& Dunlop (2001, 2002). 
The fact that four objects (three
visible on the diagram) have 
$H_\beta$-based black-hole masses which lie well below the corresponding
host-spheroid derived values is not really surprising since the $H_\beta$ 
line-width analysis is clearly capable of yielding a severe under-estimate
of black-hole mass for any object whose broad-line region is orbiting 
close to the plane of the sky (but should be incapable
of yielding a comparably serious over-estimate of black-hole mass). 
Thus for 2247$+$140, 0736$+$017, 0157$+$001, and 1012$+$008
there is a good reason for trusting the black-hole mass estimates
of $\simeq 10^9~{\rm M_{\odot}}$ given in column 3 of Table 9, more than
the lower values yielded by the $H\beta$ analysis in column 4.

Accepting this, and excluding the two low-luminosity RQQs, the third column 
of Table 9 shows that a black hole of minimum mass 
$M_{bh} > 5 \times 10^8~{\rm M_{\odot}}$ appears to be required to produce a 
nuclear luminosity corresponding to $M_R < -23$.

\begin{table*}
\begin{center}
\caption{The results of estimating black-hole mass from the $R$-band 
luminosity of the host spheroid. 
Columns two and three list the predicted galaxy
spheroid mass and central black-hole mass respectively. For comparison
with the results given in Column 3, Column 4 gives
the black-hole mass estimates as derived from the $H_{\beta}$ line width
by McLure \& Dunlop (2001). Column 5 lists
the predicted absolute $R$-band Eddington luminosity of the black-hole,
calculated according to the prescription given in Section 8.1. Column 6 gives
the ratio of the predicted Eddington luminosity to the best-fitting
nuclear model component.}
\begin{tabular}{lrrrcc}
\hline
\hline
Source & ${\rm m_{sph}/10^{11} M_{\odot}}$ & ${\rm m_{bh}/10^9
M_{\odot}}$& ${\rm m_{bh-H\beta}/10^9
M_{\odot}}$&$M_{R}$ (Eddington)&$L_{nuc}/L_{edd}$ \\
\hline
{\bf RG}\\
0230$-$027 & 4.8\phantom{0000}  &1.2\phantom{0000}  &&&\\      
0307$+$169 & 7.9\phantom{0000}  &2.0\phantom{0000}  &&&\\
0345$+$337 & 2.6\phantom{0000}  &0.7\phantom{0000}  &&&\\
0917$+$459 &10.6\phantom{0000}  &2.6\phantom{0000}  &&&\\
0958$+$291 & 3.8\phantom{0000}  &0.9\phantom{0000}  &&&\\
1215$+$013 & 2.1\phantom{0000}  &0.5\phantom{0000}  &&&\\
1215$-$033 & 3.8\phantom{0000}  &0.9\phantom{0000}  &&&\\
1330$+$022 & 5.8\phantom{0000} &1.4\phantom{0000}  &&&\\
1342$-$016 &16.7\phantom{0000}  &4.2\phantom{0000}  &&&\\
2141$+$279 & 9.3\phantom{0000}  &2.3\phantom{0000}  &&&\\  
\hline
{\bf RLQ}\\
0137$+$012 & 8.7\phantom{0000}  &2.2\phantom{0000}  &1.1\phantom{0000}&$-28.2$ &0.013\\
0736$+$017 & 5.0\phantom{0000}  &1.3\phantom{0000}  &0.3\phantom{0000}&$-27.6$ &0.035\\
1004$+$130 & 9.4\phantom{0000}  &2.3\phantom{0000}  &3.7\phantom{0000}&$-28.3$ &0.089\\
1020$-$103 & 3.8\phantom{0000}  &1.0\phantom{0000}  &0.7\phantom{0000}&$-27.4$ &0.028\\
1217$+$023 & 5.9\phantom{0000}  &1.5\phantom{0000}  &0.8\phantom{0000}&$-27.8$ &0.043\\
2135$-$147 & 3.9\phantom{0000}  &1.0\phantom{0000}  &2.6\phantom{0000}&$-27.4$ &0.049\\
2141$+$175 & 4.2\phantom{0000}  &1.1\phantom{0000}  &1.7\phantom{0000}&$-27.5$ &0.061\\
2247$+$140 & 6.5\phantom{0000}  &1.6\phantom{0000}  &0.1\phantom{0000}&$-27.9$ &0.022\\
2349$-$014 &12.2\phantom{0000}  &3.1\phantom{0000}  &1.8\phantom{0000}&$-28.6$ &0.014\\
2355$-$082 & 5.3\phantom{0000}  &1.3\phantom{0000}  &0.7\phantom{0000}&$-27.7$ &0.014\\
\hline
{\bf RQQ}\\
0052$+$251 & 2.3\phantom{0000}  &0.6\phantom{0000}  &0.6\phantom{0000}&$-$26.8   &0.107\\
0054$+$144 & 5.2\phantom{0000}  &1.3\phantom{0000}  &2.5\phantom{0000}&$-27.7$   &0.052\\
0157$+$001 &12.2\phantom{0000}  &3.1\phantom{0000}  &0.2\phantom{0000}&$-28.6$   &0.011\\
0204$+$292 & 3.6\phantom{0000}  &0.9\phantom{0000}  &0.6\phantom{0000}&$-27.3$   &0.020\\
0244$+$194 & 1.9\phantom{0000}  &0.5\phantom{0000}  &0.3\phantom{0000}&$-26.6$   &0.046\\
0257$+$024 & 0.4\phantom{0000}  &0.1\phantom{0000}  &&$-24.9$      &0.008\\
0923$+$201 & 3.4\phantom{0000}  &0.8\phantom{0000}  &2.6\phantom{0000}&$-27.2$   &0.088\\
0953$+$414 & 2.1\phantom{0000}  &0.5\phantom{0000}  &0.7\phantom{0000}&$-26.6$   &0.355\\
1012$+$008 & 6.4\phantom{0000}  &1.6\phantom{0000}  &0.02\phantom{000}&$-27.9$  &0.026\\
1549$+$203 & 1.0\phantom{0000}  &0.3\phantom{0000}  &&$-25.9$      &0.169\\
1635$+$119 & 3.2\phantom{0000}  &0.7\phantom{0000}  &0.4\phantom{0000} &$-26.9$  &0.007\\
2215$-$037 & 4.7\phantom{0000}  &1.2\phantom{0000}  &&$-27.6$      &0.010\\     
2344$+$184 & 1.3\phantom{0000}  &0.3\phantom{0000}  &&$-26.8$      &0.002\\   
\hline \hline
\end{tabular}
\end{center}

\label{tab1}
\end{table*}

\begin{figure}
\vspace{9cm}
\centering
\setlength{\unitlength}{1mm}
\includegraphics{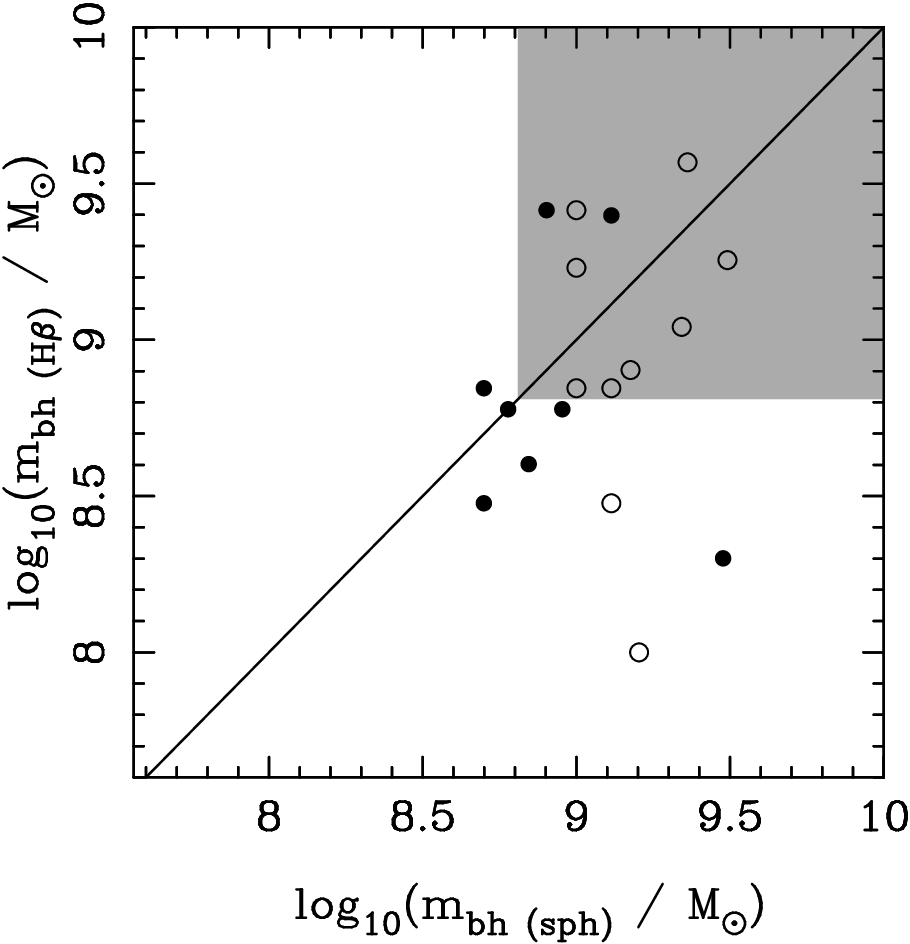}
\caption{A comparison between the black-hole masses predicted from
host-galaxy spheroid luminosity and
those determined from $H\beta$ line-width by McLure \& Dunlop (2001).
The shaded area illustrates that there is a region in which 
both approaches agree that $M_{bh} > 10^{8.8}~ {\rm M_{\odot}}$, and that this
region contains all except 2 of the RLQs (open circles), 
but excludes all except 2 of
the RQQs (filled circles).}
\label{bhcorr2}
\end{figure}

Beyond confirming the basic plausibility of the black-hole mass estimates,
the most interesting result contained in Fig 13 is that, 
as suggested by the initial results of 
McLure et al. (1999), the black-hole masses in the radio-loud quasars 
are systematically larger than in their radio-quiet counterparts.
In terms of summary statistics of central tendency, this difference, although
clear, might not appear very dramatic -- specifically, excluding the two 
low-luminosity RQQs:

\begin{tabbing}
$\langle M_{bh}/10^9~{\rm M_{\odot}}\rangle  = 1.67 \pm  0.36$ \hspace{0.5cm} \= median = 1.3\hspace{0.5cm}     \= (RG)\\
$\langle M_{bh}/10^9~{\rm M_{\odot}}\rangle  = 1.64 \pm  0.22$ \> median = 1.4    \> (RLQ)\\
$\langle M_{bh}/10^9~{\rm M_{\odot}}\rangle  = 1.05 \pm  0.24$ \> median = 0.8    \> (RQQ)
\end{tabbing}

However, the difference becomes more striking when one notices 
that there is an apparent threshold mass of $\simeq 10^9~{\rm M_{\odot}}$
{\it above} which lie 8/10 RGs and 10/10 RLQs, but {\it below} which lie 9/13
of the RQQs. In other words, using the black-hole mass estimation procedure 
outlined above, at least within this particular sample it appears that
a black-hole mass $> 10^9~{\rm M_{\odot}}$ is a necessary (albeit 
perhaps not sufficient) condition for the production of a powerful radio
source. As measured via the KS test, 
this difference between the black-hole mass
distributions of the radio-loud and radio-quiet subsamples is significant 
at the 3-$\sigma$ level, a significance which must be taken
seriously given the almost perfect matching of the distributions
of nuclear optical output demonstrated in Fig 4. 
Further support for its reality comes from the 
$H\beta$ study by McLure \& Dunlop (2001) already referred to above, in 
which it was found that 11/13 RLQs had $M_{bh} > 10^{8.8}~{\rm M_{\odot}}$
while only 4/18 RQQs had black-hole masses above this threshold. A similar 
conclusion was also reached by Laor (2000) from a study of the virial masses 
of PG quasars.

An inevitable consequence of the matching of nuclear optical luminosities
demonstrated in Fig 4, and the systematic offset in black-hole mass 
described above is that, on average, the black holes at the 
heart of the RQQs in our sample must, on average, 
be emitting more efficiently than 
their radio-loud counterparts. This is quantified in the last two columns 
of Table 9 which give estimated Eddington luminosity (converted to $R$-band
absolute magnitude) and then emitting efficiency $L/L_{Edd}$. 
The potential $R-$band Eddington luminosity of each quasar
was derived in the following manner. Initially, the host-galaxy bulge 
luminosities provided by the two-dimensional modelling were converted 
into a spheroidal mass estimate using the J$\o$rgensen et al.
mass-to-light 
ratio (Eqn 7). The linear relation between black-hole mass and 
spheroidal mass (Eqn 8) was then
used to 
estimate the central black-hole mass and resulting bolometric 
Eddington luminosity of each quasar. To calculate the corresponding
$R-$band 
magnitudes it was assumed that, on average, the 
bolometric luminosity of a quasar can be estimated via 
$L_{bol}\simeq 10\lambda L_{5100}$ (Laor et al. 1997), 
where $L_{5100}$ is the absolute 
monochromatic luminosity at 5100 \AA. The resulting estimate of the
absolute 
luminosity at 5100 \AA\, was then transformed to the central wavelength of 
the $R-$ filter (6500~\AA) assuming a power-law quasar 
spectrum ($\alpha=0.2$). Using this prescription, a quasar with a central 
black-hole mass of $10^{9}~\Msolar$ will have an Eddington-limit 
absolute $R-$band magnitude of $M_R(Edd) = -27.5$.

Observed nuclear $M_R$
is plotted against potential Eddington-limited $M_R$ in Fig 14.
As can be seen from this figure, 
the maximum luminosity produced by any quasar is
within a factor of two of the predicted Eddington limit, while the majority
appear to be radiating at $\sim 1- 10 \%$ of the Eddington
luminosity. 
Taken as a group there is a suggestion that the RQQs are radiating at a higher
percentage of the Eddington limit, but
the difference is not formally significant (KS, $p=0.29$).

\subsection{The Black-Hole--Radio Connection}

When the matched samples of RLQs and RQQs were first defined, all that 
was known about the radio properties of the RQQs was that their
5~GHz luminosities were lower than $P_{5GHz} \simeq 
10^{24}~{\rm W Hz^{-1} sr^{-1}}$
(Dunlop et al. 1993). However, with the completion of the VLA 
detection programme described in Section 4 we can now explore how the 
radio properties of {\it all} the AGN in our combined sample relate 
to their estimated black-hole masses.

In Fig 15 total radio luminosity, $P_{5GHz}$, is plotted 
against estimated black-hole mass for all the objects in our combined
sample (with the single exception of 1549$+$203, which was not observed
with the VLA to the same depth as the other objects). Also included in this
diagram are the appropriate points for 8 nearby galaxies taken from 
Franceschini et al. (1998), with the Milky Way and M87 highlighted.

Several significant trends are evident in this diagram. First it can be
seen that, within the RQQ sample, radio luminosity correlates, albeit weakly, 
with black-hole mass, although our sample doesn't span a wide dynamic range
and the correlation is not formally significant. Second, and perhaps more
interestingly, the RQQs are consistent with the locus described by the 
nearby galaxies (for which, of course, the black-hole masses are based 
directly on stellar dynamics). This simultaneously reinforces the
plausibility of our black-hole mass estimates, and suggests a common
physical origin for the radio emission from RQQs and nearby optically
more-quiescent bulges. Whatever this physical mechanism is, it must  
explain the observed strong dependence of minimum 
radio output on black-hole mass, in which the power-law index $\gamma$ 
is at least 2 (adopting $P_{radio} \propto M_{bh}^{\gamma}$). 
As noted by Franceschini et al. (1998), $\gamma = 2.2$ is the prediction
of ADAF models, while $\gamma \simeq 2$ is a generic prediction of
any mechanism in which radio output is proportional to the area of the
accretion disc.

Our results therefore provide further support for previous claims
that there is a minimum radio output from black holes and that this
is a strong function of mass. However, while this obviously implies that
the most massive black holes can never be {\it very} radio-quiet, Fig
15 graphically demonstrates that the genuine radio-loud objects lie on a quite
distinct relation. Interestingly, our radio-loud sample also
displays an (albeit weak) internal correlation between $P_{5GHz}$
and estimated black-hole mass.
Either this relationship
must be much steeper ($P \propto M_{bh}^4$) than the minimum radio-output
relation, or, as illustrated 
in Fig 15, the black-hole mass dependence of the upper envelope
may be consistent with that of the lower envelope but offset 
by several orders of magnitude.

In fact the findings of Lacy et al. (2001) strongly favour the latter option.
They report a relationship between the logarithm of
radio power and black-hole mass which has a slope of $1.4 \pm 0.2$. However, 
inspection of their Fig 2 shows that their data are also consistent 
with an upper and lower envelope with slope $\simeq 2.5$ as illustrated
here in Fig 15.

We note here that it has recently been claimed by Oshlack, Webster \& Whiting
(2001) that the radio-loud Seyfert galaxy PKS 2004$-$447 contradicts the 
existence of such an envelope, and shows that relatively low-mass black holes
can produce a powerful radio source. However, this claim is based on their 
black-hole mass estimate for this object of 
$\simeq 5 \times 10^6~{\rm M_{\odot}}$ and it seems almost certain that 
this value, derived from the relatively narrow emission lines displayed by
PKS 2004$-$447,  
is a serious under-estimate of the true mass. The reason for this is that
since this object displays optically variability and has 
a compact radio source, it seems very likely that its orbiting broad-line
region lies close to plane of the sky. In this situation (as found in our
own sample for, e.g. 1012$+$008 - see Table 9) the black-hole mass derived
from $H_{\beta}$ emission-line width can under-estimate the true value
by a factor $> 100$. In Fig 15 we have plotted the position of PKS 2004$-$447
after re-calculating its black-hole mass on the assumption that the orbital
plane of its broad-line region is oriented within $\simeq 5$ 
degrees  of the plane of
the sky (see McLure \& Dunlop 2001). 
This increases the estimated black-hole mass to $\simeq 4 \times 10^8 {\rm M_{\odot}}$, and moves PKS 2004$-$447
into complete consistency with the upper envelope 
delineated by the other available data. We have not corrected the observed
radio emission for doppler boosting - if we did then the this object would
simply move downwards in Fig 15.

\begin{figure}
\vspace{8.5cm}
\centering
\setlength{\unitlength}{1mm}
\includegraphics{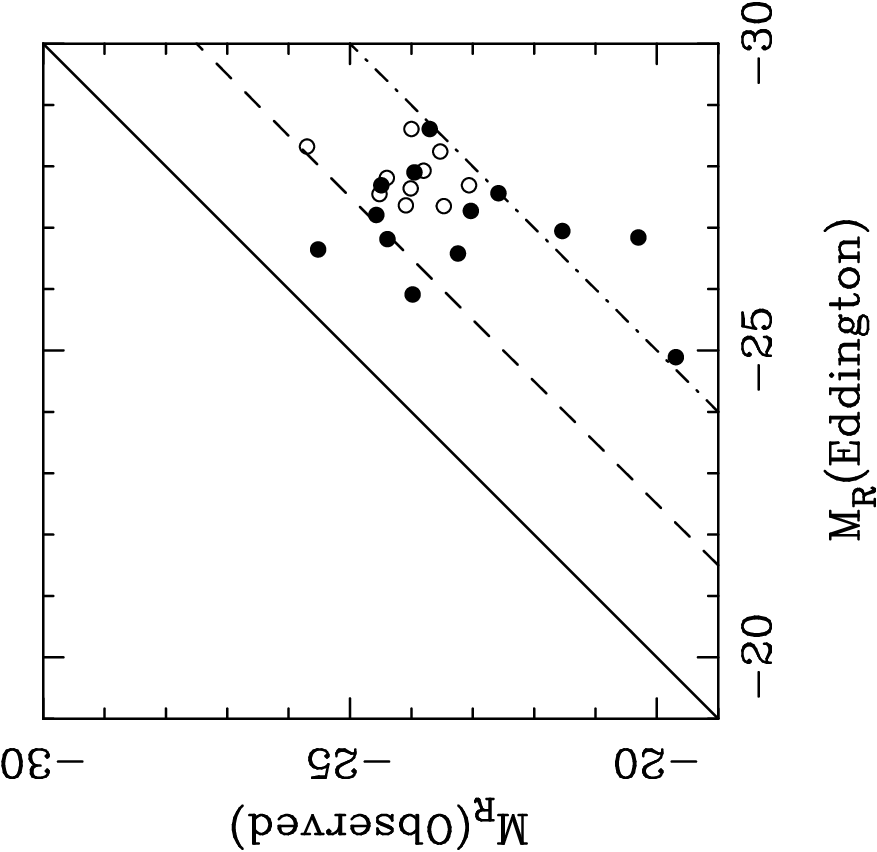}
\caption{The observed absolute magnitude $M_R$ of the nuclear
component in each quasar plotted against the absolute magnitude which
is predicted by assuming that each quasar contains a black hole of
mass $m_{bh} = 0.0025 m_{spheroid}$, and that the black hole is
emitting at the Eddington luminosity (RLQs = open circles, RQQs =
filled circles).  The solid line shows where the quasars should lie if they
were all radiating at their respective Eddington luminosities, while
the dashed line indicates $10\%$ of predicted Eddington luminosity,
and the dot-dashed line indicates $1\%$ of predicted Eddington
luminosity. The nuclear components of the radio galaxies are
not plotted because all the evidence suggests they are substantially
obscured by dust - see Section 6.7.}
\label{magorrian}
\end{figure}

Thus both our own results and other currently available data
support the existence of both a lower and an upper envelope to 
the radio luminosity which can be produced by a black hole of given mass.
Indeed Fig 15 serves to provide an elegant explanation not
only of our own findings, but also of several other well-known results in the 
literature. 

First, because we selected our radio-loud objects on the 
basis that their radio 
luminosities were greater than $log_{10}(P/{\rm W Hz^{-1} sr^{-1}}) = 24$, 
the form 
of the upper envelope in these figures can be seen to be perfectly
consistent with our finding that virtually all our radio-loud 
objects have $M_{bh} > 10^{9}~{\rm M_{\odot}}$. In the context of this diagram 
this finding can be seen to be a result of our radio-selection criterion 
rather than the existence of 
some critical black-hole mass for the production of relativistic 
jets within a galaxy mass halo. 
 
Second, the steep dependence of this apparent upper envelope on black hole
mass means that radio sources up to 
$log_{10}(P/{\rm W Hz^{-1} sr^{-1}}) =27-28$, consistent with the most 
luminous 3CR galaxies even at $z =1$ {\it can} be produced by black 
holes (and hence hosts) only a factor of 2-3 more massive, and most likely 
{\it will} be due to the steep high-mass slope of the black-hole mass function.
This provides a natural explanation of why radio galaxies spanning the top
two orders of magnitude in radio power all appear to be giant ellipticals 
with stellar masses which agree to within a factor of $\simeq 2$ 
(Jarvis et al. 2001).

Third, from Fig 15 it can be seen that the properties of
AGN hosts will inevitably become more mixed if the 
radio-luminosity threshold is move down to, say, $P_{5GHz}
\simeq 10^{22}~{\rm W Hz^{-1} sr^{-1}}$. Given the 
lower boundary in Fig 15, radio sources in this power range can still be 
produced 
by black holes as massive as $10^{9}~{\rm M_{\odot}}$ 
 living in giant ellipticals, 
but it is also clear that black holes with masses 
as low as 10$^{7.5}~{\rm M_{\odot}}$ are capable of producing 
detectable radio sources at this level.
As demonstrated 
by the properties of the lower-luminosity objects in our own sample, and by the
studies of `normal' galaxies at low redshift,
such black holes can be housed in disc-dominated
galaxies as well as discless lower-mass spheroids.

\subsection{The origin of radio loudness}
While the broad dependence of radio luminosity on 
black-hole mass may explain the relative numbers of radio-quiet
and radio-loud objects, it is also clear from Fig 15
that black hole mass cannot be invoked 
to explain the {\it range} 
of radio luminosities displayed by quasars of very similar
optical luminosity. Specifically it is clear from this figure 
that objects powered by equally massive black holes can differ 
in radio luminosity by $\simeq 4$ orders of magnitude. 

Interestingly Fig 15 also allows us to rule out the possibility that
the radio output of a black hole of given mass is driven simply
by fuel supply. This is because, while it has long been known that RQQs are 
not radio silent, it can be seen from Fig 15 that several of 
the RQQs in our sample are in fact as radio-silent as they could possibly 
be, given their estimated black-hole masses. In other words at least
some of the black holes in RQQs produce as little radio output as 
their counterparts in completely quiescent nearby galaxies, 
despite the fact that black holes in quasars are clearly in receipt of 
sufficient fuel to produce powerful {\it optical} emission.

Thus, while the results of this study cannot by themselves 
provide a definitive answer to the long standing question of the 
origin of radio loudness, they can be used to focus the argument
by excluding several possible explanations. In summary, our results can be 
used to exclude host galaxy morphology, black-hole mass, or black hole 
fueling rate as the primary physical cause of radio loudness.

We are thus forced to the conclusion that the production of powerful
relativistic radio jets is driven by some other property of the black hole 
itself, the most likely candidate being black-hole spin (e.g. Blandford
2000, Wilson \& Colbert 1995). Indeed, 
following the suggestion by 
Blandford (2000), we speculate that the evolutionary track followed
by an activated black hole may manifest itself as a near vertical descent 
in Fig 15, with an active rapidly spinning
hole appearing first as a powerful radio source
close to the upper envelope in this diagram, and then descending towards 
the lower envelope as the spin energy of the hole is exhausted.

\begin{figure}
\vspace{8.5cm}
\centering
\setlength{\unitlength}{1mm}
\includegraphics{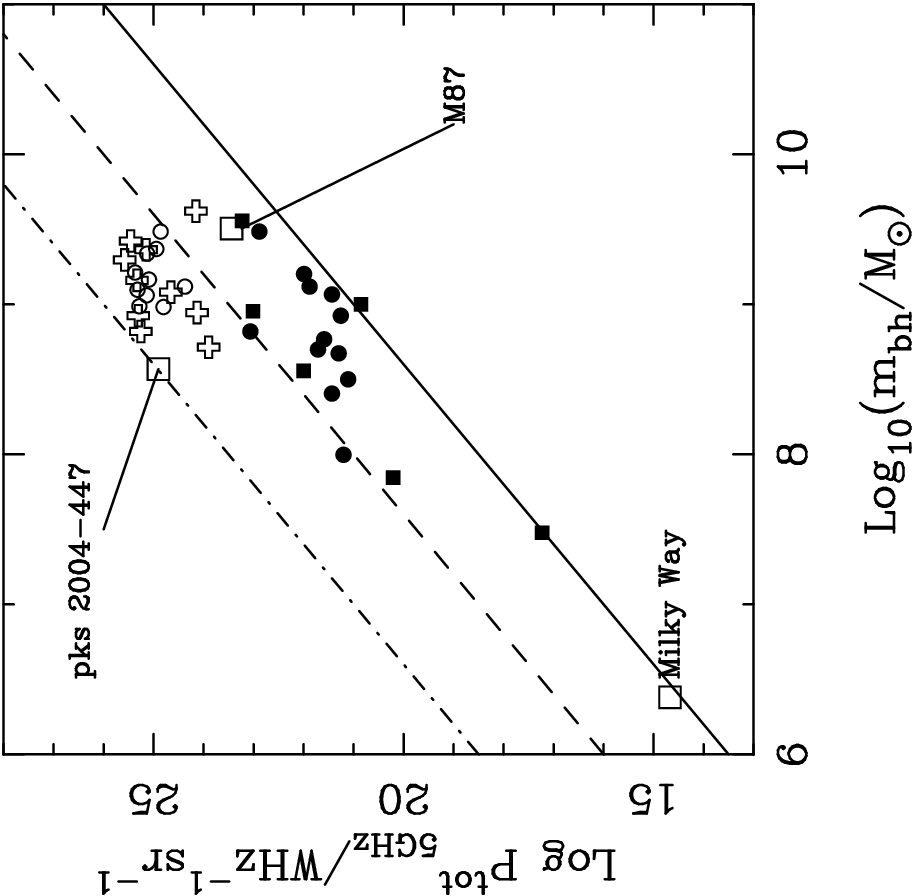}
\caption{Total radio luminosity $P_{5GHz}^{total}$ versus
black-hole mass showing 
the data on low-redshift `normal' galaxies
from Franceschini {\it et al.} (1998) (as squares, with the Milky Way and 
M87 highlighted), and the AGN from the HST sample
(RGs = crosses, RLQs = open circles, RQQs = filled circles). The
lines which bracket the data are relations of the form $P_{5GHz} 
\propto m_{bh}^{2.5}$ offset by 2.5 orders of magnitude from each other. The 
solid line therefore indicates the apparent existence of a {\it lower} limit 
to the radio output of a black hole which scales as $m_{bh}^{2.5}$ while the 
dot-dash line indicates the apparent {\it upper} limit required by our data.
While our own data are insufficient to constrain the slope of such an upper 
envelope, adoption of the same slope as the lower envelope (as done here)
is consistent with the data of Lacy et al. (2001) and is further 
reinforced by the results of McLure \& Dunlop (in preparation). 
PKS 2004$-$447 has been 
included in this plot because it has recently been claimed 
that this object contradicts the 
existence of any such mass-dependent upper envelope (Oshlack, Webster \& 
Whiting 2001).
The position of this compact and variable source on this diagram 
has been here re-calculated on the assumption that its radio axis lies 
with $\simeq 5$ degrees of the 
line of sight, but without making any corrections 
for relativistic beaming of its radio output
(see text for details). As can be seen, the resulting increase in estimated
black hole mass from $m_{bh} \simeq 5 \times 10^6~{\rm M_{\odot}}$ 
to $m_bh \simeq 4 \times 10^8~{\rm M_{\odot}}$
 is sufficient to bring it into consistency
with the upper envelope suggested by the other data.}
\label{radio_magorr}
\end{figure}

\section{Cosmological Implications}

\subsection{Relative numbers of radio-loud and radio-quiet quasars}

One interesting aspect of the different black-hole mass thresholds for 
RQQs and RLQs uncovered by this study is that this difference of a factor of 
2 in minimum black-hole mass provides a natural explanation of why radio-quiet
quasars outnumber their radio-loud counterparts by a factor of $\simeq 10$
(and why this factor may reduce with increasing optical
luminosity; Hooper et al. 1995, Goldschmidt et al. 1999).

This is because if our black-hole mass thresholds of $M_{bh} > 
5 \times 10^8~{\rm M_{\odot}}$
for RQQs, and $M_{bh} > 1 \times 10^9$~${\rm M_{\odot}}$ 
for radio-loud objects are converted back into 
absolute spheroid magnitudes in the $K$-band (using the absolute $R$-band magnitudes given in Table 4, and then converting into $K$ adopting $R-K = 2.5$ as 
justified by the results in Section 6.6) we find that RQQs can be hosted by
spheroids with $M_K < -25.3$ while radio-loud objects require 
a host spheroid with $M_K < -26$. With reference to the $K$-band luminosity
function these absolute magnitude limits correspond
to $L > 1.5 L^{\star}$ and $L > 3 L^{\star}$ respectively (Gardner et al. 1997,
Szolkoly et al. 1998, Kochanek et al. 2001) which leads to the prediction
that, in the present day universe there are $1 \times 10^{-4}$ galaxies 
per ${\rm Mpc^{3}}$ capable of hosting an RQQ brighter than $M_V = -23$,
but only $1 \times 10^{-5}$ galaxies per ${\rm Mpc^{3}}$ are 
capable of hosting an RLQ. This means 
that, assuming low-redshift quasars arise from a randomly-triggered subset 
of massive black-holes in spheroids, the factor of two difference 
in black-hole mass threshold translates into a factor of 10 difference 
in the 
number density of RQQs and RLQs due simply to the steepness of the bright 
end of the spheroid luminosity function.

\subsection{Black-hole activation fraction at ${\bf z \simeq 0}$}

We can go further and use the above numbers to estimate the activation 
fraction for massive black holes in the local universe.
First we note that the radio-loud objects in our sample have 
$P_{2.7GHz} > 1 \times 10^{25}~{\rm W Hz^{-1} sr^{-1}}$ 
(assuming a spectral index $\alpha \simeq 0.8$). Reference to the local 
radio luminosity function in Dunlop \& Peacock (1990), or to the 
more recent determination at 1.4 GHz 
from the 2dF by Sadler et al. (2001), indicates
that the present-day number density of radio sources above this threshold is
$\simeq 1 \times 10^{-8}~{\rm Mpc^{-3}}$.
Thus, in the present-day universe we find that 1 in every 1000 black holes
more massive than $1 \times 10^9$~${\rm M_{\odot}}$ (or equivalently
spheroids more luminous than $3L^{\star}$) is radio active in the present-day
universe.

Turning to the RQQ population, the present-day number density of quasars
with $M_V < -23$ can be estimated by extrapolating the QSO OLF deduced 
at $z \simeq 0.4$ by Boyle et al. (2000) back to $z = 0$ assuming luminosity
evolution $\propto (1 + z )^3$. This produces an estimated number density
of $\simeq 5 \times 10^{-8}$~${\rm Mpc^{-3}}$, which is at least consistent
with the direct determination of the low-$z$ OLF attempted
by Londish, Boyle \& Schade
(2000). Boosting this number by a factor of two to allow
for obscured quasars (i.e. adopting an opening angle of $\simeq 45^{\circ}$ 
for consistency with the correction factor which appears to apply to
the radio-loud population)
then leads to the conclusion that 1 in
every 1000 black holes more massive than $5 \times 10^8$~${\rm M_{\odot}}$ 
(or equivalently spheroids more luminous than $1.5L^{\star}$) is 
producing quasar-level optical activity.

Thus both these comparisons yield the same result, namely that in the 
present-day ($z < 0.1$) universe
1 in every 1000 massive black holes is active, and that the reason 
RQQs outnumber RLQs by a factor of 10 is a direct result of the fact 
that their respective minimum black-hole masses differ by a factor of two.

We note here that  Wisotzki, Kuhlbrodt \& Jahnke (2001) have 
recently concluded 
that the present-day black-hole activation fraction is 1 in 10000, or 1 in 5000
if the quasar LF is boosted by a factor of two as assumed above. However,
reference to their Fig 2 shows that this factor results from adopting
the spheroid luminosity function of Lin et al. (1999). If instead 
Wisotzki et al. were to adopt the Kochanek et al. (2001) luminosity function
then it is clear that (at least for the mass range probed by the present study)
Wisotzki et al.'s analysis would also yield an activation 
fraction of 0.1\%.

\subsection{Black-hole activation fraction at ${\bf z \simeq 2-3}$}

At the peak epoch of quasar activity between $z \simeq 2$ and $z \simeq 3$,
the co-moving number density of powerful radio sources with 
$P_{2.7GHz} > 1 \times 10^{25}~{\rm W Hz^{-1} sr^{-1}}$ has risen to
$\simeq 1 \times 10^{-6}$~${\rm Mpc^{-3}}$ (Dunlop \& Peacock 1990) while
that of optically-selected quasars brighter than $M_V = -23$ is 
$\simeq 5 \times 10^{-6}$~${\rm Mpc^{-3}}$ (Warren, Hewett \& Osmer 1995,
Boyle et al. 2000). Doubling the latter figure to make the same correction for obscured RQQs as applied in the previous subsection at low redshift,
leads to the conclusion that, as at $z \simeq 0$, RQQs at $z \simeq 2.5$ 
outnumber their radio-loud counterparts by a factor of $\simeq 10$. However,
at $z \simeq 2.5$ the co-moving
number density of both classes of object is enhanced by a factor of 100,
giving an activation fraction of massive black holes of 
$\simeq 10$\%, assuming that the entire massive black-hole population is 
in place by that epoch.

\subsection{The cosmological evolution of AGN}

A renewed attempt to model the cosmological evolution of AGN in the light of 
our host-galaxy results is obviously beyond the scope of this paper. However,
if one assumes that the same mass thresholds uncovered in our low-redshift
study apply at high redshift, then one simple interpretation of the 
dramatic (and very similar) evolution of radio-loud and radio-quiet AGN between
$z \simeq 2.5$ and the present-day is that it is primarily due to
density evolution, with the active fraction of massive black holes dropping
from 10\% to 0.1\% over this period. There must of course 
also be an element of luminosity evolution because pure density evolution
was excluded long ago as an acceptable representation of the evolving RLF or 
OLF. However, as suggested by Miller, Percival \& Lambert (in preparation)
if individual quasars
produce declining light curves it is inevitable that a larger fraction 
will be observed closer to peak luminosity as the activation rate rises, and 
Fig 14 indicates that at least some of the host galaxies uncovered in the
present study should be capable of 
producing quasars with $M_V \simeq -28$ if observed while accreting at the 
Eddington limit.
 
One prediction of this scenario is that, as a larger fraction of massive black holes become more efficiently fueled, the apparent 
mass difference between RLQ and RQQ
hosts should grow with increasing redshift. 
The reason for this is that, if fueling 
rates are, on average, somewhat higher at high redshift we can expect 
optically selected QSOs of a given luminosity to be produced by black holes
with, on average, lower masses than at low redshift. However, if 
the $m_{bh} > 10^9~{\rm M_{\odot}}$ mass threshold for powerful radio
activity  found in this study continues to apply at high redshift, then 
radio selection will continue to yield only the most massive black holes
residing (presumably) in the most massive spheroids.

In fact, studies of the hosts of RLQs and RQQs out to $z \simeq 2$ are already
uncovering evidence of just such a trend. Specifically, Kukula et al. (2001)
have used NICMOS on the HST to measure the rest-frame $R$-band 
luminosities of the hosts of matched samples of RLQs and RQQs at $z \simeq 1$
and $z \simeq 2$. Kukula et al. find that 
the hosts of RLQs are essentially unchanged in mass over this redshift range,
suggesting that the minimum mass threshold for powerful radio emission 
indicated 
by the upper locus in Fig 15 applies at all redshifts and is therefore 
of physical significance.
However, they also find that
the average ratio of RLQ:RQQ host luminosity  rises
from the value of $1.5$ found here at $z \simeq 0.2$, to $\simeq 2$
 at $z \simeq 1$, 
and $\simeq 3$ at $z \simeq 2$, suggesting a progressive drop in the average 
mass of RQQ hosts with increasing redshift, a finding supported
by the results of other recent studies (Ridgway et al. 2000, Rix et al. 2000).

\section{Conclusions}

In this paper we have reported the extensive results 
which follow from completion of our HST imaging programme of the host 
galaxies of low-redshift radio-quiet quasars, radio-loud quasars, and radio
galaxies. This paper represents the completion of a programme
which has spanned most of the last decade, commencing with the deep 
infrared imaging of the same sample of objects by Dunlop et al. (1993).
The depth and quality of our HST data have now allowed us to determine
accurately all the basic structural parameters of the hosts of these 3 classes
of powerful AGN. Because of this, and because of the wealth of data now 
available for this sample at other wavelengths (including the new deep VLA
observations also reported here), we have been able to address a wide range
of issues of which are hopefully of interest to workers in several
different areas of extra-galactic research.
Consequently we conclude with a detailed summary of the main conclusions 
of this study, structured to assist the interested reader in moving directly 
to the sections which may be of most relevance to their own work.
 
\subsection{Results from analysis of the HST images}

From the detailed 2-dimensional modelling of the new HST images
presented here in combination with the data reported by McLure et al. (1999)
we find that (as detailed in Sections 5 and 6):

\begin{itemize}

\item{All except the two least luminous RQQs in the sample
have bulge-dominated hosts, and in only 2 of the remaining 31 objects
(the RQQs 0052+251 and 0157+001) can we find any evidence for a 
significant disc component. Thus both radio-loud and radio-quiet 
quasars live, almost universally, in elliptical galaxies.}

\item{The hosts of all three classes of powerful AGN 
are luminous galaxies with $L > L^{\star}$, and almost always 
$L > 2L^{\star}$. The average luminosities of the RG and RLQ hosts 
are essentially identical (consistent with radio-loud unification), 
and equivalent to $4L^{\star}$. The average
luminosity of the RQQ hosts is somewhat smaller, equivalent to
$\simeq 3L^{\star}$.}

\item{The hosts of all three classes of powerful AGN are large
galaxies with typical half-light radii $r_{1/2} \simeq 10$~kpc. As with
the luminosities, the average values of host-galaxy scale-length
for the radio-loud sub-samples are indistinguishable, while the hosts
of the RQQs are typically smaller by a modest factor ($\simeq 1.5$).}

\item{The hosts of all three classes of powerful AGN display 
a surface-brightness scalelength (Kormendy) relation identical 
(in both slope and normalization) to
that displayed by inactive massive ellipticals predominantly
found in clusters.}

\item{The hosts of all three classes of powerful AGN display
a distribution of axial ratio which is indistinguishable to
that which has been long established for normal elliptical galaxies.}

\end{itemize}

\subsection{Results incorporating infrared images}

From a joint analysis of the HST optical images with existing $K$-band
images of the same sample (Dunlop et al. 1993, Taylor et al. 1996, McLure,
Dunlop \& Kukula 2000) we find that (as detailed in subsections 6.6 and 6.7):

\begin{itemize}

\item{The hosts of all three classes of powerful AGN are red galaxies,
with typical rest-frame optical-infrared colours $R-K \simeq 2.5$. 
Consistent with the results of Nolan et al. (2001), these colours 
are as expected from an evolved stellar population of age $10-13$~Gyr,
indicating that, as for quiescent massive ellipticals, the stellar mass
of an AGN host galaxy is dominated by a well-evolved stellar population 
formed at high redshift ($z > 2$).}

\item{The optical-infrared colours of the RQQ and RLQ nuclei are 
statistically indistinguishable, and lie in the range $R-K \simeq 
2-4$. For the RG nuclei, the relation between $R-K$ colour and $R$-band
luminosity is as expected under the assumption that they
all contain obscured nuclei with intrinsic optical luminosities
comparable to the quasars, reddened by dust with 
$\kappa_{\lambda} \propto \lambda^{0.95}$. This result 
strongly favours unification of RGs and RLQs via orientation
rather than time.}

\end{itemize}

\subsection{Relation of quasar hosts to normal galaxies}

From a comparison of our results with those derived from recent studies 
of the `normal' massive galaxy population, and of low-redshift 
ULIRGs and Seyferts, we find that (as detailed in Section 7):

\begin{itemize}

\item{The bulge:disc ratio of AGN hosts is a function
of AGN luminosity, with disc-dominated hosts dying out
above the traditional quasar:Seyfert boundary at $M_V (nuc) \simeq -23$.}

\item{In contrast to ULIRGs, quasar hosts lie in a region of the
fundamental plane (as judged from the photometric projection
of the FP offered by the Kormendy relation) which has been shown
to be occupied by the most massive and apparently old 
population of ellipticals which display `boxy' isophotes
and distinct kinematic cores. This provides a strong argument
against the 
possibility that the $r^{1/4}$-law luminosity profiles 
displayed by the vast majority of the quasar hosts in our sample are
the result of recent mergers between massive disc galaxies. This
result also offers little support for a strong
evolutionary connection between ULIRGS and quasars in the low-redshift
universe.}

\item{The basic structural properties of AGN hosts are indistinguishable
from those of inactive brightest cluster galaxies. Consistent 
with the environmental study of McLure \& Dunlop (2000), the largest
host galaxies in our sample are of comparable mass to the BCGs found at the
heart of Abell Class 1 or Class 2 clusters. However, the lack of
any host galaxies as massive as an Abell Class 4 cluster BCG is as 
expected given the relative rarity of such rich environments. All the 
evidence considered is consistent with the AGN hosts being drawn, 
essentially at random, from the present-day massive-elliptical
galaxy population.}

\item{There is no statistically-significant evidence that AGN hosts
display more signs of large (kpc) scale disturbance, or multiple 
nuclei than normal comparably-massive quiescent ellipticals.}

\end{itemize}

\subsection{The black-hole spheroid connection}

Combining our HST data with our new VLA data, and utilising recent 
results on the black-hole:spheroid mass relation in both quiescent 
and active galaxies, we find that (as detailed in Section 8):

\begin{itemize}

\item{Estimates of central black-hole mass for the quasars in our sample
based on host-galaxy spheroid luminosity agree well with those
derived from $H_{\beta}$ emission-line width by McLure \& Dunlop (2001).
For three objects the latter technique yields a much lower value, but this
is as expected for a small subset of objects
if the broad-line region has a predominantly disc-like structure.}

\item{Based on the relation $M_{bh} = 0.0025 M_{sph}$, we find that
all the quasars in our sample are powered by a black-hole 
of mass $m_{bh} > 5 \times 10^{8}~{\rm M_{\odot}}$, but that 
the radio-loud objects lie above an even higher mass threshold, with
$m_{bh} > 10^{9}~{\rm M_{\odot}}$.}

\item{The most efficient RQQ in our sample appears to be radiating at 
$\simeq 35$\% of the Eddington limit, but the vast majority of 
these low-redshift AGN are emitting at $1 - 10$\% of the Eddington limit.
Based on these calculations, the most massive objects in this low-redshift
sample appear capable (if fuelled at maximum efficiency)
of producing quasars with $M_V \simeq -28$, 
comparable to the most luminous objects known at high redshift.}

\item{The black-hole mass difference between RLQs and RQQs
reflects the existence an apparent upper and lower limit
to the radio output that can be produced by a black hole
of a given mass. Both the upper and lower thresholds on radio
luminosity appear to be a strong function of black mass ($\propto
m_{bh}^{2 - 2.5}$). At least some RQQs in our sample appear
to be as radio-weak as `normal' inactive comparably-massive 
spheroids found in the low-redshift universe, despite the fact
that must be receiving sufficient fuel to power the optical
quasar nucleus.}

\item{While there is a broad and clear 
trend for increasing radio luminosity with increasing black-hole mass,
it is now clear that host morphology, black-hole mass, and fuel supply 
can each be excluded as the primary physical explanation 
of why a subset of quasars are up to 4 orders of 
magnitude more radio luminous than their `radio-quiet' counterparts.
Consequently we argue that
black-hole spin is the most plausible (perhaps only) remaining
feasible explanation for the production of a powerful radio source.}

\end{itemize}

\subsection{Cosmological implications}

Considering the implications of our results for our understanding of 
the nature and evolution of AGN populations as a function of redshift, we
find that (as detailed in Section 9):

\begin{itemize}

\item{The relative numbers of radio-loud and radio-quiet quasars
can be naturally explained by the above-mentioned mass thresholds,
combined with the form of the bright end of
the elliptical galaxy luminosity
function.}

\item{The activation fraction of massive black holes is 
$\simeq 0.1$\% in the present-day universe, rising to $\simeq 10$\% 
at $z \simeq 2 - 3$, corresponding to the peak epoch of
quasar activity in the universe.}

\item{The black-hole mass threshold for powerful radio activity
can explain the trend for a growing gap with increasing 
redshift between the masses of RLQ hosts and RQQ hosts uncovered 
by recent HST-based studies of high redshift quasars.}

\end{itemize}

\section*{Acknowledgements}
Based on observations with the NASA/ESA {\it Hubble Space Telescope}, 
obtained at the Space Telescope Science Institute, which is operated by 
The Association of Universities for Research in Astronomy, Inc. under NASA 
contract No. NAS526555. 
James Dunlop, Ross McLure and Marek Kukula all acknowledge 
the support of PPARC through the awards of a Senior Fellowship, a Personal 
Fellowship and a PDRA position respectively. Marek Kukula also
acknowledges support for this work provided by the Space Telescope Science 
Institute under grant numbers 00548 \& 00573.

\newpage

\appendix

\section{New host-galaxy images and model fits}

The images, two-dimensional model fits, and model subtracted images 
for the 14 AGN for which the new HST data are reported in this paper
are presented in this appendix in Figs A1 to A14.
A grey-scale/contour
image of the final reduced F675W $R$-band image of each AGN is shown
in the top-left panel (panel A) of each figure, covering a region
of 12.5 $\times$ 12.5 arcsec centred on the target source. The 
surface-brightness level of the lowest contour is indicated in the 
top-right corner of this panel, and the grey-scale has been chosen
to highlight structure close to this limit. Higher surface-brightness
contours are spaced at intervals of 0.5 mag. arcsec$^{-2}$, and have been
super-imposed to emphasize brighter structure in the centre of the 
galaxy/quasar. Panel B in each figure shows the best-fitting two-dimensional
model, complete with the unresolved nuclear component (after convolution
with the empirical PSF described above), contoured in a manner identical 
to panel A. Panel C shows the best-fitting host galaxy 
as it would appear if the nuclear component were absent, while panel D is the
residual image which results from subtraction of the full two-dimensional
model (in panel B) from the raw $R$-band image (in panel A), in order
to highlight the presence of any morphological peculiarities such as
tidal tails, interacting companion galaxies, or secondary nuclei.
Within each figure, all four panels are displayed using the same greyscale.

\subsection{Notes on individual objects}

Here we provide a brief discussion of each of the 14 new HST images, 
with reference to other recent HST 
and ground-based data. Comparably-detailed descriptions of the other
19 objects in the sample can be found in McLure et al. (1999) and are not 
repeated here. Additional details on each object, along with a description
of the main features of our $K$-band images, can be found in Dunlop et al. 
(1993) and Taylor et al. (1996). The results of off-nuclear optical 
spectroscopy and resulting spectral model-fitting for several of the
objects in this sample can be found in Hughes et al. (2000) and Nolan et al.
(2001).

\subsubsection{The Radio Galaxies}

\vspace*{0.07in}

\noindent
{\bf0230$-$027} (PKS 0230$-$027, OD $-$050)

\vspace*{0.02in}
\noindent
The new $R$-band HST image shows this galaxy to be uniform and round with 
no sign of obvious distortion. The detail of the WF2 image shown in Fig
A1 reveals $\simeq 10$ apparent companion objects lying in a roughly 
circular formation around the galaxy, at a radius of $\simeq 30 \rightarrow 
60$ kpc. The brightest of these companions lies at a projected distance of 
$\simeq 60$~kpc to the NW. The full WF2 chip reveals there to be $\simeq 20$
faint companion objects within a radius of $\simeq 200$kpc.

\vspace*{0.07in}

\noindent
{\bf0307$+$169} (3C 079, 4C $+$16.07)

\vspace*{0.02in}
\noindent
The $R$-band image of this source shown in Fig A2 
reveals it to be a classic
brightest-cluster galaxy (BCG). Two bright companion objects can be
seen to the
North and South with three or four accompanying tidal arm features
emanating from the central galaxy. The suggestion that this source is 
in the process of undergoing merger activity is strengthened by the
overlying contours which show three distinct cores. The image of the
full WF2 chip reveals several more bright companions within a radius
of $\le100$ kpc.

This source has recently been imaged by the HST PC during the 3CR 
snapshot survey
(De Koff et al. 1996) through the F702W (wide $R$) filter. The
280-second integration presented by De Koff et al. confirms the complex
multiple structure of this object with the authors noting that the
optical and radio axes are aligned to within $15\deg$. McCarthy et al.
(1995) imaged this source both in the $R$-band and through an $H\alpha$
emission-line filter, detecting a curving filament of extended
$H\alpha$ emission stretching some 12$\asec$ to the NW.

\vspace*{0.07in}

\noindent
{\bf1215$-$033} (PKS 1215$-$033)

\vspace*{0.02in}
\noindent
The $R$-band image of this source shows it to be a uniform round galaxy
with no obvious signs of interaction or disturbance. There are two
companion objects lying just on the edge of the detail shown in
Fig A3, to the SW and ESE at projected distances of
$\simeq60$ kpc.

\vspace*{0.07in}

\noindent
{\bf1215$+$013} (PKS 1215+013)

\vspace*{0.02in}
\noindent
The $R$-band image of this source shown in Fig A4 
reveals numerous faint companion
objects around the central galaxy, the brightest lying $\simeq30$ kpc
to the South. The overlying contours for this source suggest some sort
of disturbance immediately to the North of the galaxy core.

\vspace*{0.07in}

\noindent
{\bf1330$+$022} (3C 287.1, 4C $+$02.36)

\vspace*{0.02in}
\noindent
Numerous companion objects are seen in the new $R$-band image of this
source shown in Fig A5, 
the brightest of which can be seen some $25$ kpc to the NE. 
There is a linear tidal
feature $\simeq35$ kpc to the SW. The overlying contours
for this source show there to be an apparent second nucleus at less than
$1\asec$ separation to the WNW 

This object was in the sample imaged in the $R$-band during the 3CR
 snapshot survey
(De Koff et al. 1996). The 280-second image presented by De Koff et al.
confirms the presence of the second nucleus. 3C 287.1 has been shown
to have a power-law X-ray spectrum by Crawford \& Fabian (1995).

\vspace*{0.07in}

\noindent
{\bf1342$-$016} (PKS 1342$-$016, MRC 1342$-$016)

\vspace*{0.02in}
\noindent
The new HST image shows this galaxy to be large, luminous and
uniform. A bright foreground star is present $\simeq13\asec$ to the
SWW, on the edge of the frame shown in Fig A6.  No
obvious companion objects are present in the detail shown in Fig A6
although the full WF2 chip image shown in Fig 13 reveals there to 
be a large
number of companion objects, consistent with a moderately rich cluster.

\subsubsection{The Radio-Loud Quasars}

\vspace*{0.07in}
\noindent
{\bf1020$-$103} (PKS 1020$-$103, UT 1020$-$103)

\vspace*{0.02in}
\noindent
The $R$-band image of this source shown in Fig A7 reveals it to
have a comparatively small and faint host, with an obvious PA of $\simeq 
110$ degrees. There is a triangle of three faint companions directly to
the South. This object was identified as an X-ray source in the ROSAT 
All-Sky Survey (Brinkmann et al. 1997).

\vspace*{0.07in}
\noindent
{\bf1217$+$023} (PKS 1217$+$02, ON 029)

\vspace*{0.02in}
\noindent
The $R$-band image of this relatively nuclear-dominated object shown
in Fig A8 clearly
reveals an elliptical-looking host galaxy with an apparent PA of
$\approx100\deg$. There is a group of four faint apparent companion objects
running North-South at a projected distance of $\simeq40$ kpc.

\vspace*{0.07in}
\noindent
{\bf2135$-$147} (PKS 2135$-$14, PHL 1657)

\vspace*{0.02in}
\noindent
The image of this source shown in Fig A9 
shows a large companion galaxy at a projected
distance of $\simeq30$ kpc ESE and an apparent close-in companion, or
secondary nucleus, at a distance of $\simeq10$ kpc. A recent analysis of the
spectrum of the close-in companion by Canalizo
\& Stockton (1997) has shown that this object is actually a foreground
star.
This object has also been imaged in the $V$-band with HST by BKS. 
Their analysis also reveals the underlying host to be best matched by
an elliptical galaxy model. This quasar was identified as an X-ray
source in the ROSAT All-Sky Survey (Brinkmann et al. 1997).
 
\vspace*{0.07in}
\noindent
{\bf2355$-$082} (PKS 2355$-$082, PHL 6113)

\vspace*{0.02in}
\noindent
The new $R$-band HST image of this source in Fig A10
reveals the presence of a
apparently early-type host with a PA of $\approx180\deg$. A group of
five small companion objects can be seen to the NE. This object was 
identified as an X-ray source in the ROSAT All-Sky Survey 
(Brinkmann et al. 1997).

\subsubsection{The Radio-Quiet Quasars}

\vspace*{0.07in}

\noindent
{\bf0052$+$251} (PG 0052$+$251)

\vspace*{0.02in}
\noindent
Although not very obvious in the grey-scale representation used in Fig A11,
the host galaxy of this quasar can be seen to have clear spiral
structure in the raw $R$-band image. There are two 
spiral arms present
to the East and West of the nucleus with the Eastern arm being more
extended and apparently terminating in a companion object. The
overlying contours for this image reveal it to
have a highly-luminous nuclear component. 

This object has recently been imaged in the $H$-band by McLeod \& Rieke
(1994) and in $J$-, $H$- and $K$-band by Hutchings \& Neff (1997). Hutchings
\& Neff performed an analysis of the numbers, magnitudes and colours of the
companion objects of this source, covering an angular extent
comparable to the full WF2 image. They
identify some 22 companion
objects brighter than $K=19.6$, the vast majority of which have colours
consistent with mature stellar populations. Comparing their
near-infrared images with previous optical studies, Hutchings \& Neff
note that there is no evidence for tidal structure made up of old
stars, but conclude that the host is in a phase of secondary star
formation possibly induced by interaction with one of the close group
of companions. 0052+251 was also included in the $V$--band HST imaging
programme of Bahcall, Kirhakos \& Schneider, who list the host morphology
as spiral and identify many of the knots seen in the Eastern arm with
H{\sc ii} regions. Interestingly Bahcall, Kirhakos \& Schneider 
comment that the
inner regions of the surface-brightness profile of 0052+251 are well
matched by an $r^{1/4}$ law, consistent with the best-fit elliptical
host from the $K$-band imaging of Taylor et al. (1996), and suggestive
that there may be a strong bulge component to the host galaxy. This is of 
course precisely what we have found through the construction pf a combined
disc+bulge model for the galaxy as detailed in Section 5. Despite the fact 
that the eye is drawn to the high surface-brightness spiral arms in the HST image, we find that in fact 70\% of its total $R$-band light is contributed by
the spheroidal component.
Finally, we note that this
quasar has also been identified as an X-ray source in the ROSAT All-Sky
Survey (Yuan et al. 1998).

\vspace*{0.07in}

\noindent
{\bf0204$+$292}

\vspace*{0.02in}
\noindent
In the new $R$-band image of this quasar the host galaxy appears to be
elliptical with a well-defined position angle of
$\approx90\deg$. There are two companion objects on the sub-image
shown in Fig A12, $\simeq70$ kpc to the WSW and
$\simeq45$ to the NW respectively. This quasar was identified as an
X-ray source in the ROSAT All-Sky Survey (Yuan et al. 1998).

\vspace*{0.07in}

\noindent
{\bf1549$+$203} (LB 906, 1E 15498$+$203)

\vspace*{0.02in}
\noindent
It is immediately obvious from the new $R$-band image of this object
that the host galaxy is small, and relatively faint compared to the
nuclear component. Using a different grey-scale to that used in 
Fig A13, there is a suggestion of spiral-like features
to the NW and SE of the nucleus, with the SE arm terminating at the
apparent companion object which can be seen $\simeq20$ kpc to the E of
the nucleus. A large, luminous elliptical galaxy can just be seen on
the edge of the frame to the SW, with numerous fainter companions
visible inside a radius of $\simeq60$ kpc. The full WF2 image would 
appear to show numerous companion objects although, as pointed out by
 Taylor et al. 1996, the density of the environment of
this quasar is uncertain due to the presence of a nearby foreground
cluster at $z\simeq0.14$ (Stocke et al. 1983).
The first electronic images of this quasar were presented by
Hutchings \& Neff (1992). They imaged the object in both the $V$- and
$I$-bands, detecting what looked like a bar structure running North-South
through the nucleus, while noting that the surface-brightness profile
of 1549+203 was not exponential. This quasar was identified as an X-ray 
source in the ROSAT All-Sky Survey (Yuan et al  1998).

\vspace*{0.07in}

\noindent
{\bf2215$-$037} (EX 2215$-$037)

\vspace*{0.02in}
\noindent
The image of this quasar shown in Fig A14 
reveals the host to be compact and apparently
undisturbed. A small apparently unresolved companion is detected
only $\simeq10$ kpc to the SW of the quasar nucleus. A large
elliptical galaxy can be seen some $65$ kpc to the North with a second
apparently unresolved companion $\simeq50$ kpc to the East. 
This object has been previously imaged on the HST PC using the F702W
(wide $R$) filter by Disney et al. (1995). Using a two-dimensional
cross-correlation modelling technique,  Disney et al. found the host galaxy
to be excellently matched by an elliptical galaxy model. Their wide-$R$
image confirms the existence of the unresolved companion at
$\simeq2\asec$ separation from the nucleus and suggests that the
environment of 2215$-$037 resembles a poor cluster. This suggestion is
supported by the image of the full WF2 chips which shows $\ge10$
companion objects inside a radius of $\simeq100$ kpc. This quasar was
identified as an X-ray source in the ROSAT All-Sky Survey (Yuan et al. 1998).

\clearpage

\begin{figure*}
\vspace{16cm}
\centering
\setlength{\unitlength}{1mm}
\includegraphics{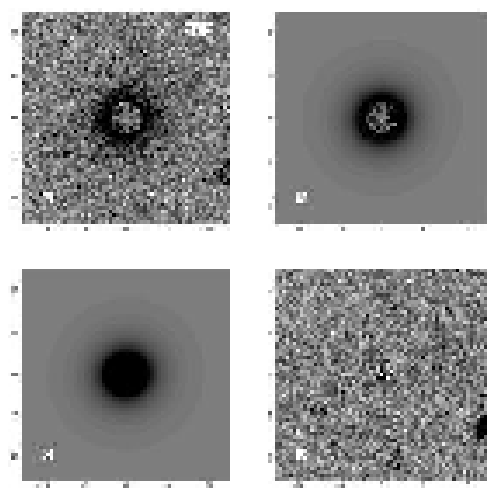}
\caption{The radio galaxy 0230$-$027}
\end{figure*}

\begin{figure*}
\vspace{16cm}
\centering
\setlength{\unitlength}{1mm}
\includegraphics{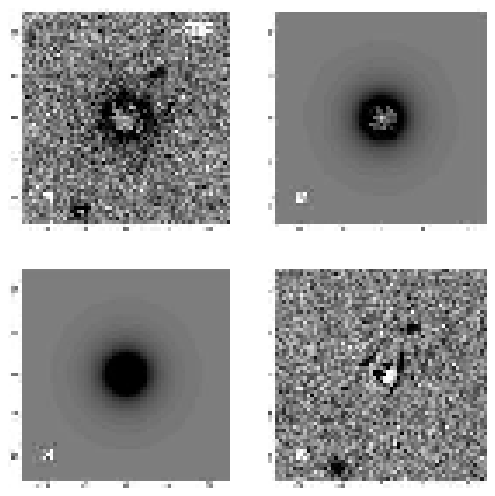}
\caption{The radio galaxy 0307$+$169}
\end{figure*}

\begin{figure*}
\vspace{16cm}
\centering
\setlength{\unitlength}{1mm}
\includegraphics{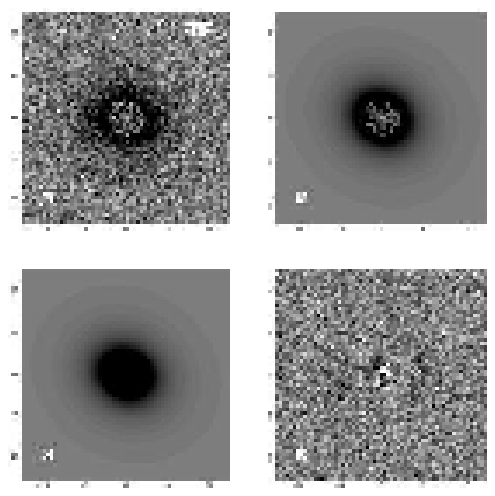}
\caption{The radio galaxy 1215$-$033}
\end{figure*}

\begin{figure*}
\vspace{16cm}
\centering
\setlength{\unitlength}{1mm}
\includegraphics{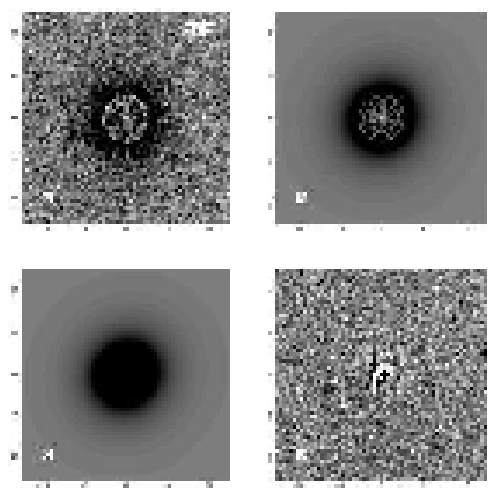}
\caption{The radio galaxy 1215$+$013}
\end{figure*}

\begin{figure*}
\vspace{16cm}
\centering
\setlength{\unitlength}{1mm}
\includegraphics{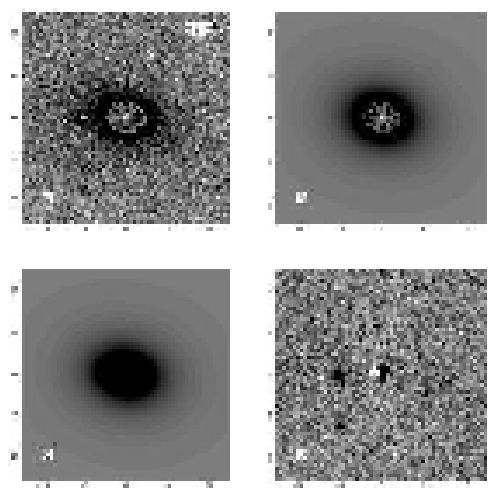}
\caption{The radio galaxy 1330$+$022}
\end{figure*}

\begin{figure*}
\vspace{16cm}
\centering
\setlength{\unitlength}{1mm}
\includegraphics{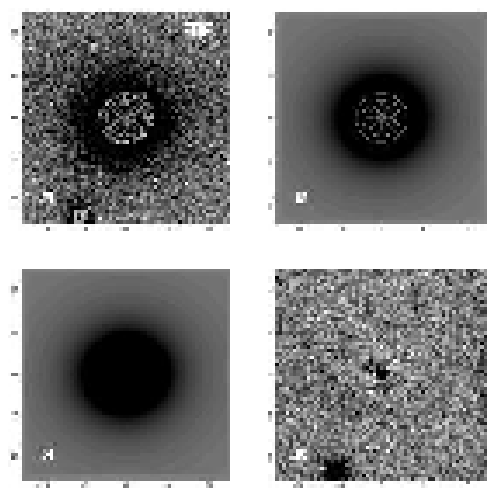}
\caption{The radio galaxy 1342$-$016}
\end{figure*}

\begin{figure*}
\vspace{16cm}
\centering
\setlength{\unitlength}{1mm}
\includegraphics{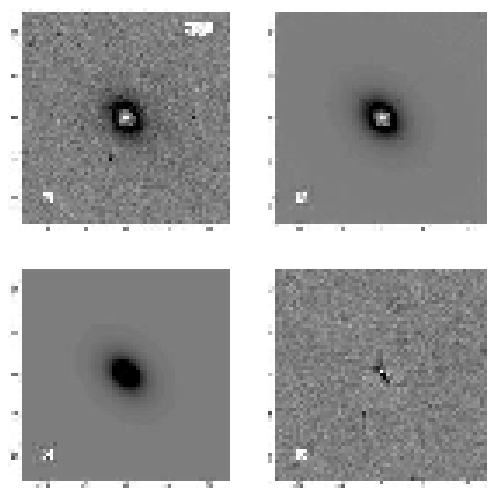}
\caption{The radio-loud quasar 1020$-$103}
\end{figure*}

\begin{figure*}
\vspace{16cm}
\centering
\setlength{\unitlength}{1mm}
\includegraphics{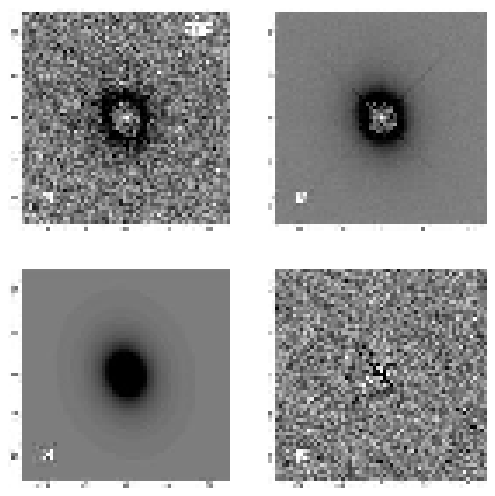}
\caption{The radio-loud quasar 1217$+$023}
\end{figure*}

\begin{figure*}
\vspace{16cm}
\centering
\setlength{\unitlength}{1mm}
\includegraphics{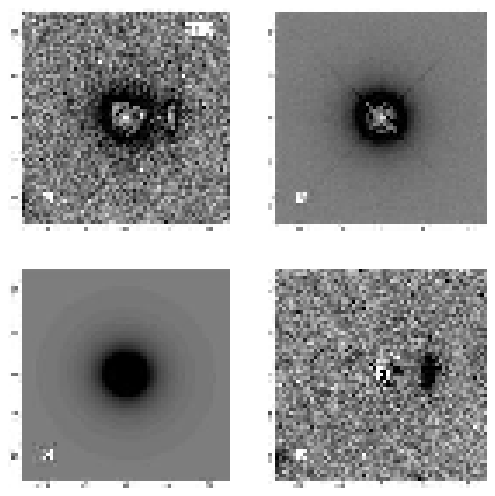}
\caption{The radio-loud quasar 2135$-$147}
\end{figure*}

\begin{figure*}
\vspace{16cm}
\centering
\setlength{\unitlength}{1mm}
\includegraphics{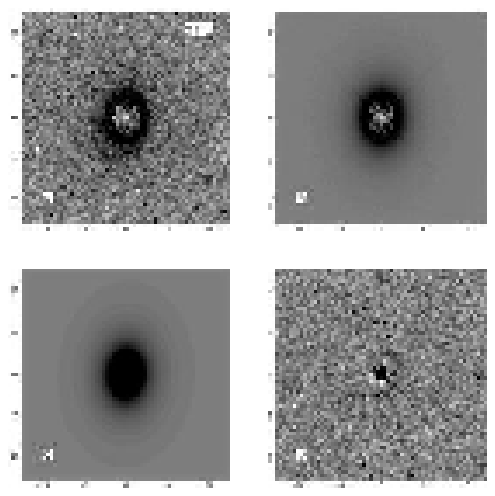}
\caption{The radio-loud quasar 2355$-$082}
\end{figure*}

\begin{figure*}
\vspace{16cm}
\centering
\setlength{\unitlength}{1mm}
\includegraphics{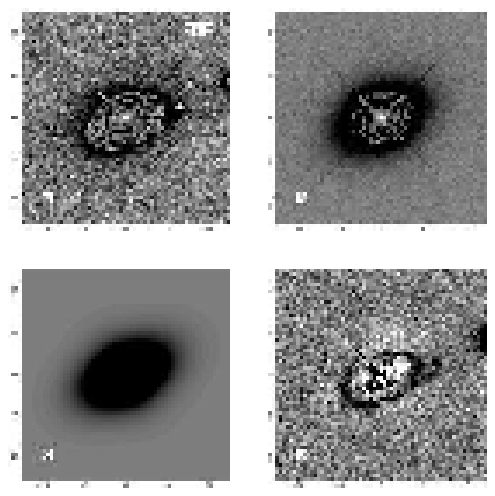}
\caption{The radio-quiet quasar 0052$+$251}
\end{figure*}

\begin{figure*}
\vspace{16cm}
\centering
\setlength{\unitlength}{1mm}
\includegraphics{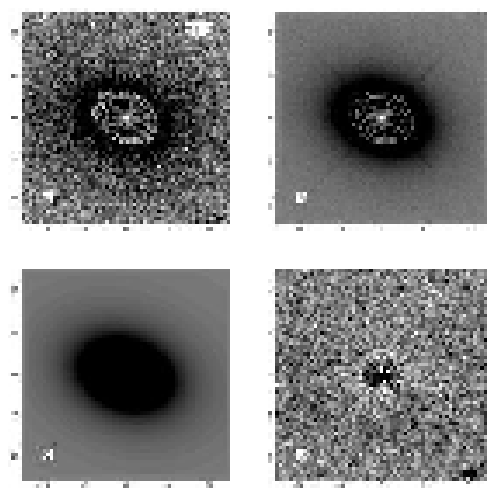}
\caption{The radio-quiet quasar 0204$+$292}
\end{figure*}

\begin{figure*}
\vspace{16cm}
\centering
\setlength{\unitlength}{1mm}
\includegraphics{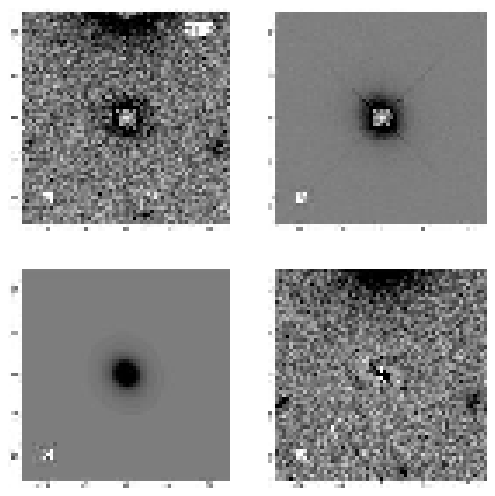}
\caption{The radio-quiet quasar 1549$+$203}
\end{figure*}

\begin{figure*}
\vspace{16cm}
\centering
\setlength{\unitlength}{1mm}
\includegraphics{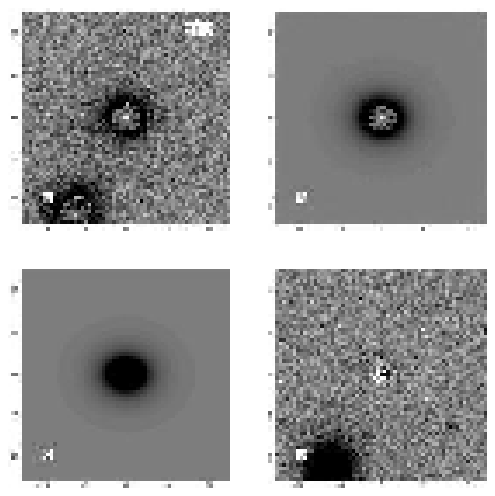}
\caption{The radio-quiet quasar 2215$-$037}
\end{figure*}

\clearpage

\section{Luminosity Profiles}

In this appendix we provide the observed and best-fitting
model luminosity profiles for 14 AGN for which the 
new HST data are reported in this paper.
The profiles are followed out to a radius of 10$^{\prime \prime}$, which
is representative of the typical outer radius used in the image modelling.
In each figure the azimuthally averaged data are indicated by circles, the azimuthal average of the best-fitting 2-dimensional model is shown by 
the solid line, and the dotted line indicates the contribution made to 
the surface-brightness profile by the point-source component of the model
(after convolution with the PSF).

\newpage

\begin{figure}
\centering
\setlength{\unitlength}{1mm}
\vspace{6.7cm}
\includegraphics{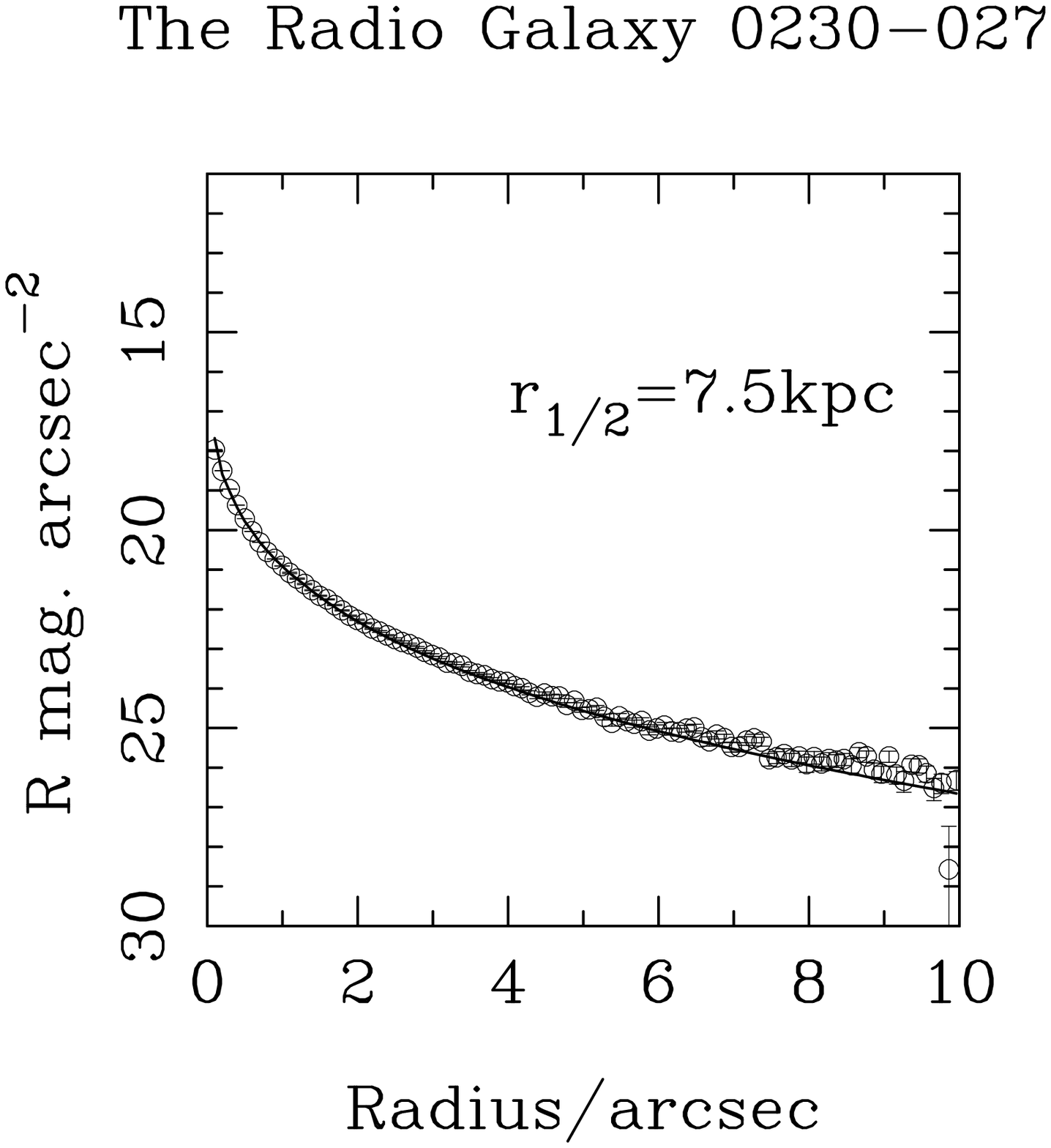}
\caption{}
\end{figure}

\begin{figure}
\centering
\setlength{\unitlength}{1mm}
\vspace{6.7cm}
\includegraphics{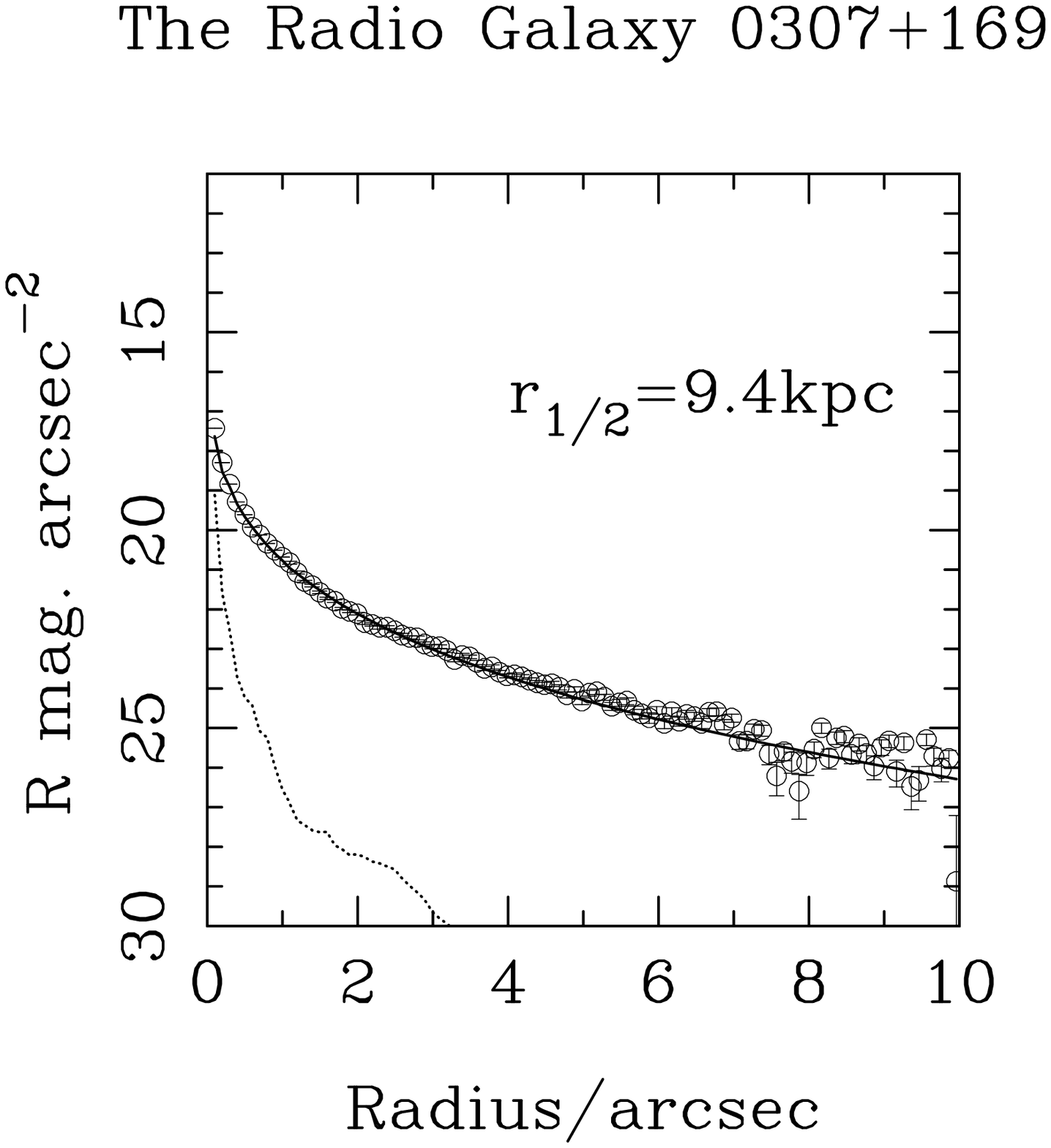}
\caption{}
\end{figure}

\begin{figure}
\centering
\setlength{\unitlength}{1mm}
\vspace{6.7cm}
\includegraphics{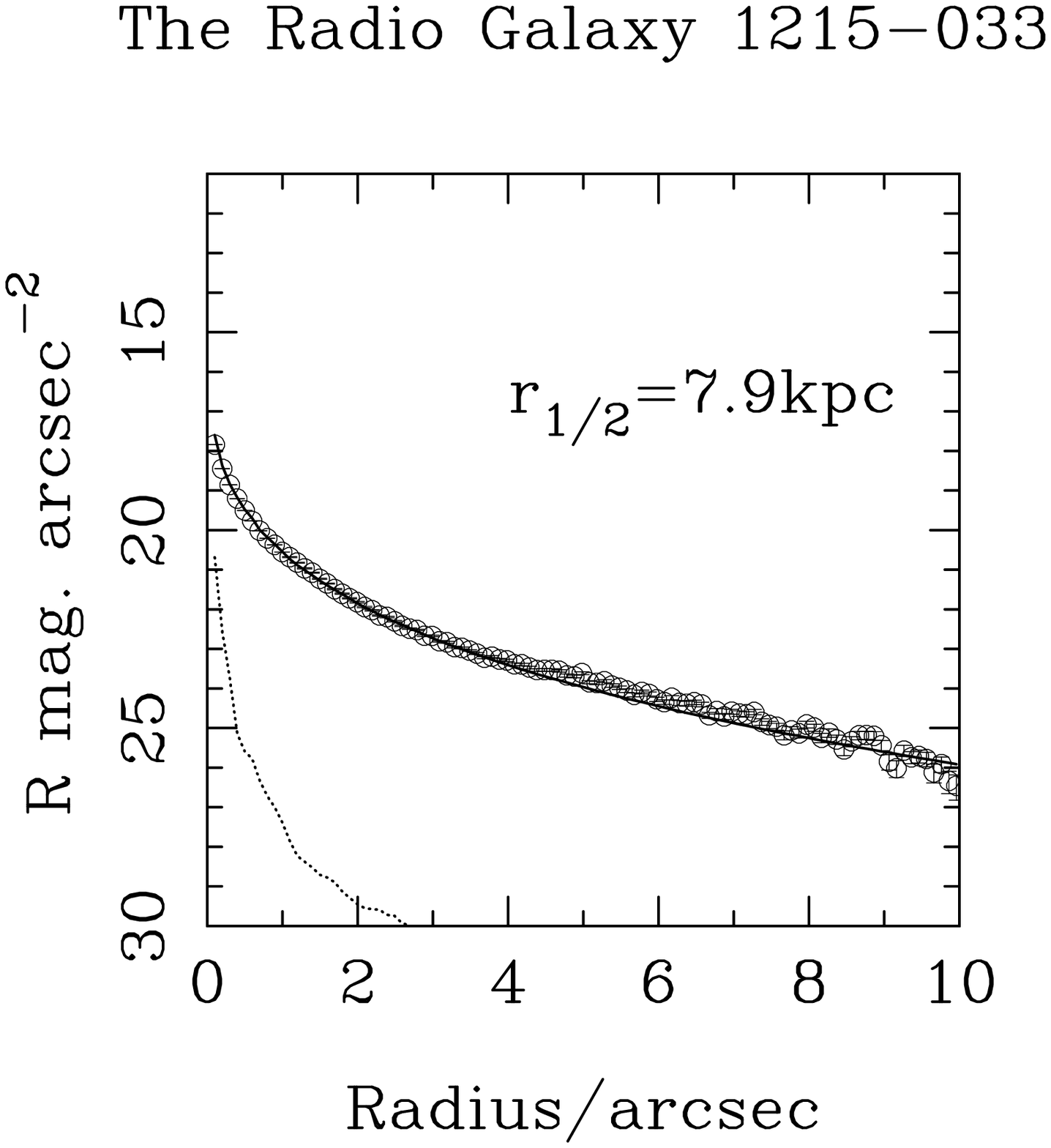}
\caption{}
\end{figure}

\clearpage

\begin{figure}
\centering
\setlength{\unitlength}{1mm}
\vspace{6.7cm}
\includegraphics{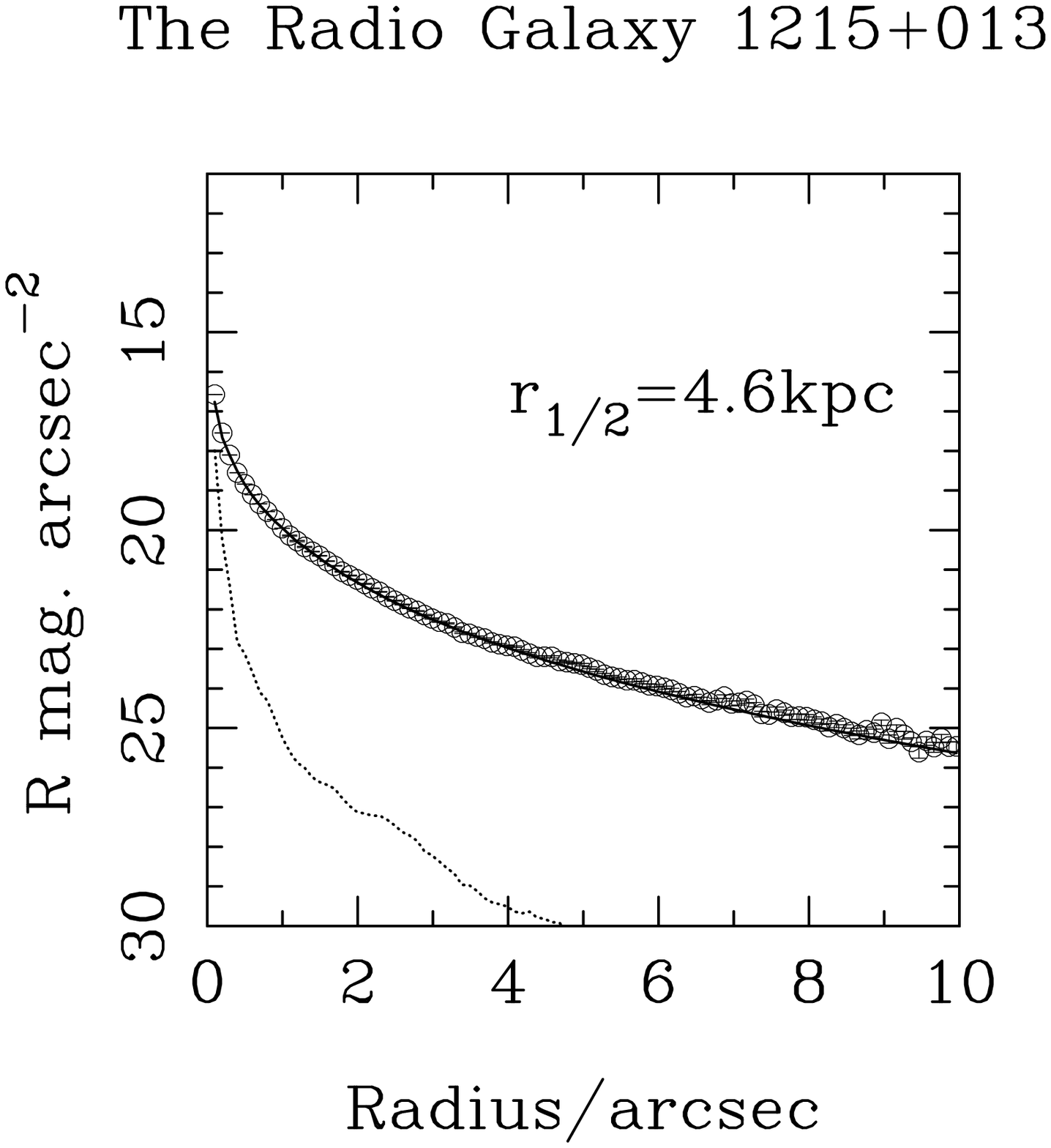}
\caption{}
\end{figure}

\begin{figure}
\centering
\setlength{\unitlength}{1mm}
\vspace{6.7cm}
\includegraphics{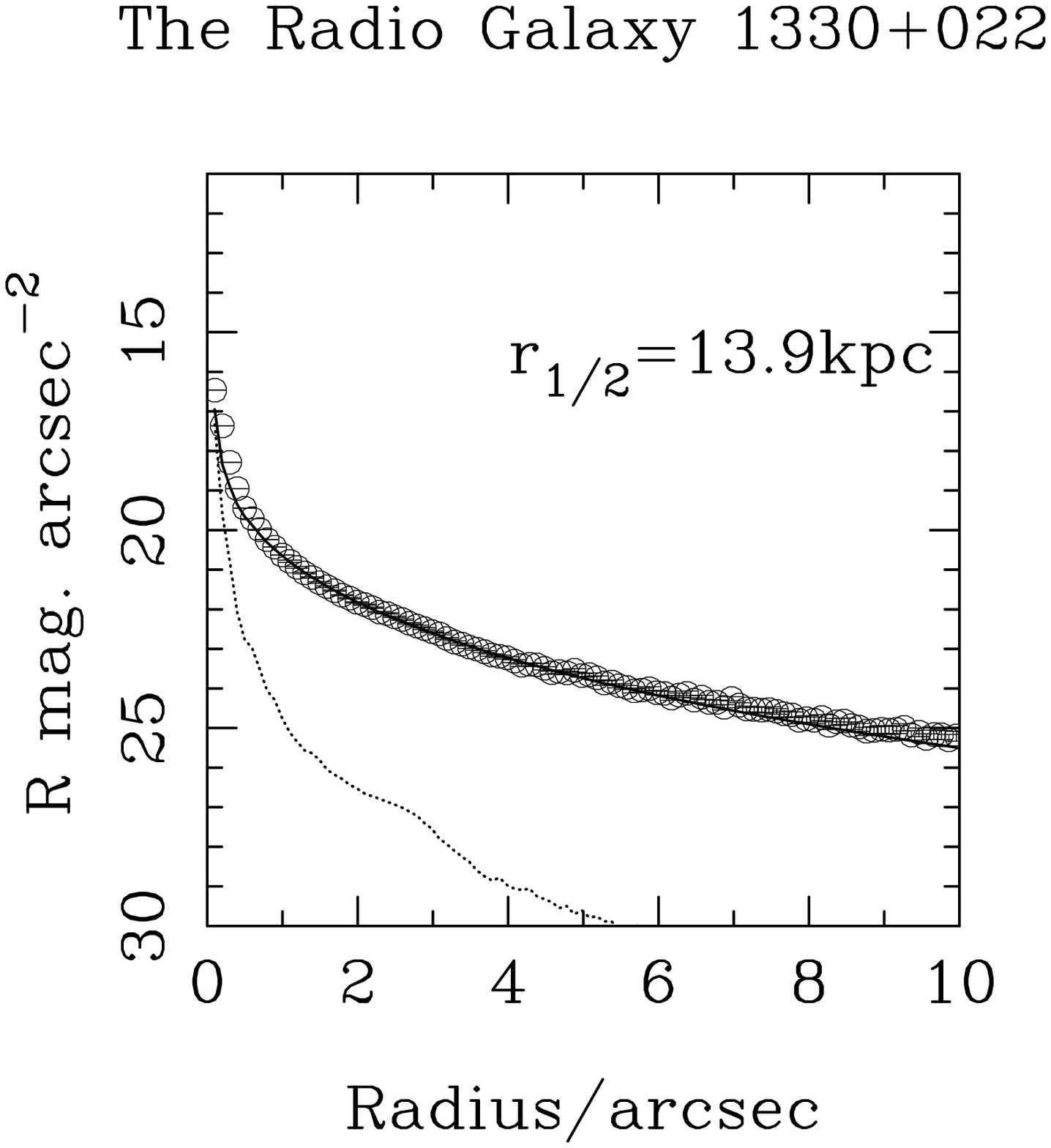}
\caption{}
\end{figure}

\begin{figure}
\centering
\setlength{\unitlength}{1mm}
\vspace{6.7cm}
\includegraphics{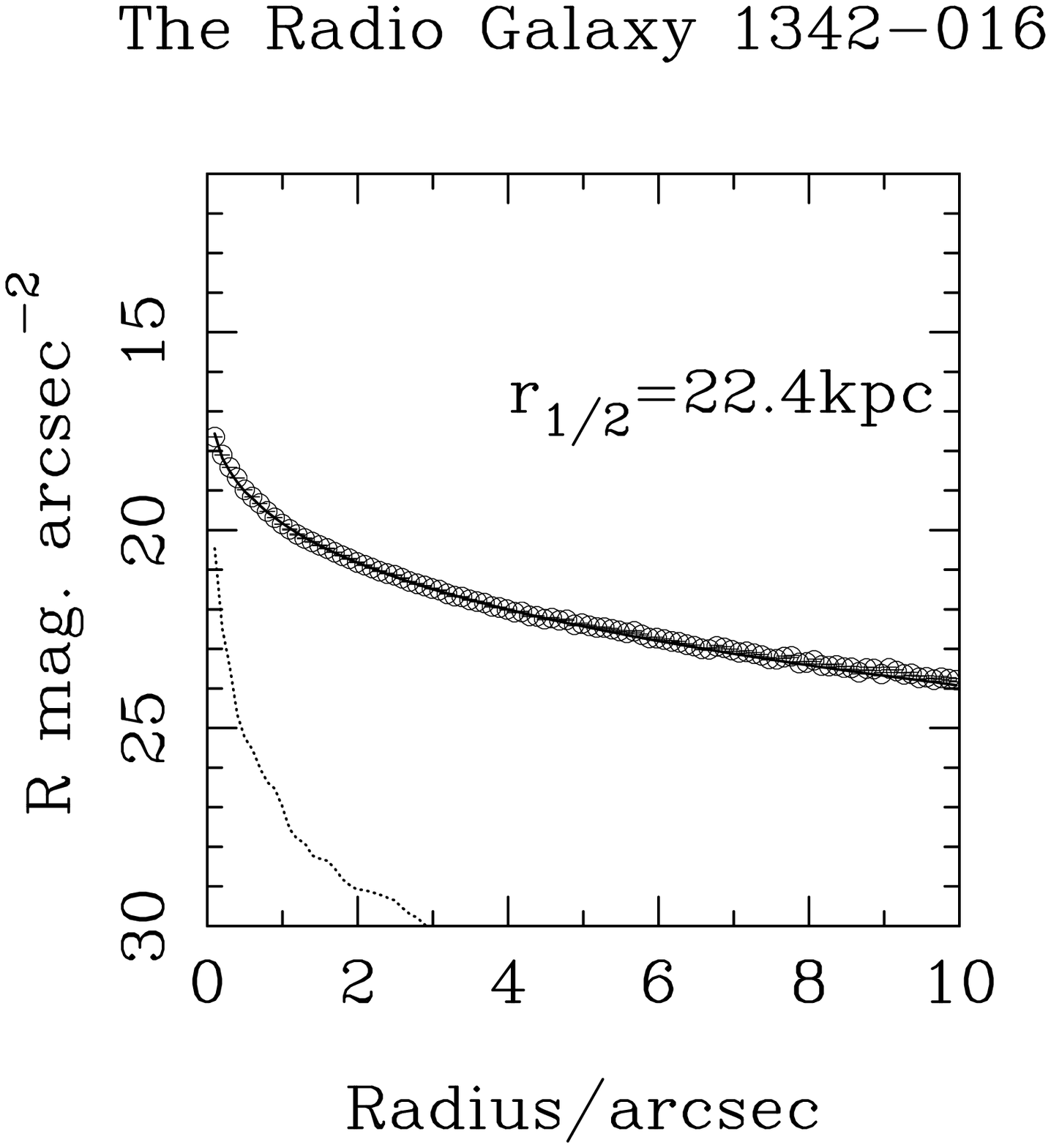}
\caption{}
\end{figure}

\begin{figure}
\centering
\setlength{\unitlength}{1mm}
\vspace{6.7cm}
\includegraphics{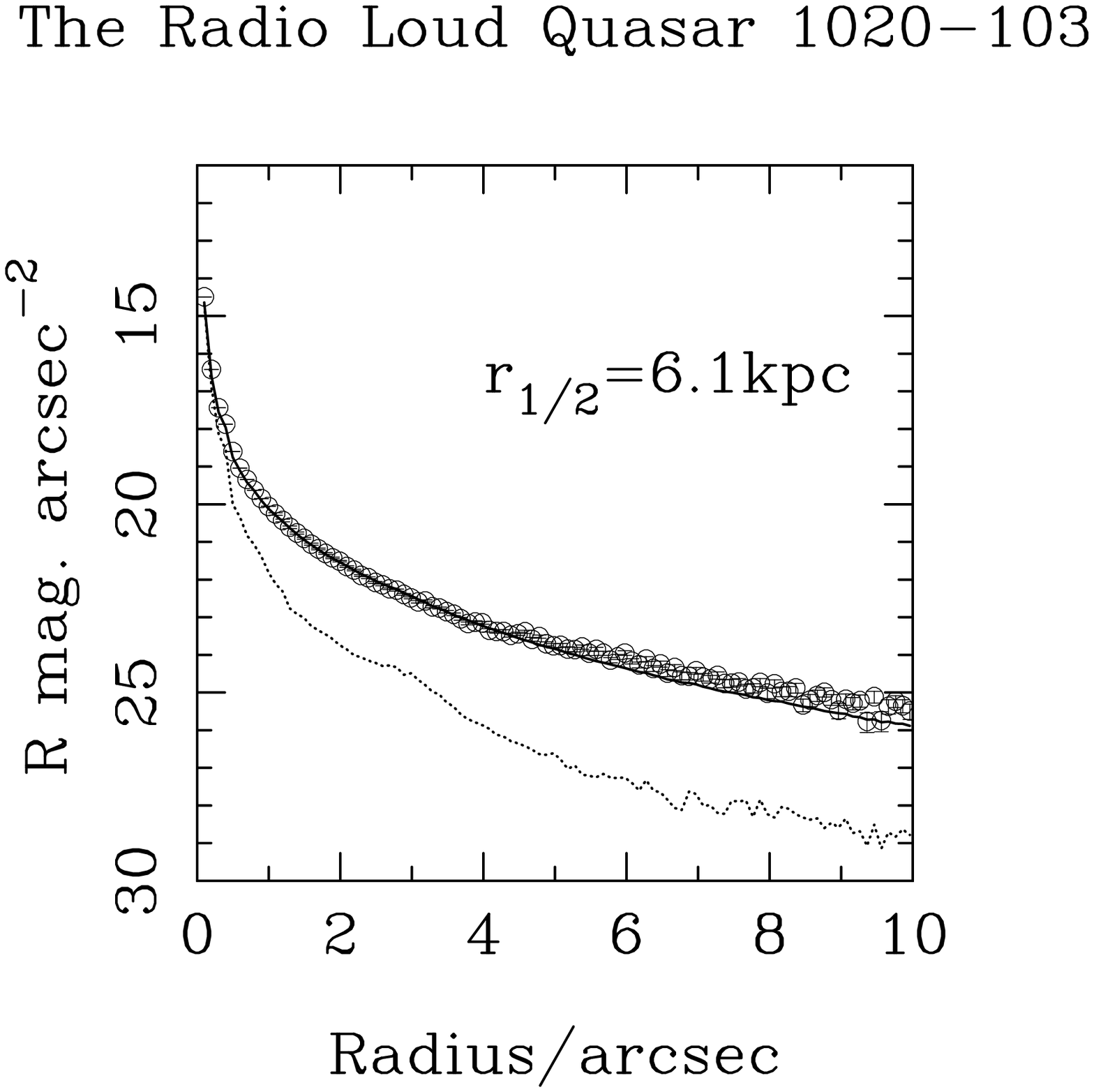}
\caption{}
\end{figure}

\begin{figure}
\centering
\setlength{\unitlength}{1mm}
\vspace{6.7cm}
\includegraphics{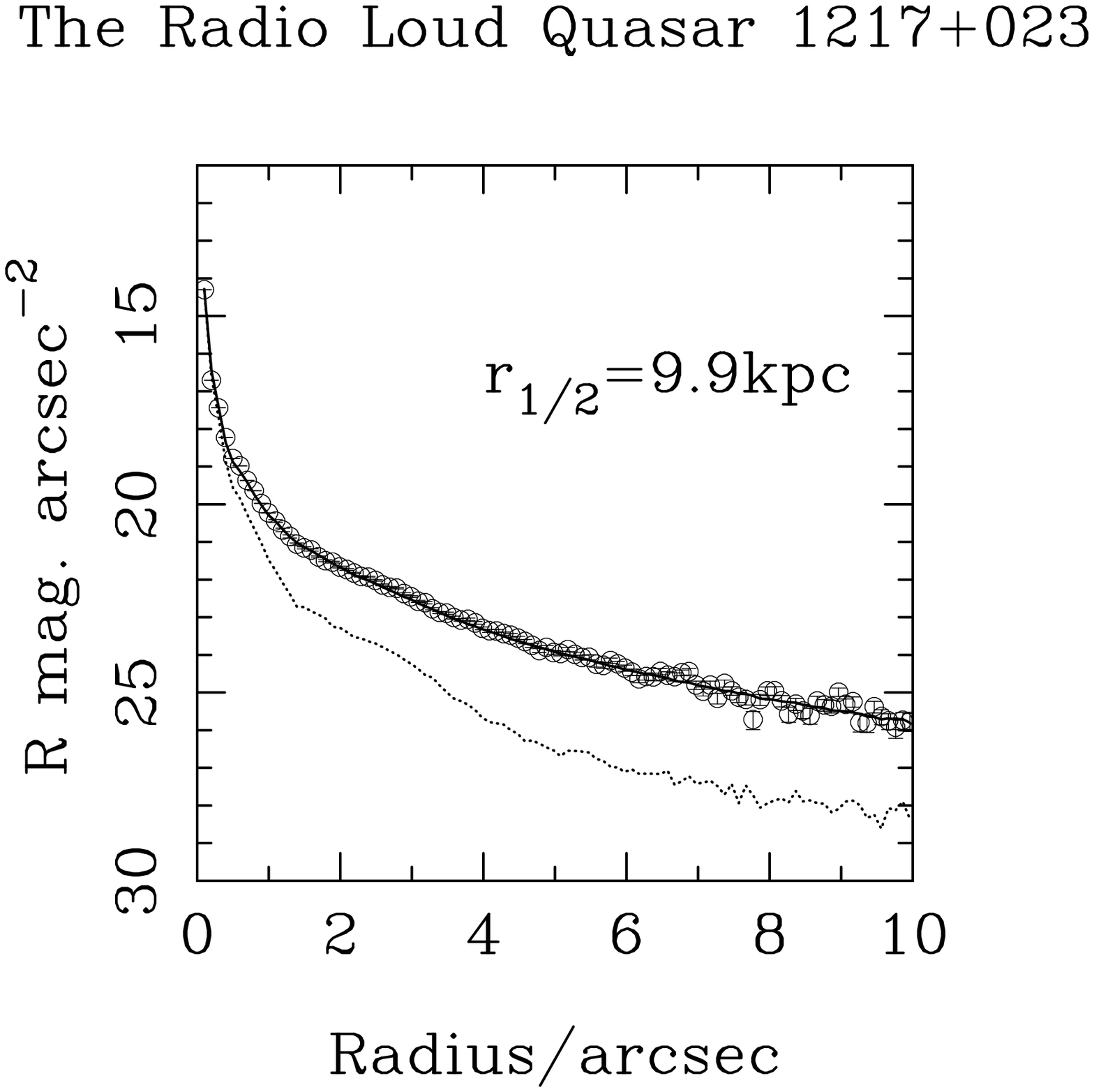}
\caption{}
\end{figure}

\begin{figure}
\centering
\setlength{\unitlength}{1mm}
\vspace{6.7cm}
\includegraphics{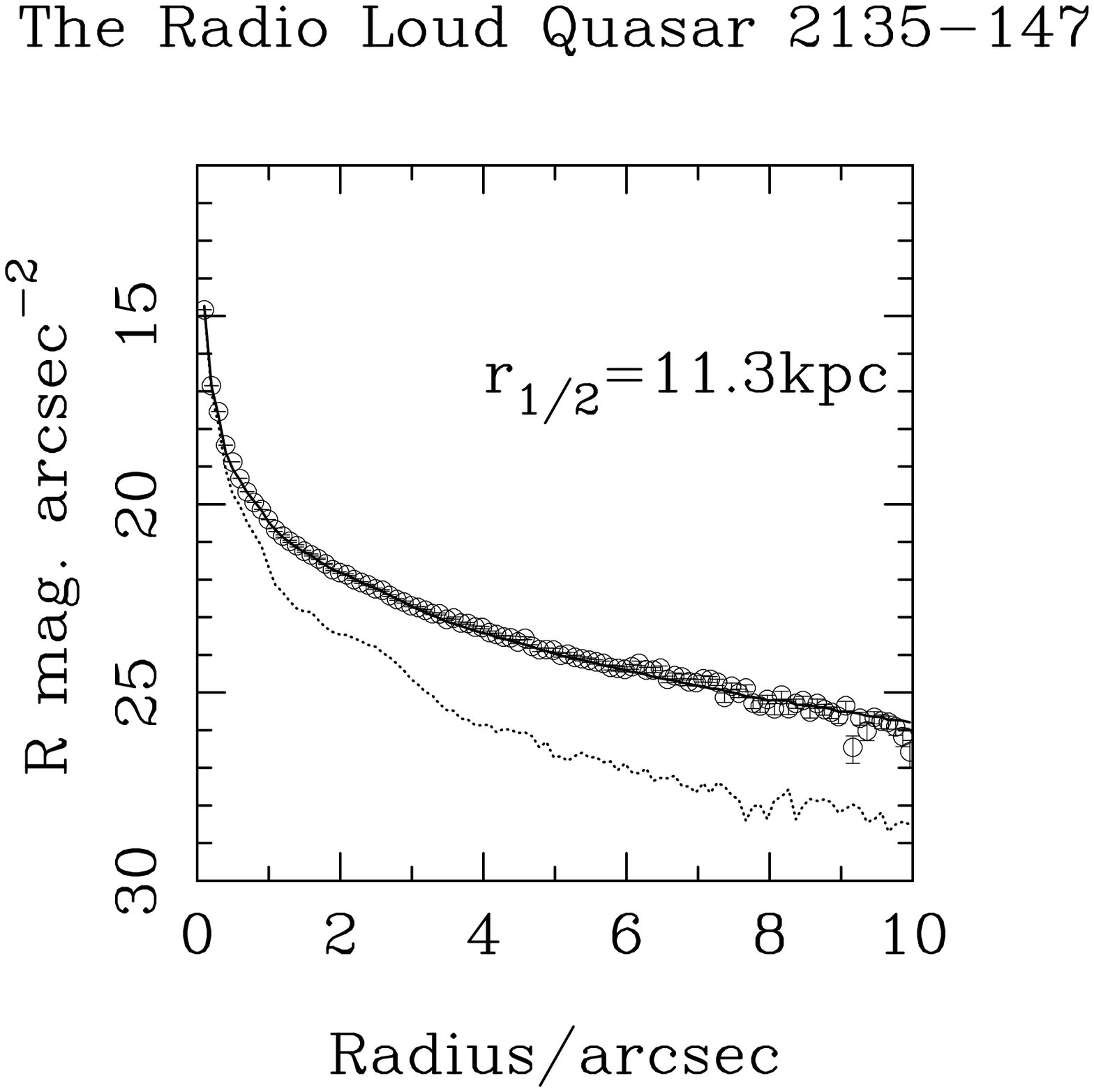}
\caption{}
\end{figure}

\clearpage

\begin{figure}
\centering
\setlength{\unitlength}{1mm}
\vspace{6.7cm}
\includegraphics{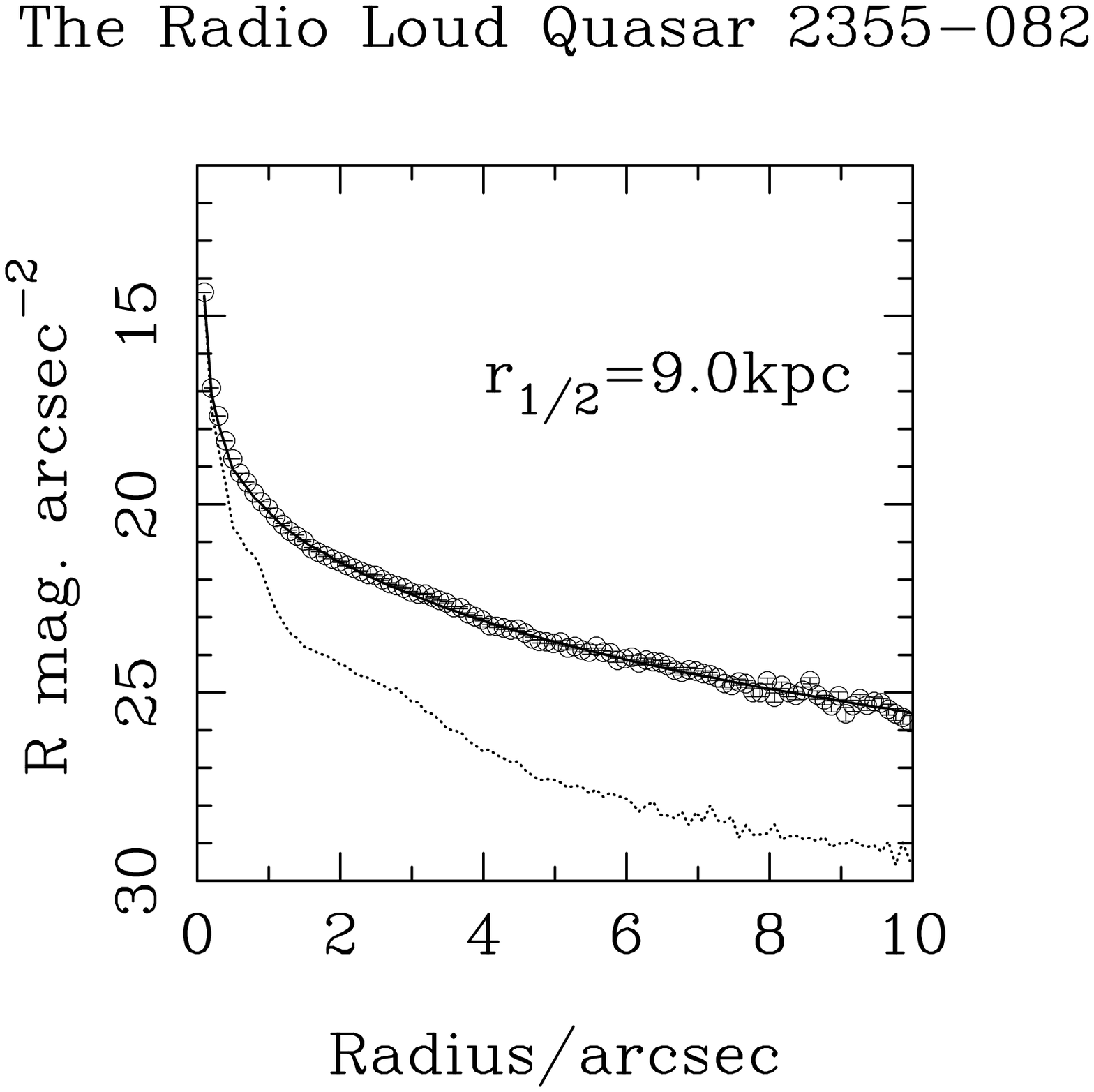}
\caption{}
\end{figure}

\begin{figure}
\centering
\setlength{\unitlength}{1mm}
\vspace{6.7cm}
\includegraphics{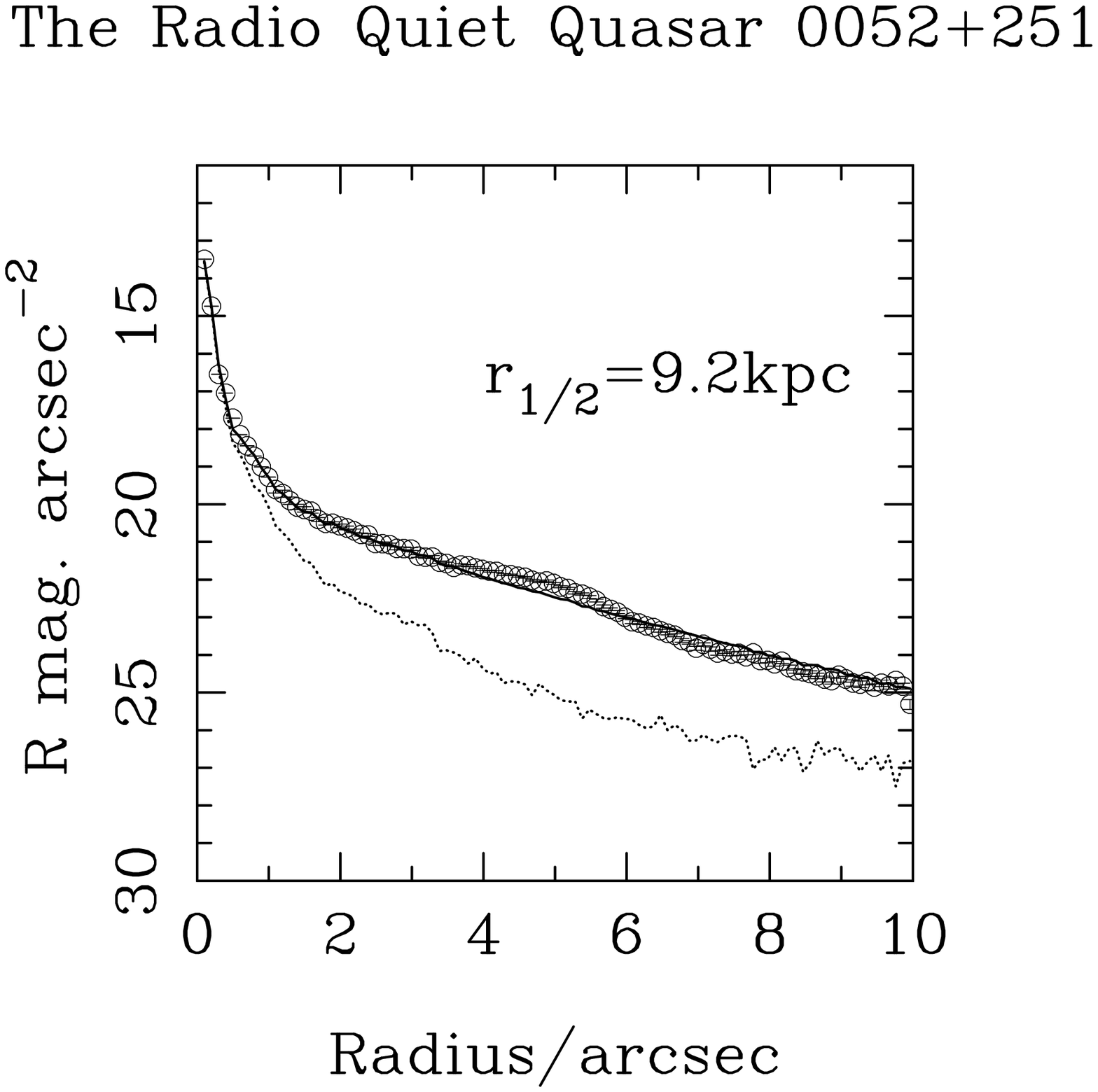}
\caption{}
\end{figure}

\begin{figure}
\centering
\setlength{\unitlength}{1mm}
\vspace{6.7cm}
\includegraphics{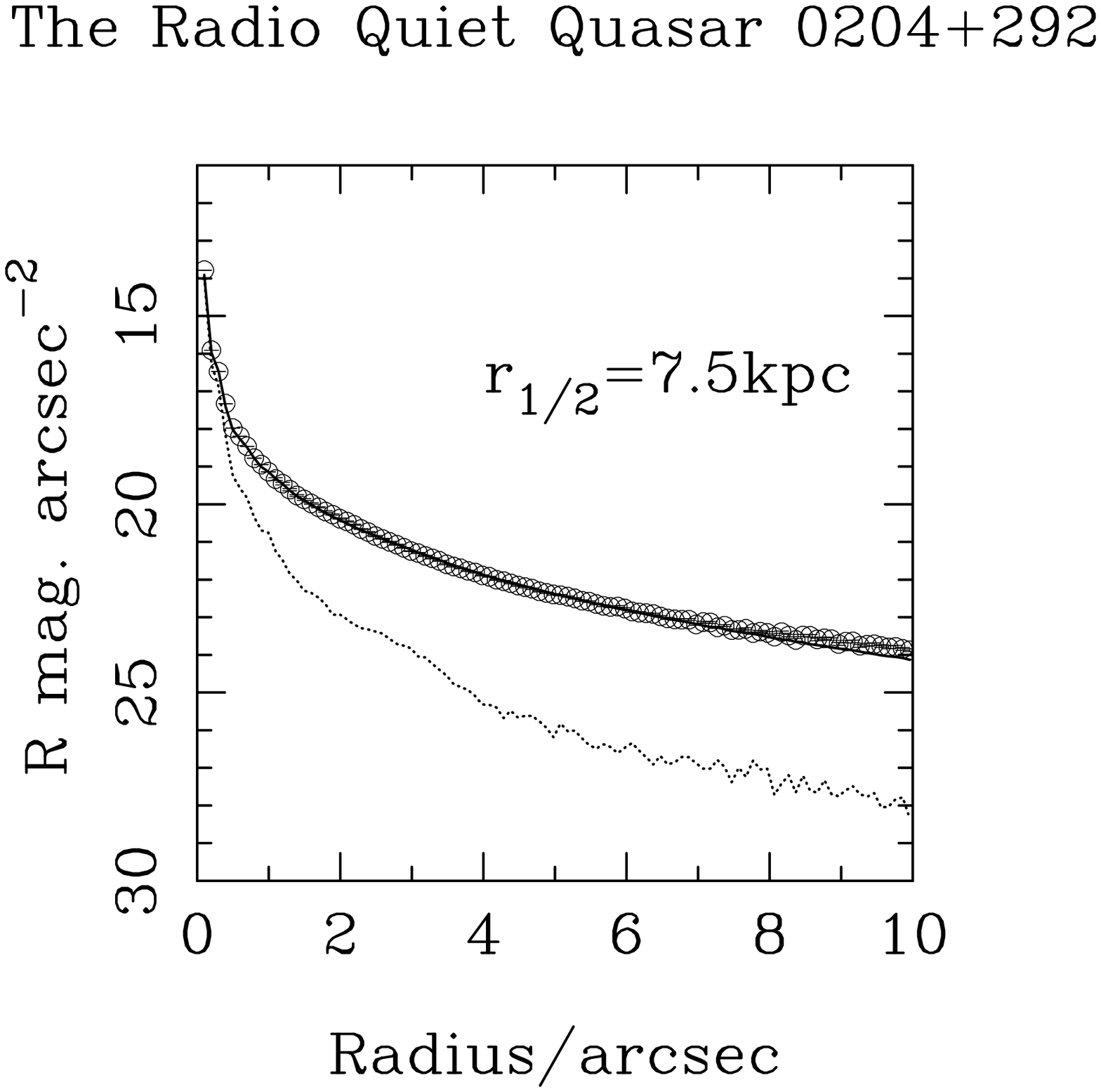}
\caption{}
\end{figure}

\begin{figure}
\centering
\setlength{\unitlength}{1mm}
\vspace{6.7cm}
\includegraphics{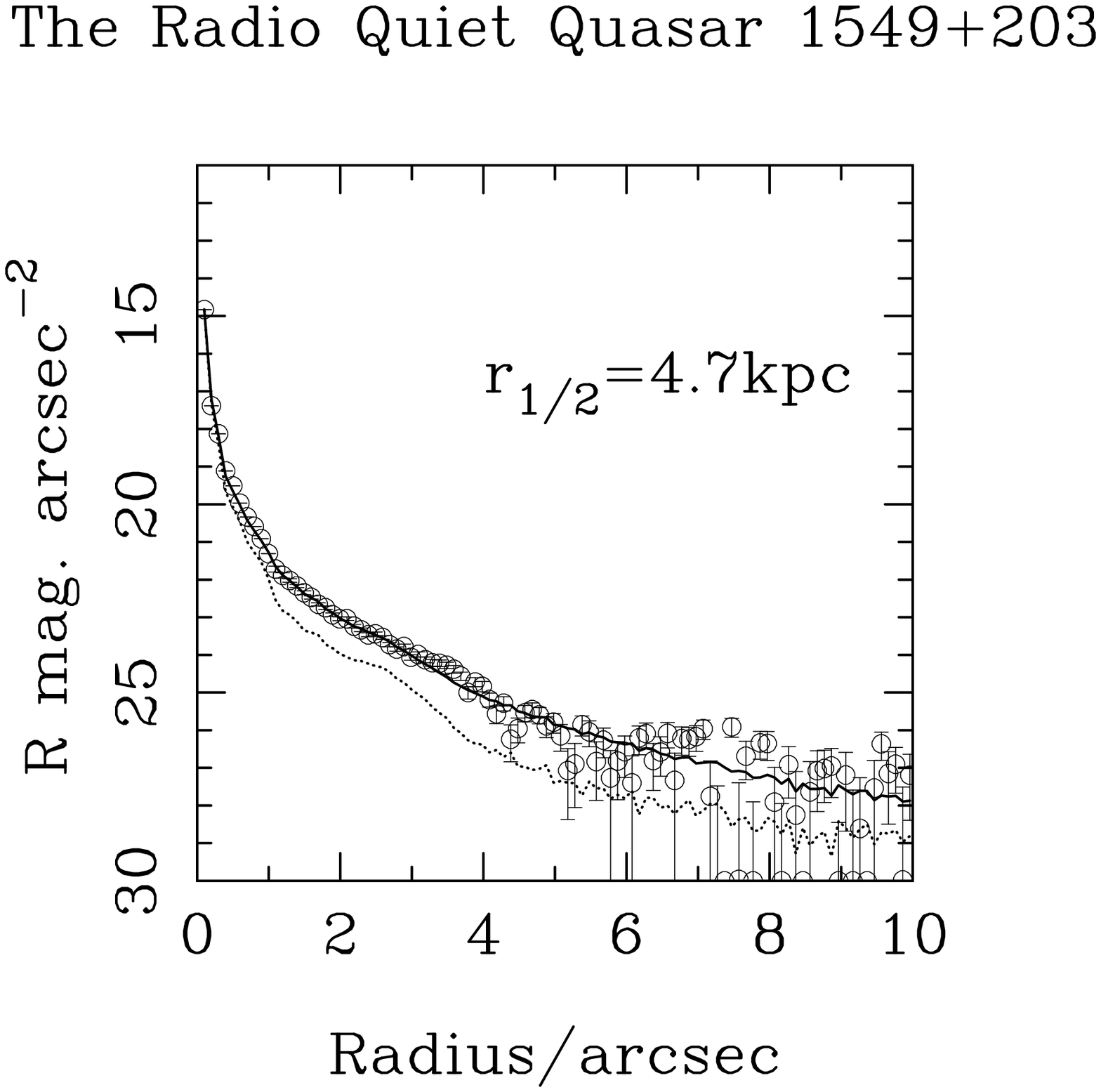}
\caption{}
\end{figure}

\begin{figure}
\centering
\setlength{\unitlength}{1mm}
\vspace{6.7cm}
\includegraphics{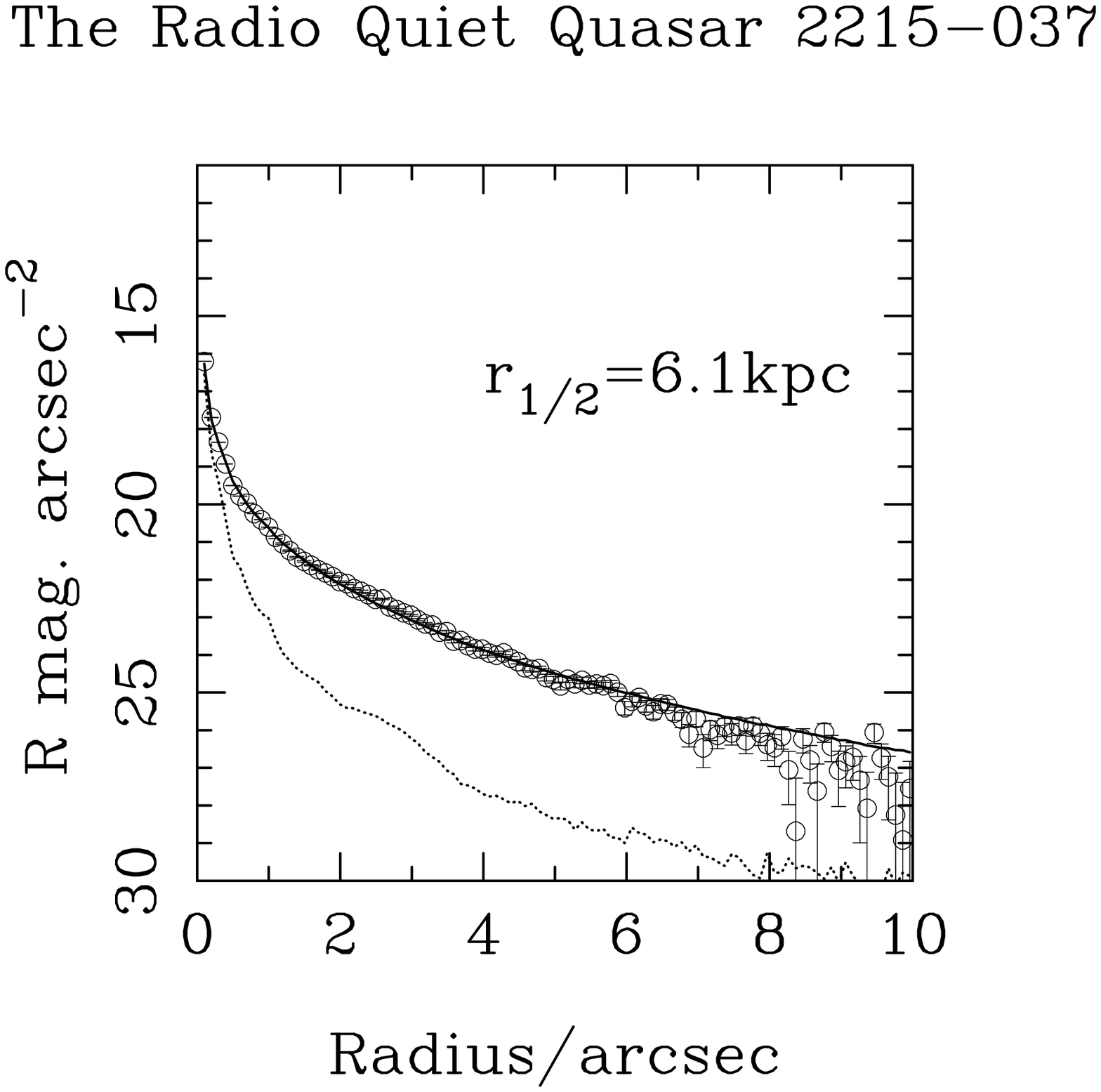}
\caption{}
\end{figure}

\end{document}